\documentclass[fleqn,usenatbib]{mnras}
\usepackage[T1]{fontenc}
\usepackage[english]{babel}
\usepackage{txfonts}

\usepackage{graphicx}	

\def\l{\ensuremath{\lambda}}

\def\t{$\times$}

\def\cd{cycle d$^{-1}$}
\def\md{mag\,d$^{-1}$}

\newcommand{\Msun}{$M_{\sun}$}

\newcommand{\kms}{km\,s$^{-1}$}
\newcommand{\myemail}{vitaly@neustroev.net}
\newcommand{\iraf}  {{\sc iraf}}

\newcommand {\bc}{\begin {center}}
\newcommand {\ec}{\end {center}}
\newcommand {\be}{\begin {equation}}
\newcommand {\ee}{\end {equation}}
\newcommand {\beq}{\begin {eqnarray}}
\newcommand {\eeq}{\end {eqnarray}}

\newcommand{\Halpha} {H$\alpha$}
\newcommand{\Hbeta}  {H$\beta$}
\newcommand{\Hgamma} {H$\gamma$}
\newcommand{\Hdelta} {H$\delta$}

\newcommand{\CIII} {C\,{\sc iii}}
\newcommand{\CIV}  {C\,{\sc iv}}

\newcommand{\HeI}  {He\,{\sc i}}
\newcommand{\HeII} {He\,{\sc ii}}

\newcommand{\LiI}  {Li\,{\sc i}}

\newcommand{\NaI}  {Na\,{\sc i}}
\newcommand{\NaD}    {Na\,D}
\newcommand{\NIII} {N\,{\sc iii}}

\newcommand{\SSSJ} {SSS J122221.7$-$311525}
\newcommand{\SSS} {SSS J122222}

\title[The superoutburst of \SSSJ]
{
 The remarkable outburst of the highly evolved post-period-minimum dwarf nova \SSSJ
\thanks{This paper includes data gathered with the 6.5 m Magellan Telescopes located at Las Campanas Observatory, Chile.}
}

\author[V.~V.~Neustroev et al.]{V.~V.~Neustroev$^{1,2}$\thanks{E-mail:
\myemail},
T.~R.~Marsh$^3$, S.~V.~Zharikov$^4$, C.~Knigge$^5$, E.~Kuulkers$^6$,
\newauthor
J.~P.~Osborne$^7$, K.~L.~Page$^7$, D.~Steeghs$^3$, V.~F.~Suleimanov$^{8}$, G.~Tovmassian$^4$,
\newauthor
E.~Breedt$^{3}$, A.~Frebel$^{9}$, Ma.~T.~Garc{\'i}a-D{\'i}az$^4$, F.-J.~Hambsch$^{10,11}$, H.~Jacobson$^{9}$,
\newauthor
S.~G.~Parsons$^{12}$, T.~Ryu$^{13,14}$, L.~Sabin$^4$, G.~Sjoberg$^{15,11}$, A.~S.~Miroshnichenko$^{16}$,
\newauthor
D.~E.~Reichart$^{17}$, J.~B.~Haislip$^{17}$, K.~M.~Ivarsen$^{17}$, A.~P.~LaCluyze$^{17}$, J.~P.~Moore$^{17}$
\\
$^{1}$Finnish Centre for Astronomy with ESO (FINCA), University of Turku, V\"{a}is\"{a}l\"{a}ntie 20,
FIN-21500 Piikki\"{o}, Finland\\
$^{2}$Astronomy research unit, PO Box 3000, FIN-90014 University of Oulu, Finland\\
$^{3}$Department of Physics, University of Warwick, Gibbet Hill Road, Coventry CV4 7AL, UK\\
$^{4}$Instituto de Astronomia, Universidad Nacional Autonoma de Mexico, Apdo. Postal 877, Ensenada, Baja California, 22800 Mexico\\
$^{5}$School of Physics and Astronomy, University of Southampton, Southampton, SO17 1BJ, UK\\
$^{6}$European Space Astronomy Centre (ESA/ESAC), Science Operations Department, 28691 Villanueva
de la Ca\~{n}ada, Madrid, Spain\\
$^{7}$Department of Physics \& Astronomy, University of Leicester, University Rd, Leicester, LE1 7RH, UK\\
$^{8}$Institut f\"ur Astronomie und Astrophysik, Universit\"at T\"ubingen, Sand 1, D-72076 T\"ubingen, Germany\\
$^{9}$Department of Physics and Kavli Institute for Astrophysics and Space Research, Massachusetts Institute of Technology,
Cambridge, MA 02139, USA\\
$^{10}$Vereniging voor Sterrenkunde, Oude Bleken 12, B-2400 Mol, Belgium\\
$^{11}$American Association of Variable Star Observers, 49 Bay State Road, Cambridge, MA 02138, USA\\
$^{12}$Departmento de F{\'i}sica y Astronom{\'i}a, Universidad de Valpara{\'i}so, Avenida Gran Bretana 1111,
Valpara{\'i}so, 2360102, Chile\\
$^{13}$SOKENDAI, The Graduate University for Advanced Studies, 2-21-1 Osawa, Mitaka, Tokyo 181-8588, Japan\\
$^{14}$National Astronomical Observatory of Japan, 2-21-1 Osawa, Mitaka, Tokyo 181-8588, Japan\\
$^{15}$The George-Elma Observatory, New Mexico Skies, 9 Contentment Crest, \#182, Mayhill, NM 88339, USA\\
$^{16}$University of North Carolina at Greensboro, Greensboro, NC 27402, USA \\
$^{17}$Department of Physics and Astronomy, University of North Carolina, Chapel Hill, NC 27599, USA
}

\date{Accepted 2017 January 11. Received 2016 December 23; in original form 2016 May 29}

\pubyear{2017}

\begin{document}
\label{firstpage}
\pagerange{\pageref{firstpage}--\pageref{lastpage}}
\maketitle

\begin{abstract}
We report extensive 3-yr multiwavelength observations of the WZ~Sge-type dwarf nova \SSSJ\ during its
unusual double superoutburst, the following decline and in quiescence. The second segment of the
superoutburst had a long duration of 33 d and a very gentle decline with a rate of 0.02 mag~d$^{-1}$,
and it displayed an extended post-outburst decline lasting at least 500 d. Simultaneously with the
start of the rapid fading from the superoutburst plateau, the system showed the appearance of a strong
near-infrared excess resulting in very red colours, which reached extreme values ($B$$-$$I$$\simeq$1.4) about
20 d later. The colours then became bluer again, but it took at least 250 d to acquire a stable
level. Superhumps were clearly visible in the light
curve from our very first time-resolved observations until at least 420 d after the rapid fading
from the superoutburst. The spectroscopic and photometric data revealed an orbital period of 109.80~min and
a fractional superhump period excess $\lesssim$0.8 per cent, indicating a very low mass ratio $q\lesssim$0.045.
With such a small mass ratio the donor mass should be below the hydrogen-burning minimum mass limit.
The observed infrared flux in quiescence is indeed much lower than is expected from a cataclysmic variable
with a near-main-sequence donor star. This strongly suggests a brown-dwarf-like nature for the donor and
that \SSSJ\ has already evolved away from the period minimum towards longer periods, with the donor now
extremely dim.
\end{abstract}


\section{Introduction}

In the accreting white dwarf (WD) systems known as cataclysmic variable stars (CVs -- for a comprehensive
review, see \citealt{Warner}), accretion is the dominant factor in their discovery through line emission
and outbursts. The donor components in CVs are low-mass main-sequence stars or brown dwarfs which
lose matter via the inner Lagrangian point. In the absence of a strong magnetic field, the material
transferred from the donor star forms an accretion disc around the WD. Dwarf novae are an important
subset of CVs with relatively low mass-transfer rates, the discs of which can suffer thermal instabilities,
resulting in outbursts for which CVs are named \citep[for review of the disc-instability model see,
e.g.,][]{Cannizzo93,Osaki96,Lasota}. The recurrence time of the outbursts is usually weeks or a few months
and their amplitude is mostly less than 6 mag. However, the outbursts in different dwarf novae have
different amplitudes, light-curve shapes, durations and recurrence times. Based on the variety of these
parameters, a few sub-types have been assigned to the dwarf nova class.

SU~UMa-type dwarf novae have short orbital periods, usually below the 2--3~h CV period gap, and show
two types of outbursts: short (normal) outbursts lasting a few days, and less frequent superoutbursts
which have a slightly larger amplitude and a longer duration of about two weeks.
At the short period end of the CV orbital period distribution, accretion rates are
low, and some systems only outburst once every several years or even decades. Such systems compose an
extreme subgroup of the SU~UMa-type dwarf novae, called WZ~Sge-type stars (for a modern and comprehensive
review of the WZ~Sge-type objects, see \citealp{KatoWZ}). Their other peculiar properties are very large
superoutburst amplitudes exceeding 6 mag and the lack or great rarity of normal outbursts.

A unique property of
superoutbursts is the appearance of superhumps, low-amplitude modulations with a period of a few percent
longer than the orbital one. The superhump phenomenon is well explained by the tidal instability of the
accretion disc \citep{Vogt82,Whitehurst,Osaki89}. This instability grows when a disc expands beyond the
3:1 resonance radius that causes the disc to become quasi-elliptical and precess, initiating superhumps.
The period of superhumps $P_{sh}$ is known to depend on the orbital period $P_{orb}$ and the mass ratio
$q=M_{2}/M_{wd}$, where $M_{2}$ and $M_{wd}$ are the masses of the donor and WD, respectively.
An empirical $P_{sh}$--$q$ relation (see, e.g., \citealt{PattersonEps05})
is an important tool to estimate $q$ without the recourse to spectroscopy. In particular, it was shown
that the distribution of the mass ratios of the donor and WD in WZ~Sge-type stars shows a sharp
peak between 0.07 and 0.08 \citep{KatoWZ}.

The CVs with such small mass ratios are of a special interest because a significant number of `period
bouncers' should be present among them. According to standard evolutionary theory, CVs evolve from
longer to shorter orbital periods until a minimum period is reached ($\approx$80 min), when the donor
star becomes of a substellar mass and partially degenerate, resulting in a kind of a brown-dwarf-like
object. Systems that have passed beyond $P_{min}$ are evolving back towards longer periods
\citep[see][and references therein]{Kolb93,Howell97,Gaensicke,KniggeCVevol}. This has long been predicted to be
the `graveyard' and current state of 70 per cent of all CVs \citep{Kolb93}, however only about a dozen more
or less robust candidates for such period bouncer systems have been identified until now, out of about
1000 of known CVs \citep[for a fairly recent compilation see][]{PattersonDist}.

One such system, called \SSSJ\footnote{Also known as SSS130101:122222$-$311525, 1RXS J122221.5$-$311545
and OT J122221.6$-$311525.} (hereafter \SSS), was discovered on 2013 January 1 at $V$=12.3 by the Catalina
Real Time Survey (CRTS; \citealt{Drake}), and was subsequently found to have been in outburst by 2012
December 16 \citep[with unfiltered CCD magnitude 11.8;][]{Levato}. The large outburst amplitude (the
average quiescent magnitude of the counterpart was found to be $\approx$19; \citealt{Drake}), the appearance
of its optical spectrum \citep{MarshATel} and the detection of an `S-wave' sinusoidal component in the
emission lines with a period of about 80 to 95 min allowed \citet{KuulkersATel} to propose the WZ~Sge-type
dwarf nova classification for \SSS. \citet{KatoSSS} discussed some of the outburst properties of \SSS,
studied the superhump period changes in the system and
found that \SSS\ has a very small mass ratio, $q$$<$0.05. They also concluded that the relatively long orbital
period $P_{orb}$ (which was not known, but should be close to the average superhump period of $\sim$111 min)
and low $q$ make this object a perfect candidate for a period bouncer. The confirmation of the period bouncer
status would be of great importance due to the long orbital period of \SSS, which would make this binary the
most evolved CV \citep{KniggeCVevol}.

For more than three years we have been performing extensive X-ray -- ultraviolet (UV) -- optical -- near-infrared
(NIR) photometric and spectroscopic observations of \SSS. The observations were taken from 2013 January 5 (four
days after the discovery) until 2016 April 17, during the superoutburst, the following decline and in quiescence.
These data provide crucial insights into the properties of \SSS. We found that a number of observed
features are quite unusual when compared with other WZ~Sge-type systems. Although most of these features are
not completely unique, their union in one object makes \SSS\ a remarkable case of the WZ-Sge-type system.
Our unprecedentedly detailed, multiwavelengths observations allow us to shed more light on mechanisms
underlying the system's extraordinary properties. In this paper, we concentrate on the long-term evolution
of the binary as revealed by the UV--optical--NIR data.

\begin{table*}
\caption{The averaged magnitudes of \SSS\ on 2013 February 5 (the plateau stage of the superoutburst) and on 2016 April 7--17
         (quiescence). The superoutburst $U$ magnitude is given in the Bessel filter, whereas for the quiescent state we
         show the {\it Swift} $u$ magnitude.}
\begin{center}
\begin{tabular}{lccccccccccc}
\hline
 State       & $uvw2$   &  $uvm2$  & $uvw1$   & $u$ / $U$&   $B$    &   $V$    &   $R_c$  &  $I_c$   &   $J$    &   $H$    &   $Ks$    \\
\hline
Superoutburst&          & 11.08(5) &          & 11.91(5) & 12.65(4) & 12.69(3) & 12.68(3) & 12.77(4) &          &          &           \\
 Quiescence  & 17.65(8) & 17.53(9) & 17.57(7) & 17.97(7) & 19.05(5) & 19.00(5) & 18.72(5) & 18.76(5) & 18.47(5) & 18.31(7) & 17.77(23) \\
\hline
\end{tabular}
\end{center}
\label{Tab:Magnitudes}
\end{table*}

\section{Observations}

During our observing campaign we used many ground- and space-based telescopes equipped with different
instruments. All details on the observations and data reduction can be found in Appendix~\ref{Sec:Appendix}.
Here we describe shortly the data used in our analysis.

Our optical time-resolved photometric observations started on 2013 January 8, $\sim$7~d
after the discovery announcement by CRTS \citep{Drake}. From then on, the observations were conducted
almost every clear night until 2013 April 21. After a short interruption, the observations were resumed
on 2013 May 4 and continued until 2013 June 18, the end of visibility of the source. Thus,
about 100 nights of time-resolved photometric data were taken during this interval. We restarted
the observations on 2014 January 13, when the star became visible again, and finished on 2014 June 28.
A total of 35 nights of data were obtained during 2014. In the beginning of 2015 we obtained seven
additional sets of time-resolved data. In January--April of 2016 we took several single-shot observations.
The final observation was taken on 2016 April 17.
The observations were taken with a 0.40-m f/6.8 Optimized Dall Kirkham (ODK) telescope of the Remote
Observatory Atacama Desert (ROAD) in Chile, a 0.35-m Celestron C14 robotic telescope at New Mexico Skies
in Mayhill, New Mexico, two 0.4-m robotic PROMPT telescopes located in Chile \citep{PROMPT}, and with the
0.84-m telescope at the Observatorio Astron\'{o}mico Nacional at San Pedro M\'{a}rtir (OAN SPM) in Mexico.
A significant part of our photometric data was taken with
Johnson--Cousins $BV(RI)_C$ filters, and other observations were performed with the $V$ or $R$ filters, or
unfiltered. Table~\ref{ObsPhotTab} provides the journal of these observations.

On 2016 April 7 we conducted a near-infrared observation using the Infrared Survey Facility (IRSF)
at Sutherland, South Africa. The IRSF consists of a 1.4~m telescope and Simultaneous Infrared Imager for
Unbiased Survey (SIRIUS: \citealt{SIRIUS}), which can obtain $J$, $H$ and $Ks$-band images
simultaneously.

The optical spectroscopic observations of \SSS\ were performed on 16 nights between 2013 January 5 and
2013 May 5. Additional observations were taken on 2015 May 5, and on 2016 February 9 and 10. The spectra
were obtained with the 4.2-m William Herschel Telescope on La Palma, the 2.1-m telescope at the OAN SPM,
the 6.5-m Magellan Clay Telescope at Las Campanas Observatory near La Serena in Chile, and with the
VLT-UT1 and VLT-UT2 telescopes at Paranal Observatory, Chile. Spectra were obtained in both single-shot
and time-resolved manner with different instrumental setups and exposure times. Table~\ref{Tab:SpecObs}
provides the journal of the optical spectroscopic observations.

The {\it Swift} X-ray satellite \citep{Swift} started observing \SSS\ on 2013 January 6, 5.8 d after
the discovery announcement \citep{KuulkersATel}. For each observation, data were collected using both the
X-ray Telescope (XRT; \citealt{SwiftXRT}) and the UV/Optical Telescope (UVOT; \citealt{SwiftUVOT}) with
the $uvm2$ filter in position. In this paper we present only the UV/Optical part of the observations.
The observations were
obtained approximately every 1--3 d until 2013 July 1. Two additional observations were performed on
2014 June 26 and 2015 January 16. A final data set was collected on 2016 April 17. This UVOT observation was
carried out in all six available filters.

In Section~\ref{Sec:LongTerm} we show that \SSS\ had returned to quiescence in the beginning of 2015.
In Table~\ref{Tab:Magnitudes} we summarize the averaged magnitudes of \SSS\ in all the available filters,
obtained on 2016 April 7--17 (quiescence) and on 2013 February 5 (the superoutburst).

\begin{figure*}
\centering
\includegraphics[width=17cm]{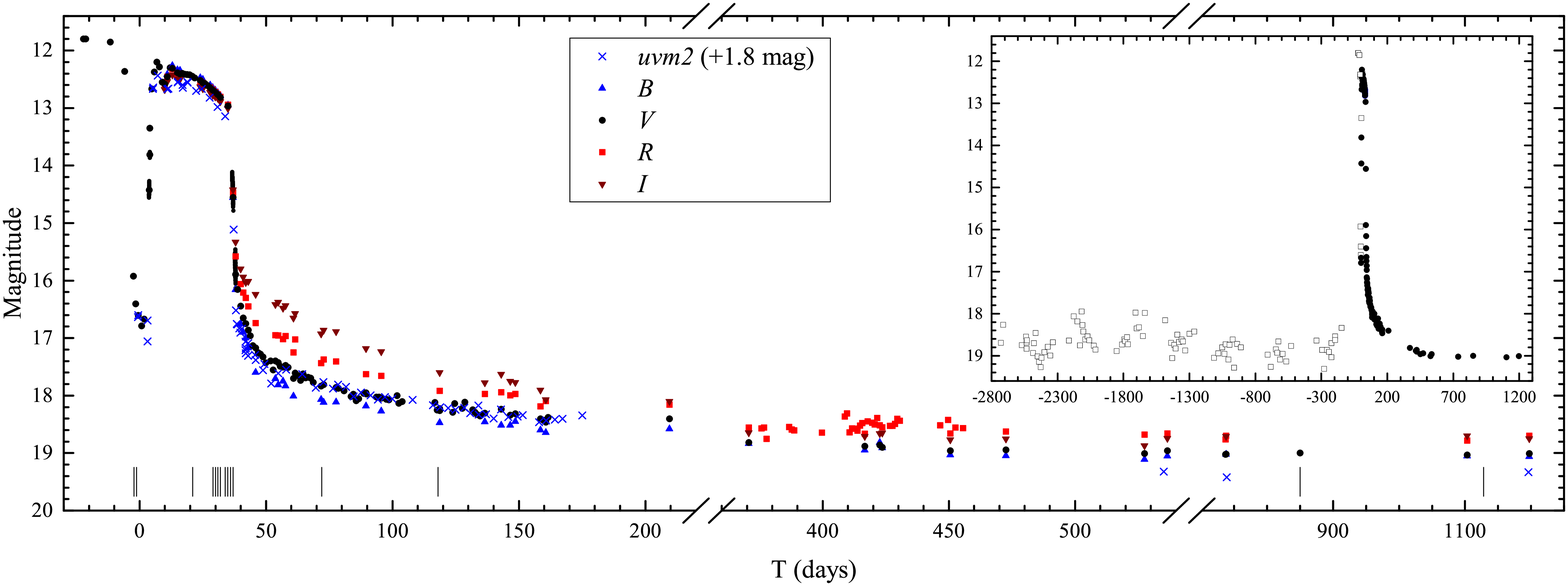}
\caption{The light curve of the superoutburst of \SSS\ in all the available optical and UV filters.
Each point is the 1-d average of observations in the corresponding filter. Short vertical lines
mark epochs of optical spectra. In the inset the overall light curve is shown which includes the
pre-outburst 1-d averaged data from the CRTS observations (the open squares). $T$ is the
number of days elapsed since HJD 245\,6300.0.
}
\label{Fig:LC}
\vspace{3 mm}
\includegraphics[width=17cm]{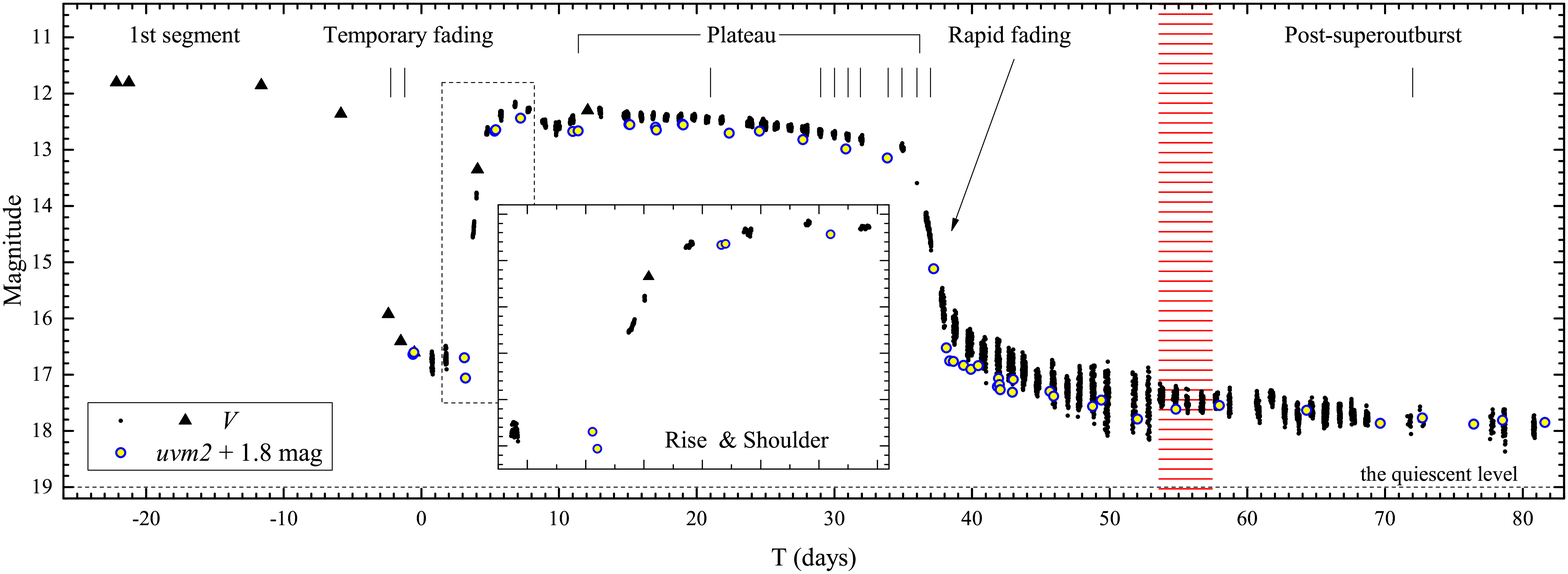}
\caption{A superoutburst segment of the light curve in the $V$ and $uvm2$ filters. Black
         points are the individual observations in the $V$ filter. The triangles represent the
         optical data taken from the literature and other sources.
         Blue/yellow symbols are the observations in the
         $uvm2$ filter, shifted by 1.8 mag for clarity. Short vertical lines mark epochs of optical
         spectra. The different stages of the
         superoutburst are indicated. The red vertical ribbon marks the
         days of the reddest optical spectra. In the inset the rise and the shoulder in the beginning
         of the second segment of the outburst are shown.
         }
\label{Fig:LC2}

\vspace{3 mm}

\includegraphics[height=6.8cm]{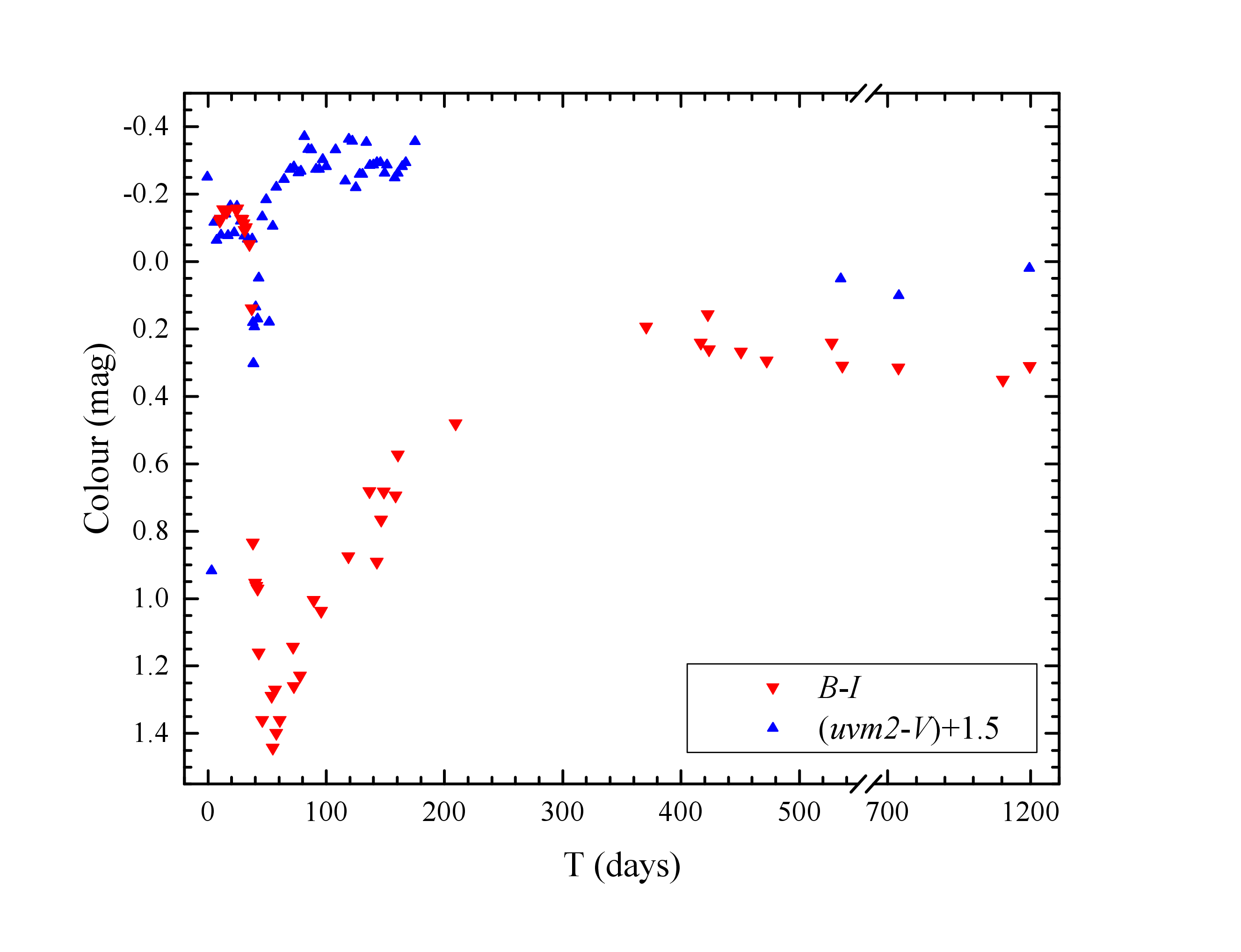}
\includegraphics[height=6.8cm]{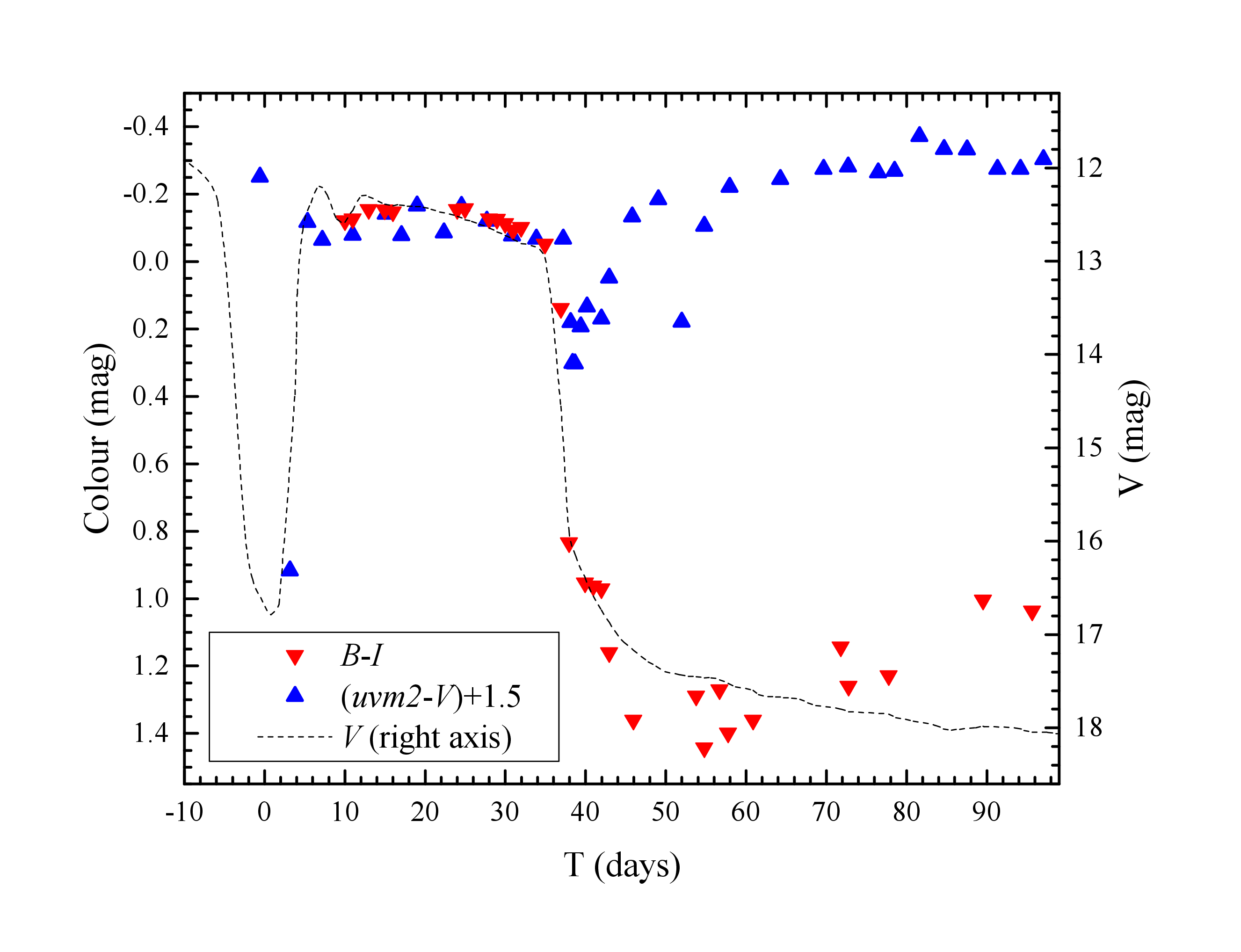}
\caption{Variations of optical and UV colour indices. Left: the entire period of our observations.
Right: a superoutburst segment is shown together with the $V$-band light curve.}
\label{Fig:Colours}
\end{figure*}


\begin{figure}
\includegraphics[width=8.5cm]{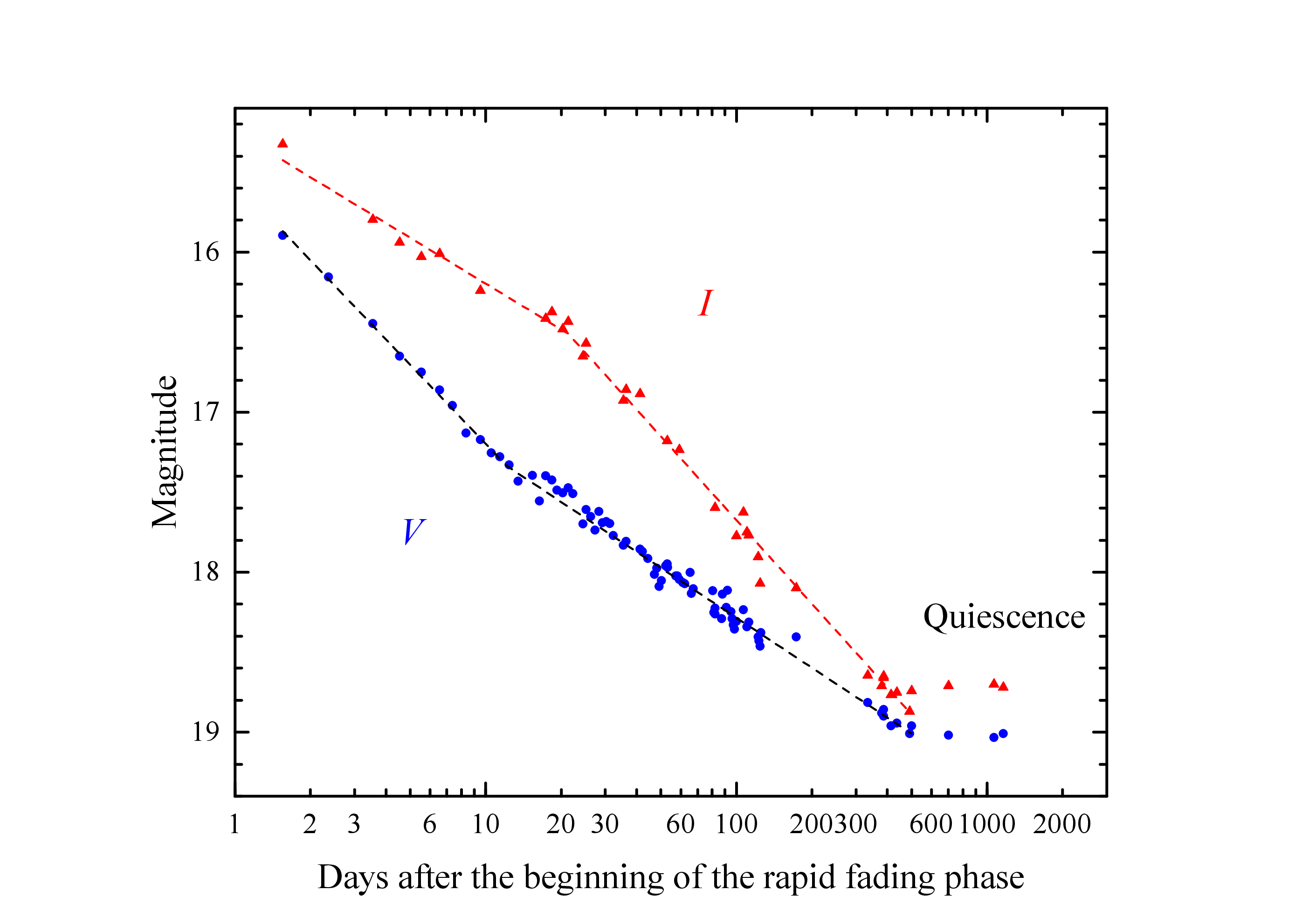}
\caption{The $V$ and $I$ photometry during the decline stage of the superoutburst, plotted with
         a logarithmic time-scale. The days are counted from the beginning of the rapid fading phase at $T$=36.4.}
\label{Fig:DeclineLog}
\end{figure}

\section{Optical and UV photometry}
\subsection{Long-term light curve and the colour evolution}
\label{Sec:LongTerm}

Throughout the paper, we will denote time in units of days since 2013 January 8, when we began our
time-resolved photometric observations. Thus, $T$=0 corresponds to HJD 245\,6300.0.

Fig.~\ref{Fig:LC} shows the light curves of \SSS\ containing all our observations in the optical
and UV filters taken over about three years. In the inset of this figure we also show the
overall light curve which includes the pre-outburst 1-d averaged data from the Catalina Real-time
Transient Survey (CRTS) observations\footnote{http://nesssi.cacr.caltech.edu/DataRelease/} \citep{CRTS}
begun 10 years ago from 2005 August 5. The pre-outburst data reveal a half-magnitude variability between
$\sim$18.5 and $\sim$19 mag on few years' time-scale, but show no indication of previous outbursts.

The most notable feature of the superoutburst is its unusual double structure and long duration
(Fig.~\ref{Fig:LC2}). Unfortunately, the first segment of the superoutburst was scarcely observed.
The most recent observations prior to the superoutburst were obtained on 2012 August 13 ($T$=$-$147),
when the system was at $V$$\sim$18.3. The earliest observations of \SSS\ in outburst were obtained on
$T$=$-$22 ($\sim$11.8 mag; \citealt{Levato}). Subsequent photometry showed that the object
had still been near maximum light until at least $T$=$-$5.8 ($V$=12.3; \citealt{Drake}),
after that \SSS\ had faded to magnitude $\sim$15.5 on $T$=$-$2.4 and $\sim$16 on $T$=$-$1.5.
Thus, the duration of the superoutburst's first segment was at least 16 d.

When fading, the transient did not reach its quiescent level, and rebrightened again on $T$=3,
entering the second segment of the superoutburst \citep{NeustroevATel}. We began our photometric observations
on $T$=0, near the minimum light of this temporary fading stage ($V$=16.8).
During the rise the optical flux increased at a rate of $\sim$2.8 \md. The outburst's second
segment had reached the maximum of $V$=12.2 on $T$=7. After a short `dip' by
$\sim$0.4 mag exhibited during $T$=9--10, \SSS\ reached another local maximum of $V$=12.3 on $T$=13
and then monotonically faded very slowly. On $T$=37 we observed a rapid fading with
a rate of 1.31$\pm$0.01 \md, followed by a slow decline. Thus, the duration of the second segment of
the superoutburst was about 33 d. We note that though the peculiar pattern of the light curve of \SSS\
is not unprecedented and similar double superoutbursts have been seen in a few other CVs (e.g.,
OT~J184228.1+483742, see fig.~1 in \citealt{KatoSSS}) and transient low-mass X-ray binaries
(e.g., XTE~J1118$+$480, \citealt{KuulkersLMXB}), yet the superoutburst of \SSS\ was the longest
of any other known dwarf novae.
During 33 d of the second segment of the superoutburst \SSS\ faded by only $\sim$0.7 mag. The average
fading rate during $T$=13--20 was just 0.006$\pm$0.001 \md, after that it increased to 0.032$\pm$0.001 \md\
($T$=20--35). Such a slow, gradual decline is atypical for SU~UMa stars, whose plateau slope is usually
about 0.11$\pm$0.01 \md\ \citep{Warner}. See Section~\ref{Sec:DiscWZ} for further discussion of atypical
properties of the light curve.

The shoulder in the beginning of the second segment of the superoutburst resembles a so-called precursor
outburst, a normal outburst which is often observed shortly before the superoutburst \citep{MarinoWalker},
and which is thought to trigger it \citep{Osaki89,OsakiKato2013}. We note, however, that in contrast
to \SSS, the maximum light of precursors is usually lower by 0.5--1 mag than that of the following
superoutbursts (see e.g. the very detailed high-fidelity {\it Kepler} light curves of several superoutbursts
of two SU~UMa stars; \citealt{CannizzoKepler}), thus the origin of the shoulder in \SSS\ might be
different from that of the precursors discussed above.

\begin{figure*}
\centering
\includegraphics[width=17cm]{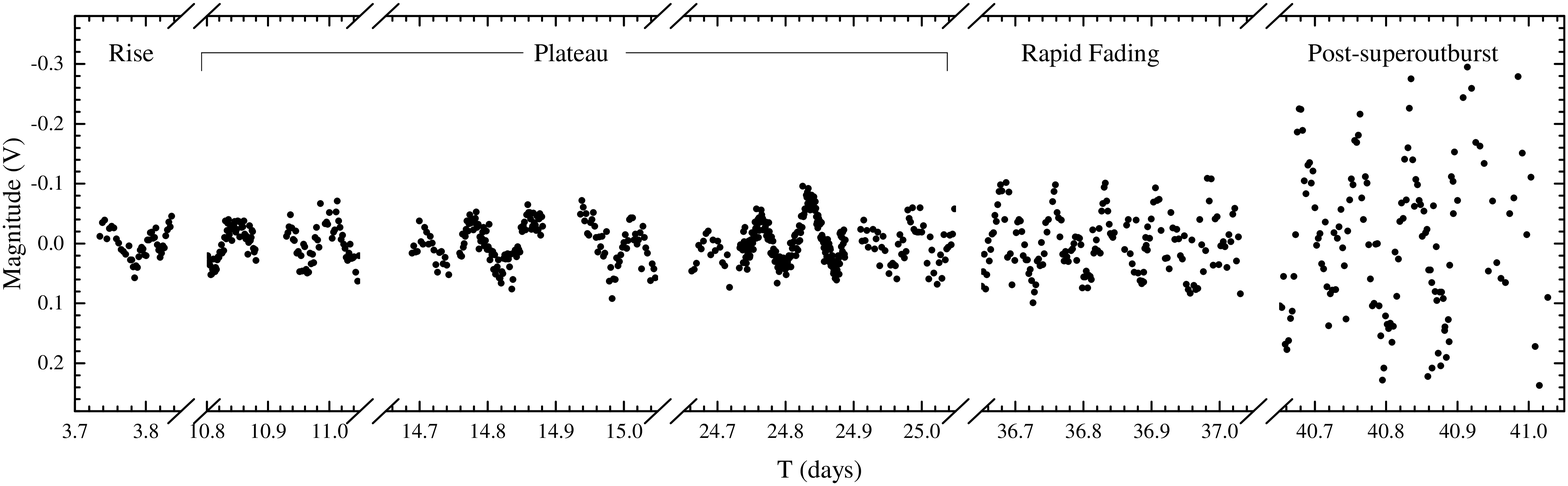}
\caption{Six light-curve samples from different stages of the superoutburst of \SSS\ (with the means,
and linear trends subtracted for each night).
}
\label{Fig:LC-2013}
\end{figure*}

The shapes of the optical and UV light curves are qualitatively similar. In particular, the data
do not show any apparent signs of a UV-delay in the early rising phase of the second segment of the
superoutburst\footnote{A UV-delay, a lag as long as one day between the rise to outburst in the optical
flux and the rise in UV, is observed in some dwarf novae (see \citealt{Verbunt87} and \citealt{Warner}, and
references therein), the cause of which is still not fully understood. We note, however, that the observed
rise in \SSS\ had occurred after the temporary fading stage, from the level of about 2~mag brighter than that
in quiescence. Thus, it is not clear if the discussion of the existence or non-existence of the UV-delay in
such conditions is relevant.}. Note, however, that at the very beginning of the optical rise the UV flux
showed a drop by $\sim$0.25 mag (see the inset in Fig.~\ref{Fig:LC2}). We do not have simultaneous optical
photometry which would enable us to verify such a behaviour, although the detailed {\it Kepler} observations
of many CV superoutbursts \citep[e.g.,][]{CannizzoKepler} do not seem to show such drops at optical wavelengths.
On the other hand, the drops of the UV flux in the beginning of outbursts were reported
for several DNe \citep{WuPanek,Verbunt87}. They, however, were different in the sense that they
occurred after an initial rise, and after the drop the rise resumed. In \SSS\ this UV drop occurred before
the rise.

Fig.~\ref{Fig:LC} displays an apparent difference in the
post-outburst behaviour of the transient in different wavelengths. The magnitude of the decline during
the rapid fading phase decreases greatly towards longer wavelengths, resulting in
strong colour evolution (Fig.~\ref{Fig:Colours}). Fig.~\ref{Fig:DeclineLog} shows the optical
photometry during and after the rapid fading phase, plotted with a logarithmic time-scale. For clarity,
we show only the $V$ and $I$ light curves, but the $B$ and $R$ look similar. We found that all the light
curves are well described by a broken power law as a function of time, with the time origin being at
$T$=36.4 and a slope-break some 20 d later. The power-law indices of the decline are given in
Table~\ref{Tab:DeclineInd}. It is unclear how common is such a (broken) power-law decline among CVs
and LMXBs, because we found no papers, either observational or theoretical, reporting on it.

\begin{table}
\caption{The power-law indices (columns 2 and 3) of the broken power-law decline of \SSS\ in the
         $B$, $V$, $R$ and $I$ bands (in flux units), and the number of days (column 4) elapsed
         between the beginning of the rapid fading phase at $T$=36.4 and the slope-break.}

\begin{center}
\begin{tabular}{cccc}
\hline
 Band &     Before the break &  After the break  &  Break time after $T$=36.4 \\
\hline
 $B$  & $-0.58\pm$0.03 & $-0.33\pm$0.01 & $\sim$23 \\
 $V$  & $-0.64\pm$0.01 & $-0.43\pm$0.01 & $\sim$18 \\
 $R$  & $-0.54\pm$0.02 & $-0.43\pm$0.02 & $\sim$24 \\
 $I$  & $-0.38\pm$0.03 & $-0.70\pm$0.01 & $\sim$24 \\
\hline
\end{tabular}
\end{center}
\label{Tab:DeclineInd}
\end{table}

From Fig.~\ref{Fig:DeclineLog} it is also clearly seen that \SSS\ has
not fully recovered from the outburst even 500~d after the rapid fading phase because it
was still fading with the same rate. Only our observations obtained on and after
$T$=736  revealed that the data do not follow the decline trend anymore,
confirming that the transient had finally returned to quiescence. The following observations
showed nearly constant magnitudes. Thus, the total amplitude of the outburst was $\sim$7 mag and
its post-outburst decline lasted no less than 500~d.

During the superoutburst the colour indices\footnote{In order to calculate the values of the $uvm2$-$V$
indices, we interpolated the $V$ magnitudes to the times of the UV observations.}
demonstrated dramatic changes with time, and the colour
curve has an extraordinary shape (Fig.~\ref{Fig:Colours}), never observed before in such detail.
At the superoutburst plateau the optical colour indices varied slowly around 0, their
average values were quite usual for a dwarf nova in outburst \citep{Warner}: $B$--$V$=$-0.04\pm$0.01,
$V$--$R$=$0.00\pm$0.01, $R$--$I$=$-0.09\pm$0.01. With the start of the decline, the object had begun
reddening. The colour indices reached the extreme values some 20 d later, around the time of
the slope-break of the broken power-law decline, at $T$$\simeq$54--58: $B$--$V$=$0.32\pm$0.05,
$V$--$R$=$0.48\pm$0.03, $R$--$I$=0.55$\pm$0.03.
These colours are redder than in most dwarf novae in quiescence
\citep{Echevarria1,Echevarria2, KatoColours}.
The colours then turned back to becoming bluer. This trend continued for several hundred days.
Only the data obtained after $T$$\sim$472 revealed nearly stable,
relatively blue colours: $B$--$V$=$0.06\pm$0.05, $V$--$R$=$0.30\pm$0.03, $R$--$I$=$-0.07\pm$0.10.
Although they are not so atypical for WZ Sge-type dwarf novae \citep{KatoColours}, the colours
are significantly bluer than in many of them in quiescence (compare e.g. $B$--$I$=0.43 in \SSS\
and $B$--$I$=0.75 in V455~And; \citealt{FS_Aur}).

The reddening of optical light after the superoutburst has been previously detected in several WZ~Sge-type
stars \citep{KatoWZ}. For example, a similarly strong cool continuum source was reported for EG~Cnc by
\citet{PattersonEGCnc}, which also showed the red colours ($B$--$I$$\simeq$1.4). However, multicolour
observations after the active stage of superoutbursts are rare and sporadic, and no detailed study
of the above phenomenon has ever been presented.

In contrast to the optical wavelengths, the UV data showed redder colours at the superoutburst plateau,
and bluer during the post-outburst stage (Fig.~\ref{Fig:Colours}). However, the data obtained after the long
gap in our observations (after $T$$\sim$535) show that in quiescence the $uvm2$-$V$ index dropped again to
the level even lower than during the plateau, reaching a nearly stable value of $-$1.44$\pm$0.10 mag.
We also note three red jumps in the $uvm2$-$V$ curve (see the right-hand panel of Fig.~\ref{Fig:Colours}).
Besides the one in the beginning of the rise, mentioned above, another reddening episode had begun immediately
after the beginning of the rapid fading phase, almost simultaneously with the optical colours (but possibly
after a short delay). However, $uvm2$-$V$ reached the maximum value in less than a day and returned to the
superoutburst level in a week. On $T$$\simeq$52--55, the days of the reddest optical spectrum, the UV flux
dropped again, producing another red jump in the $uvm2$-$V$ curve.

\begin{figure}
\includegraphics[width=8.5cm]{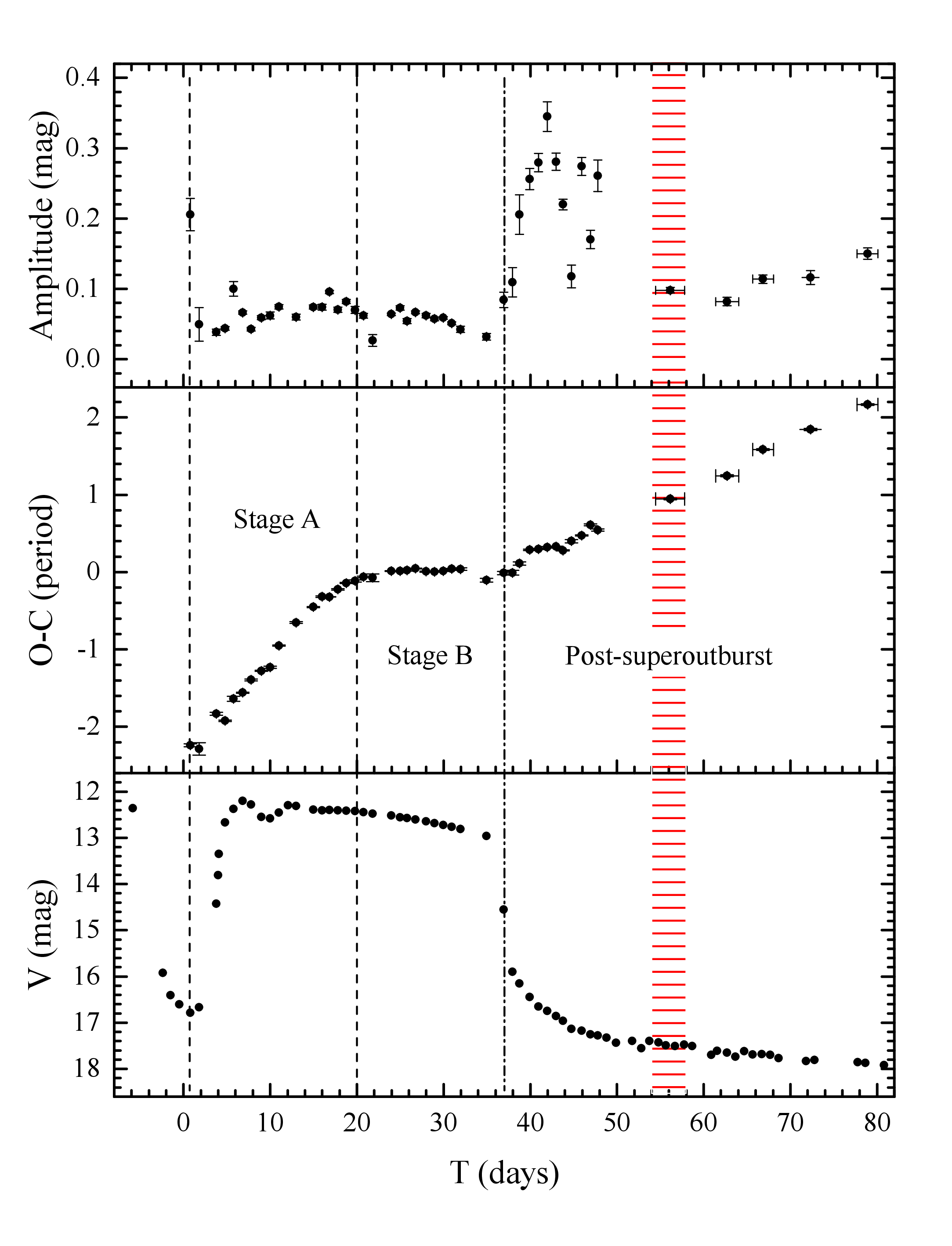}
\caption{Top: the amplitudes of superhumps in \SSS. Middle: the $O-C$ diagram of superhumps.
A period of 110.09~min was used for the calculations. Bottom: the $V$ light curve.
The red vertical ribbon marks the days of the reddest optical spectra.
The stages of the superhump evolution are marked following \citet{KatoSSS}.
}
\label{Fig:O-C}
\end{figure}

\subsection{Development of superhumps}
\label{Sec:Superhumps}

\subsubsection{Superoutburst superhumps}
\label{Sec:SuperoutburstStage}

Photometric modulations, known as superhumps, are clearly visible in the light curve of \SSS\ from our
very first time-resolved observations. Fig.~\ref{Fig:LC-2013} shows six light-curve samples which
illustrate the development of superhumps during the superoutburst stage.
It is known that the period of superhumps $P_{sh}$ in SU~UMa stars is not stable, but changes during a
superoutburst in a complex way. An extensive study of SU~UMa superhumps showed that the evolution of
their periods is often composed of three distinct stages \citep[see][and their later papers]{Kato1}.
These complex period changes can be explained by variation of the accretion disc radius during a
superoutburst \citep{HiroseOsaki1990}, which is possibly accompanied by variable gas pressure effects
in the disc \citep{OsakiKato2013}. The evolution of
the superhump period in \SSS\ has been investigated by \citet{KatoSSS}. Here we extend their analysis.
In order to examine the evolution of superhumps, we determined the times of superhump minima and their
amplitudes by fitting a sine wave to nightly light curves. The resultant $O-C$ diagram and the superhump
amplitudes are shown in the middle and top panels of Fig.~\ref{Fig:O-C}, respectively.

\begin{figure*}
\includegraphics[width=8.5cm]{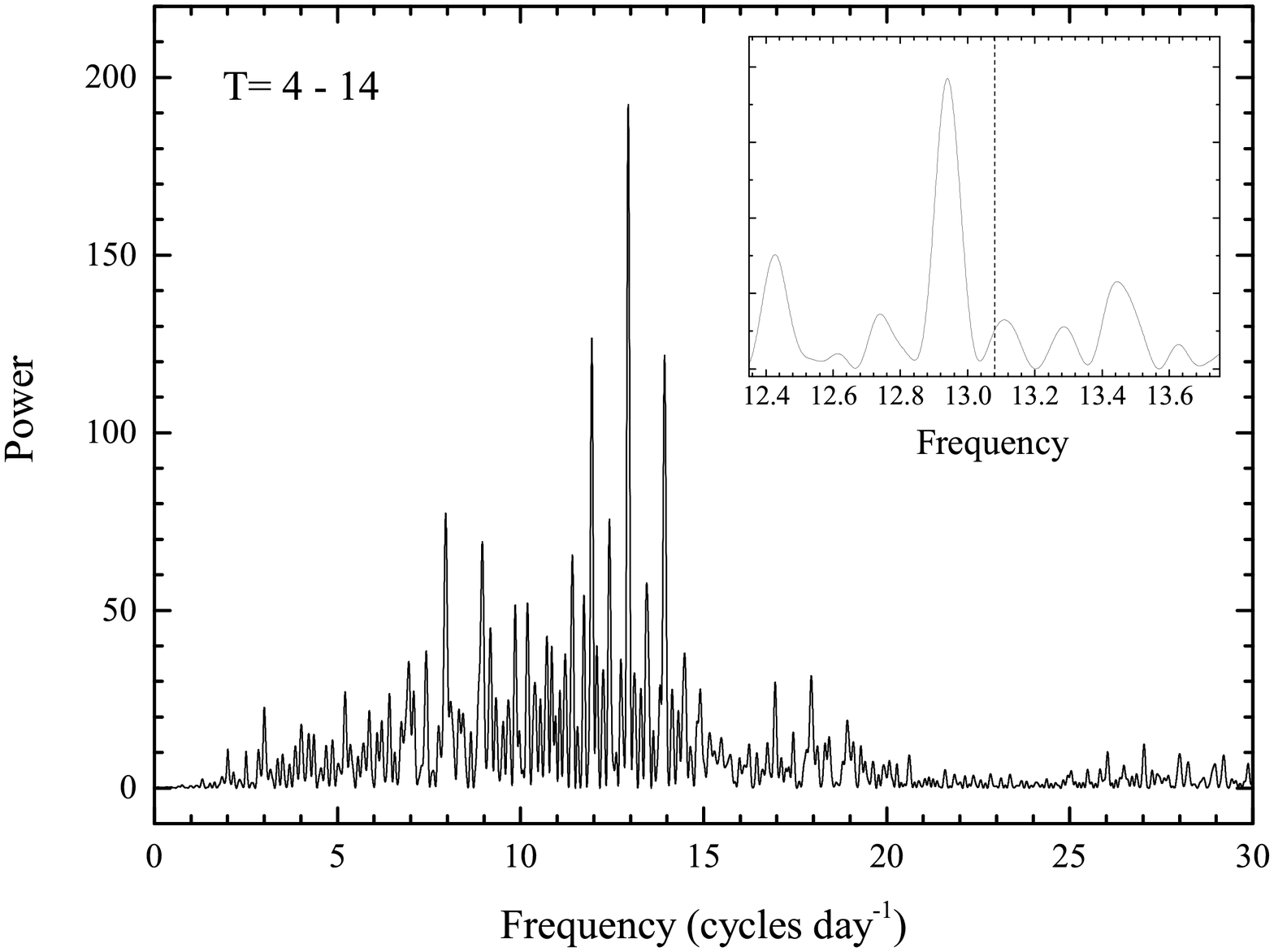}
\includegraphics[width=8.5cm]{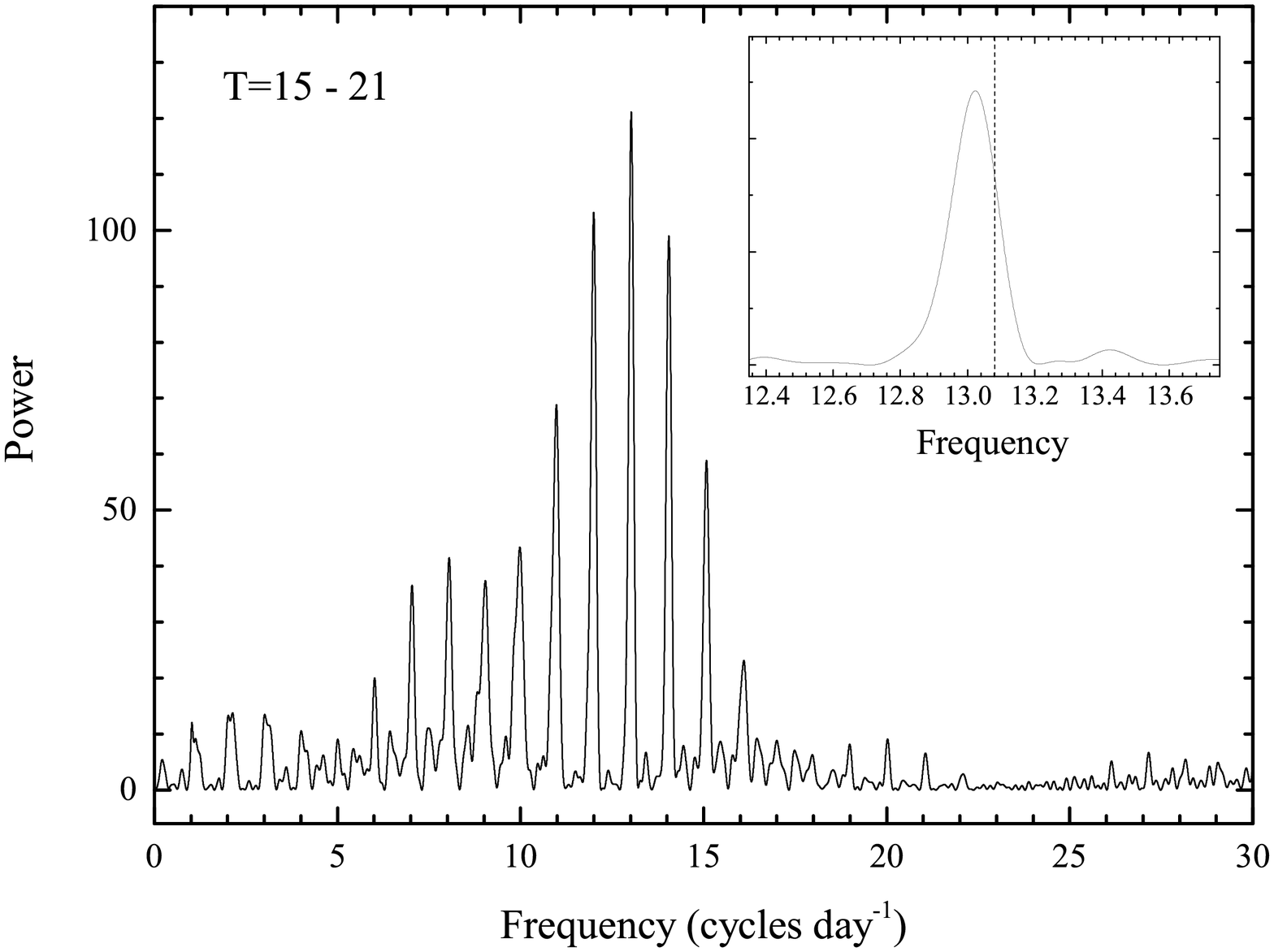}\\
\includegraphics[width=8.5cm]{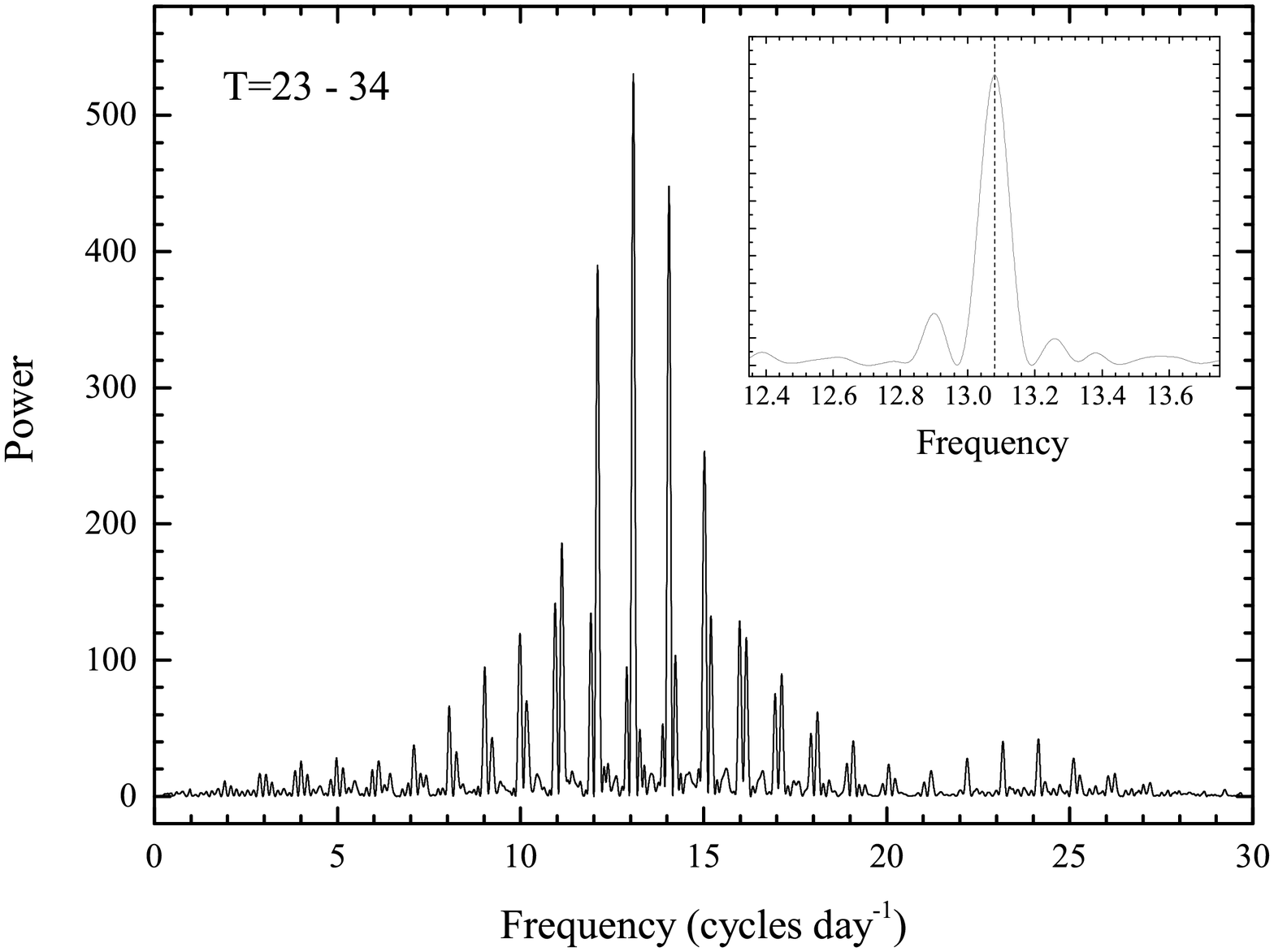}
\includegraphics[width=8.5cm]{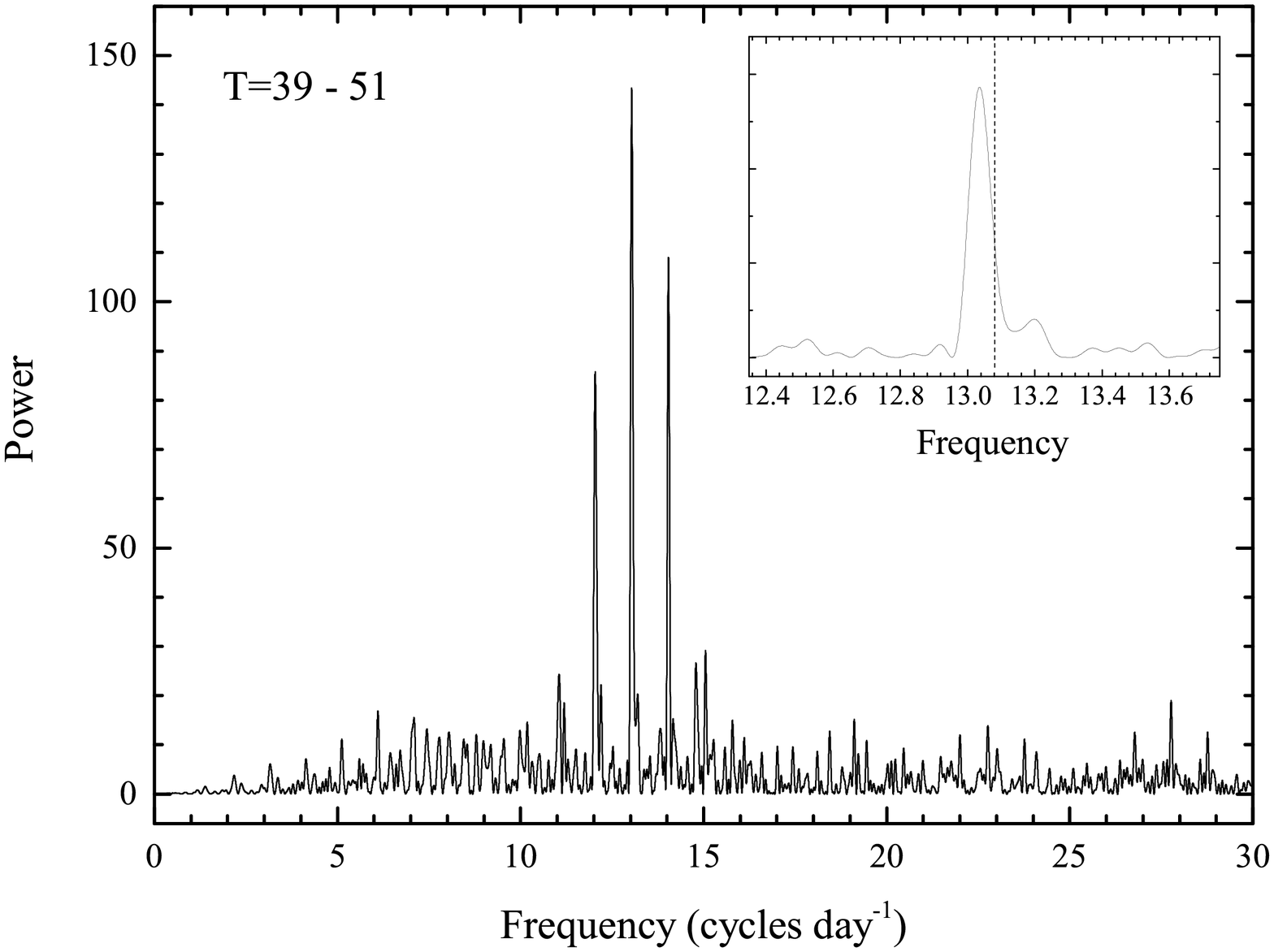}
\caption{Lomb--Scargle periodograms of the $V$ data combined among $T$=4--14, 15--21, 23--34 and 39--51.
         The inset shows the enlarged region around the frequency 13.08 \cd\ -- the averaged frequency
         of superhumps during the superoutburst stage B, marked by the vertical dashed line.}
\label{Fig:PSOutburst}
\vspace{5 mm}
\includegraphics[width=4.3cm]{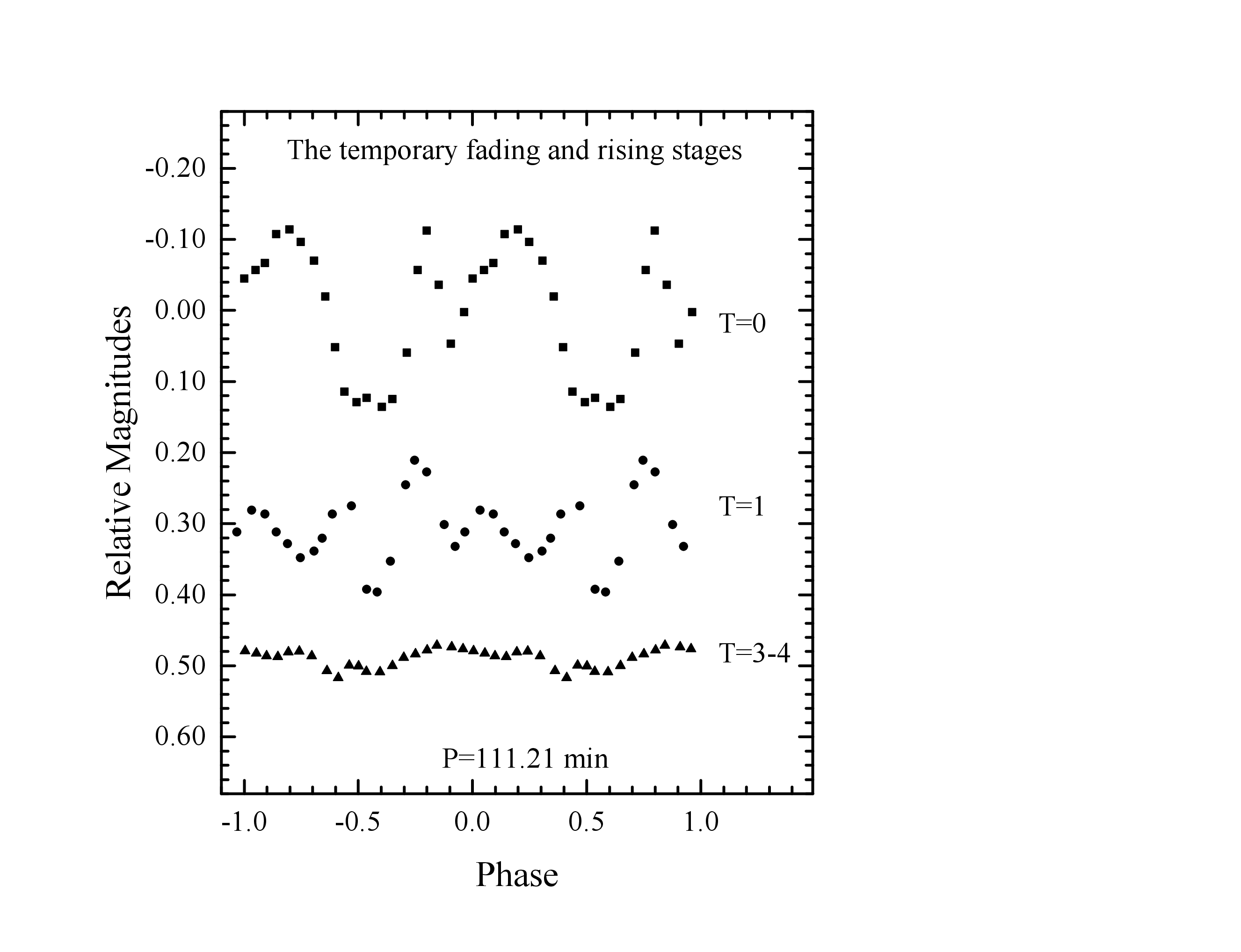}
\includegraphics[width=4.3cm]{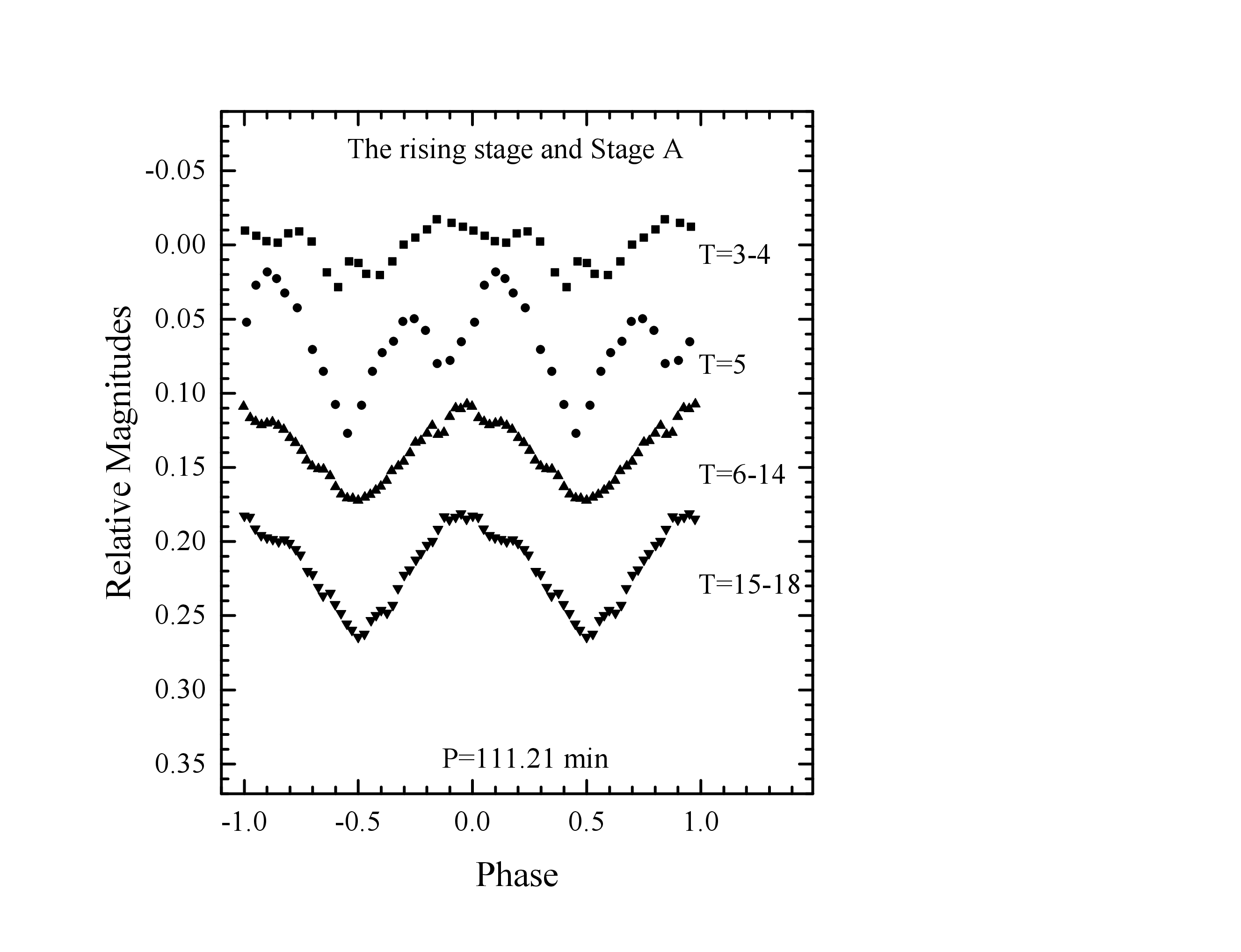}
\includegraphics[width=4.3cm]{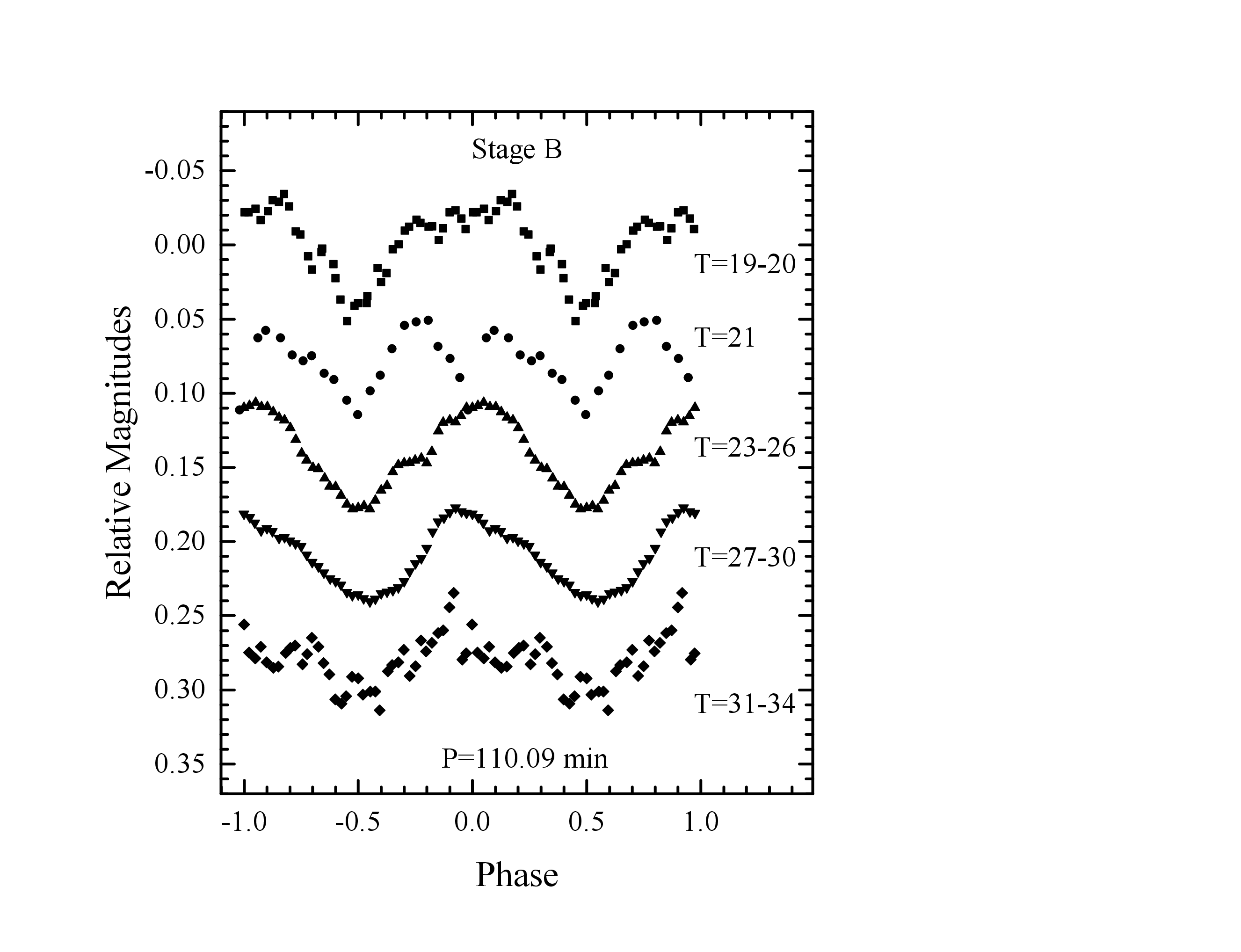}
\includegraphics[width=4.3cm]{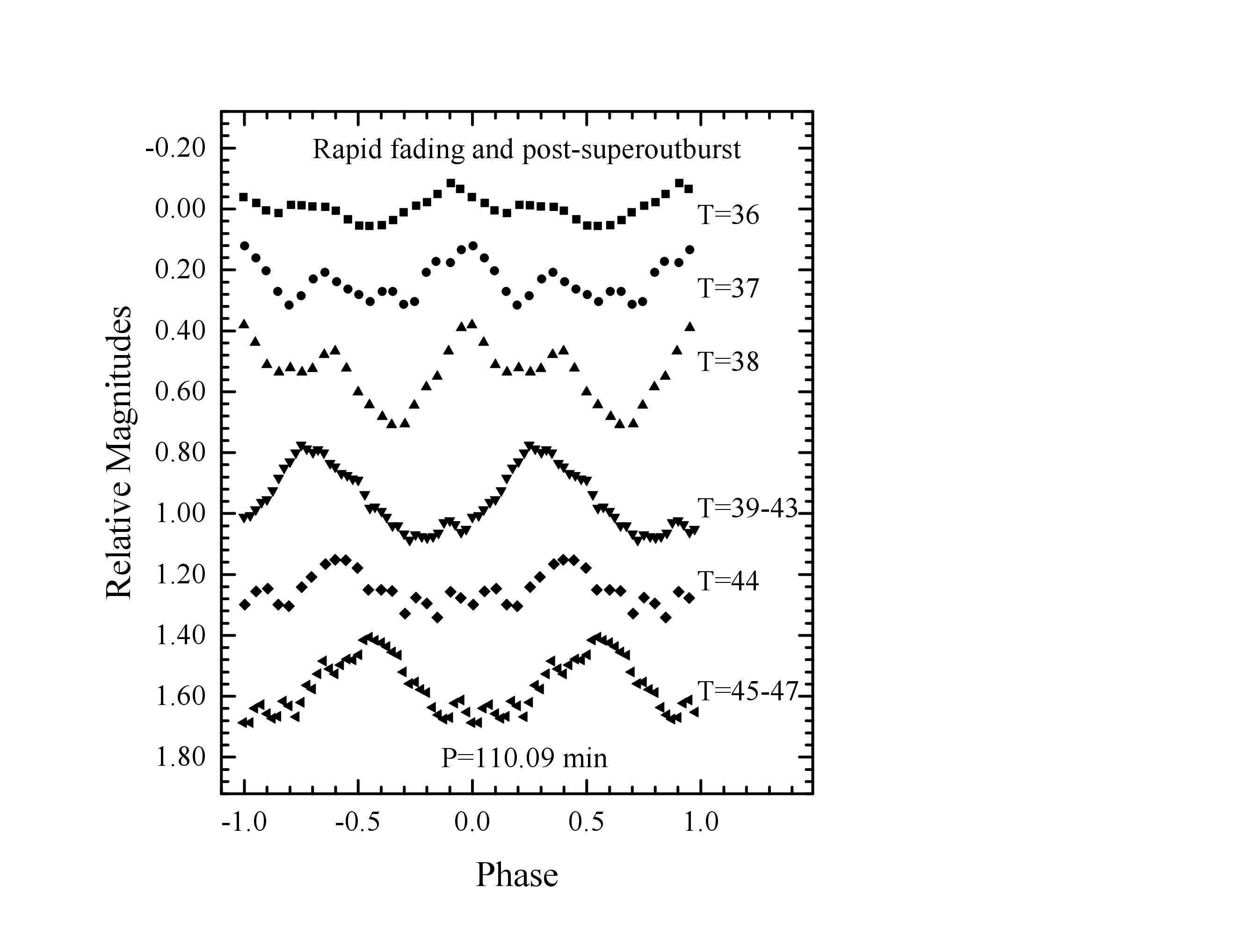}
\caption{Variation of superhump profiles during the superoutburst of \SSS.
         }
\label{Fig:SuperhumpsProfiles}
\end{figure*}

\begin{figure*}
\includegraphics[width=8.5cm]{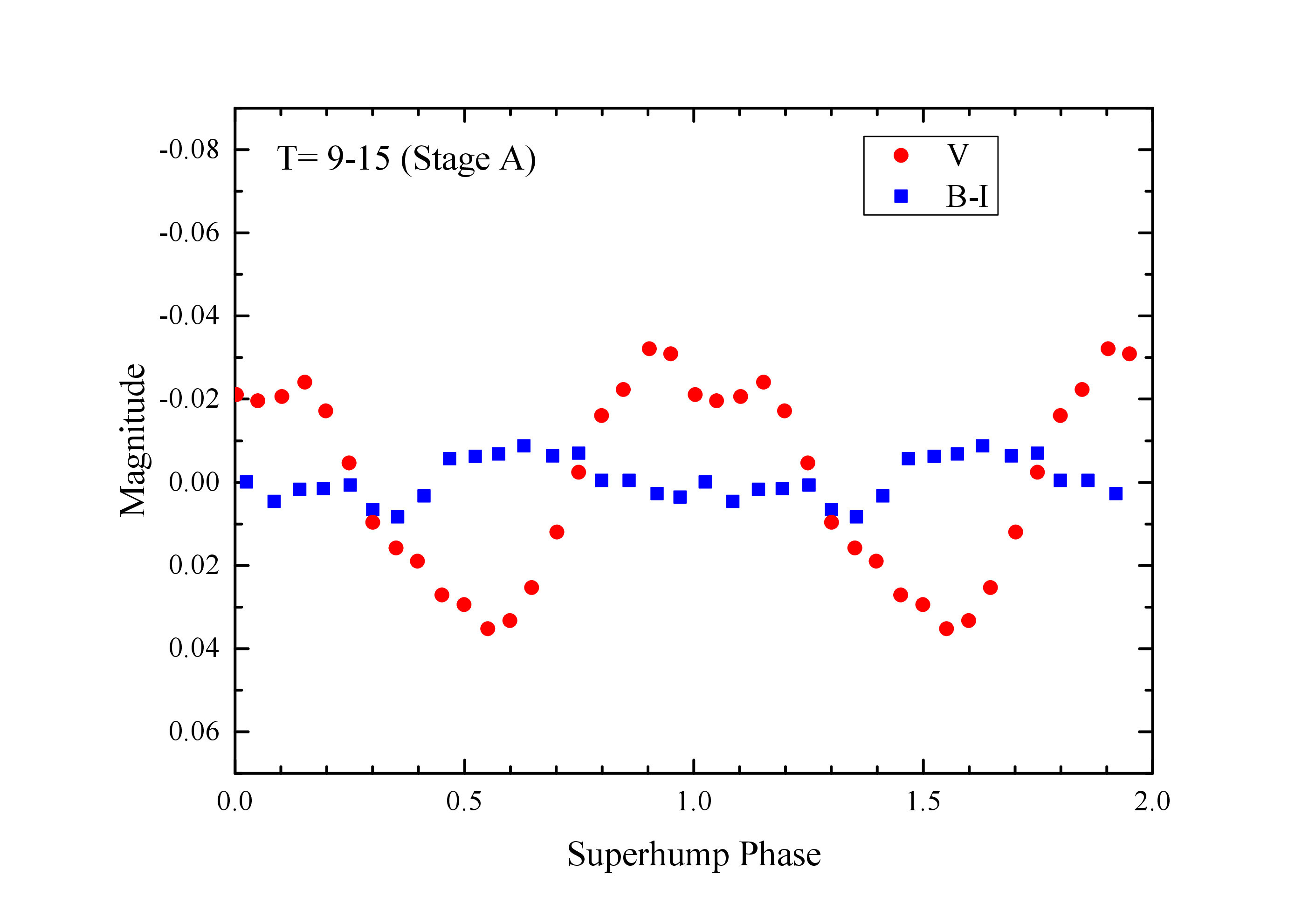}
\includegraphics[width=8.5cm]{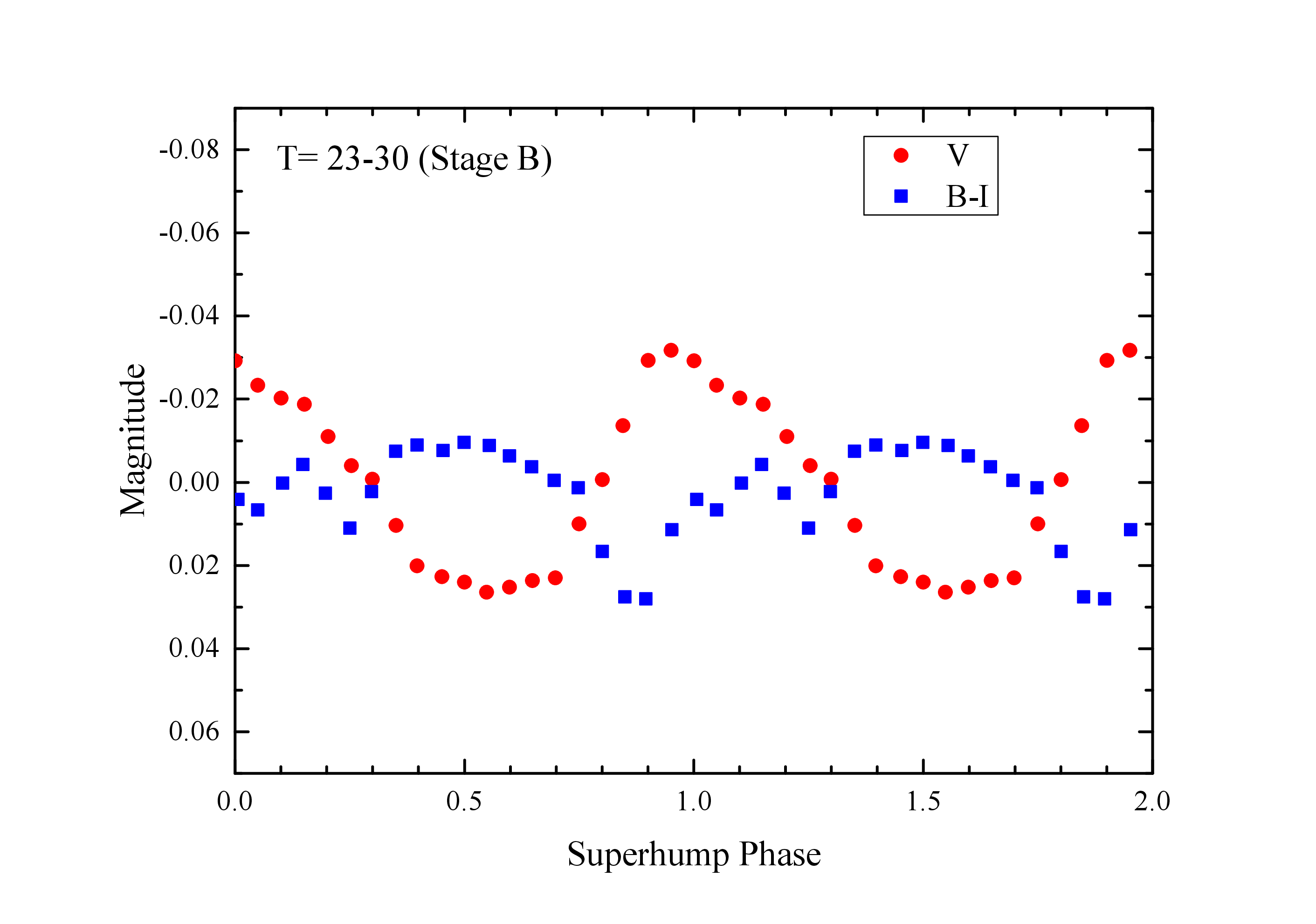} \\
\includegraphics[width=8.5cm]{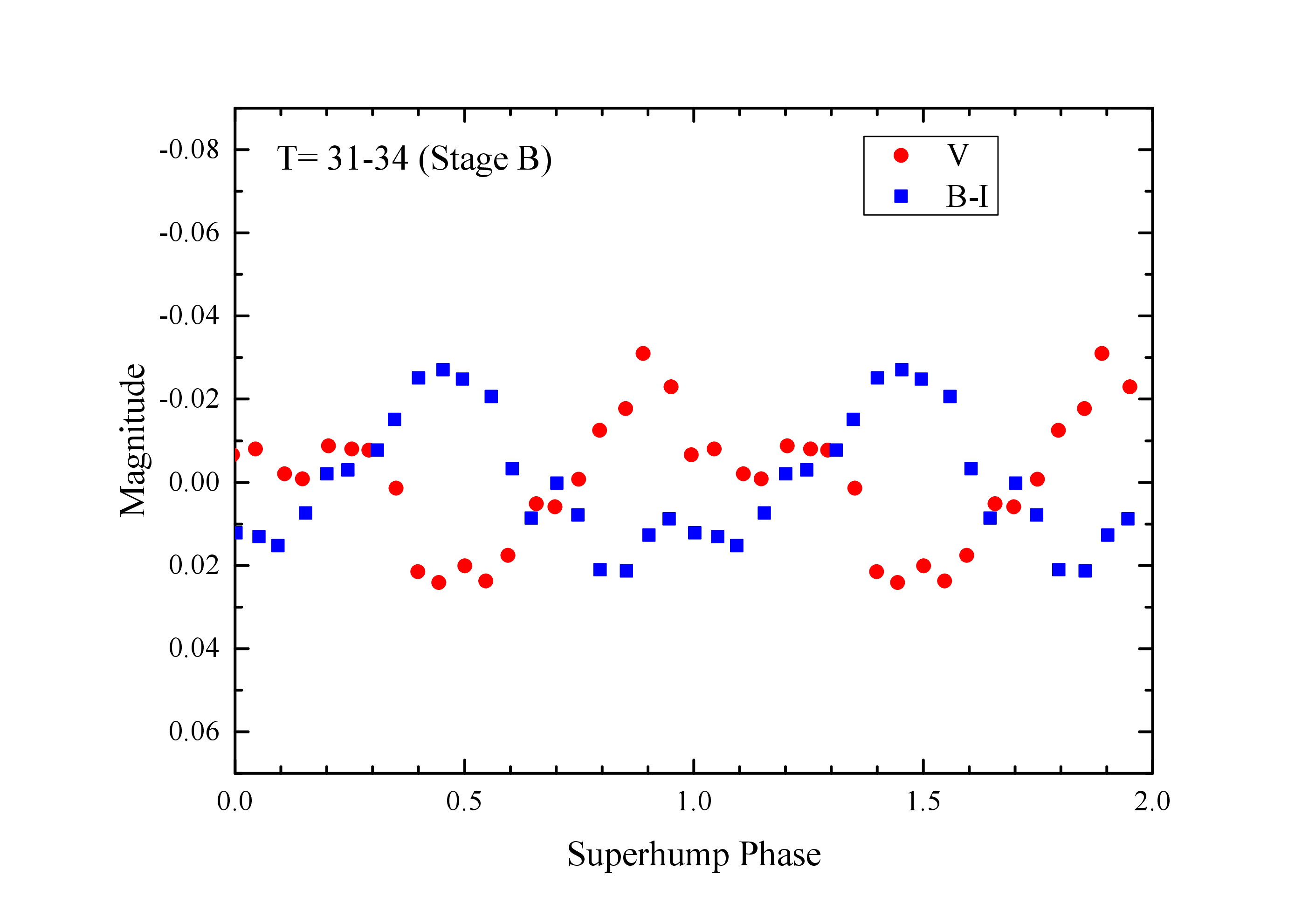}
\includegraphics[width=8.5cm]{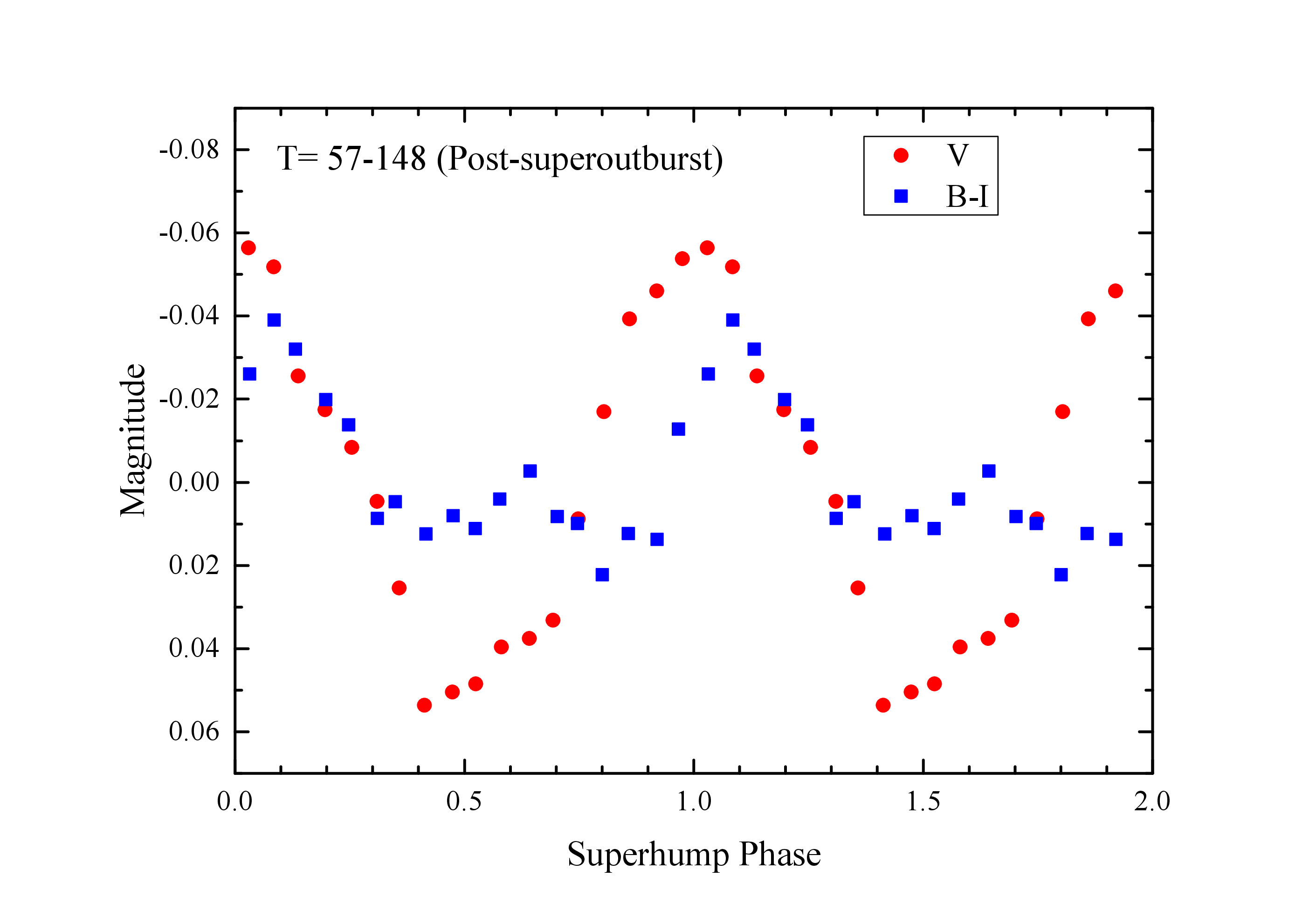}
\caption{Phase-averaged $V$ light curves and $B$--$I$ colour variations of superhumps at
         different stages of the superoutburst of \SSS.}
\label{Fig:SuperhumpsColours}
\end{figure*}

Although the superhump period in \SSS\ evolved rather differently to other SU~UMa dwarf novae
(see \citealt{KatoSSS} for a discussion of possible causes), nevertheless
our analysis confirmed that the superhumps experienced three stages of their evolution as seen
from the $O-C$ diagram. The latter shows three distinct segments, which we identified and marked in
Fig.~\ref{Fig:O-C} as stages A, B and post-superoutburst, following \citet{KatoSSS}. Fig.~\ref{Fig:PSOutburst}
shows the Lomb--Scargle periodograms calculated for different time intervals.
The linear rise of the stage A segment of the $O-C$ diagram (until $T$$\sim$20) indicates an approximately
constant period. The superhump amplitudes grew
from $\sim$0.04~mag at the beginning of the stage to $\sim$0.07~mag at its end. Using the data for
$T$=4--14 and 15--21, we determined the periods of superhumps for these dates to be 111.26(4) and
110.64(6)~min, respectively. The periods and their errors were computed using a linear fit to
the corresponding segment of the $O-C$ diagram. During the following $\sim$18 d
(21$\lesssim$$T$$\lesssim$38) of stage B the superhumps also had a nearly
stable, but shorter period of 110.10(2)~min determined using the data for $T$=23--34. To the end of
this stage the amplitudes decreased to $\sim$0.04~mag, but started growing again after the transient
had faded.
The third segment of the $O-C$ diagram ($T$$\gtrsim$38) has an uneven, stepped structure, indicating
sudden changes of the period. The duration of segments with a stable period is about 4~d. The superhump
amplitude reached $\sim$0.3~mag, but showed short-duration decreases at the time of the period changes.
Using the data for $T$=39--51, we determined the mean period of superhumps to be 110.40(7)~min.
The periods of superhumps we find are consistent with those reported by \citet{KatoSSS}.
The variation of superhump profiles during the superoutburst is shown in Fig.~\ref{Fig:SuperhumpsProfiles}.
Most of the time the modulations had a nearly symmetric, sinusoidal profile, although there were several
occasions when the superhump profiles transformed from single-humped to double-humped and back.

Our multicolour photometry shows that the superhumps were redder at their light maximum
(Fig.~\ref{Fig:SuperhumpsColours}). Such a behaviour is in general agreement with previous reports
\citep{SchoembsVogt,Hassall,Isogai2015}. This can be explained by assuming that the superhump light source
(SLS) is an expanded low-temperature area in outer regions of the accretion disc \citep[see, e.g.,][and
references therein]{Warner,Uemura}. We note, however, that not only the period, amplitude and profiles
of superhumps had changed over the course of the superoutburst. Fig.~\ref{Fig:SuperhumpsColours}
displays a notable difference in the colour profiles of superhumps at different stages of superhump evolution.
During the stage A the superhump colour variability was rather weak, but it significantly increased after
the transition to the outburst stage B. This indicates that the accretion disc at the superoutburst plateau
stage experiences more complex evolution than just the variation of its radius, suggesting substantial
expanding of the SLS area or/and lowering of its temperature.

\begin{figure*}
\begin{center}
\hbox{
\includegraphics[width=8.7cm]{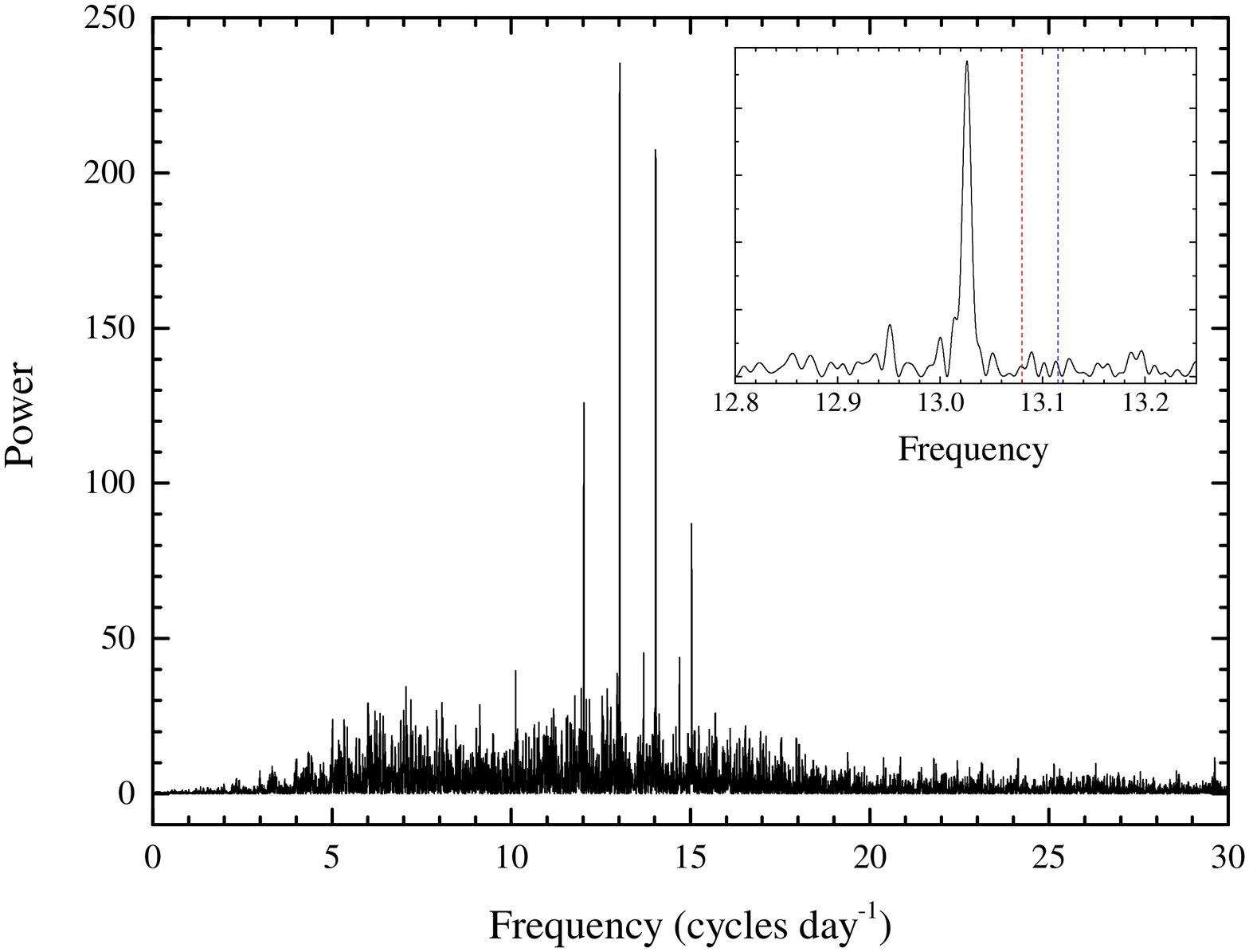}
\includegraphics[width=8.7cm]{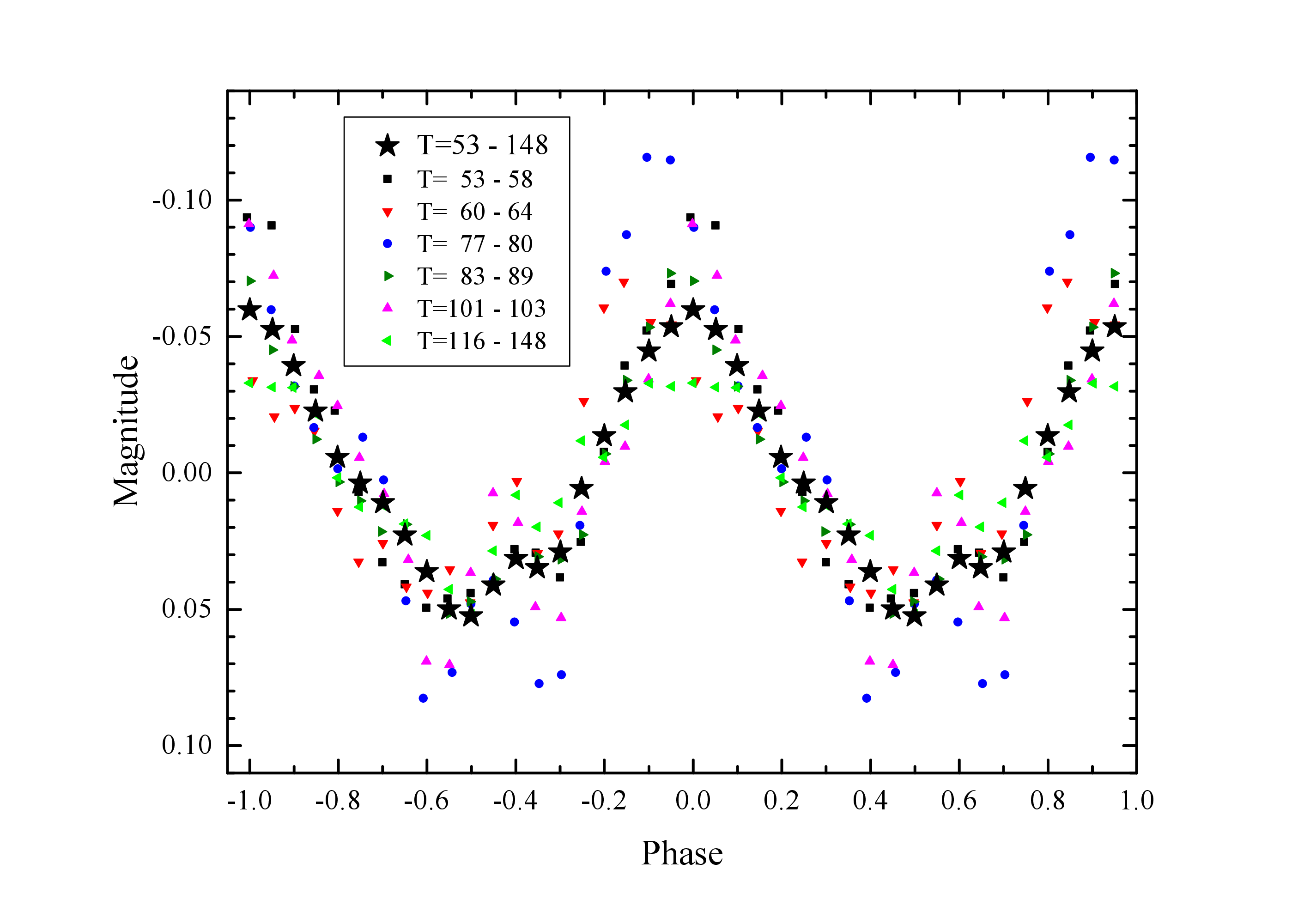}
}
\end{center}
\caption{Left: Lomb--Scargle periodogram of the $V$ data combined between $T$=53 and 148. The inset shows
         the enlarged region around the frequency 13.026 \cd\ -- the frequency of post-superoutburst
         modulations observed in 2013. The vertical dashed lines mark the orbital (blue) and
         stage B superhump (red) frequencies of 13.11 and 13.08 \cd, respectively.
         Right: the entire set (large stars) and short subsets (small symbols) of the $V$ data between
         $T$=53 and 148, folded according to ephemeris $HJD_{max} = 245\,6300.0610(3)+0.07676887(5) \cdot E$.}
\label{Fig:PostOutburst}

\begin{center}
\hbox{
\includegraphics[height=6.5cm]{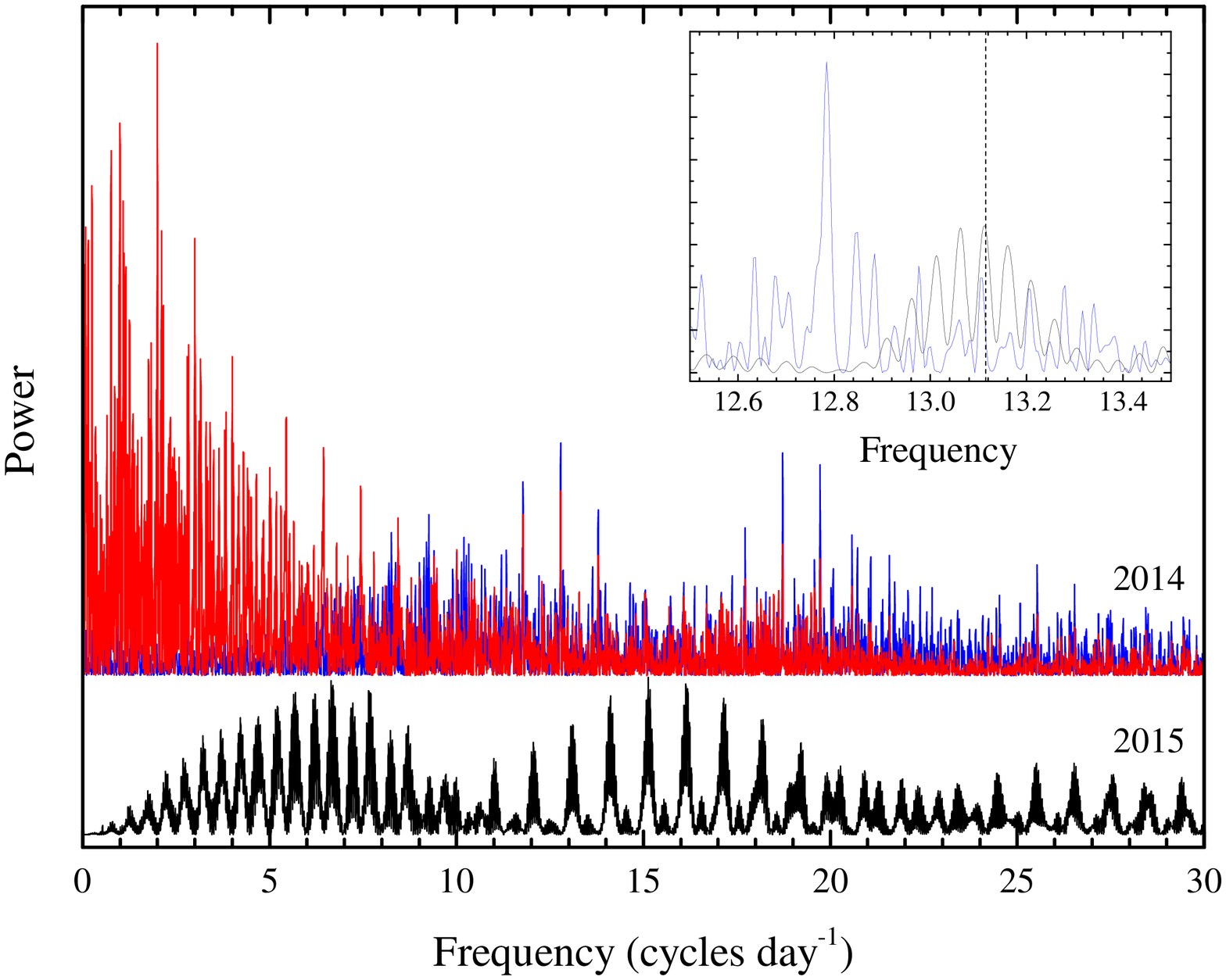}
\hspace{5 mm}
\includegraphics[height=6.5cm]{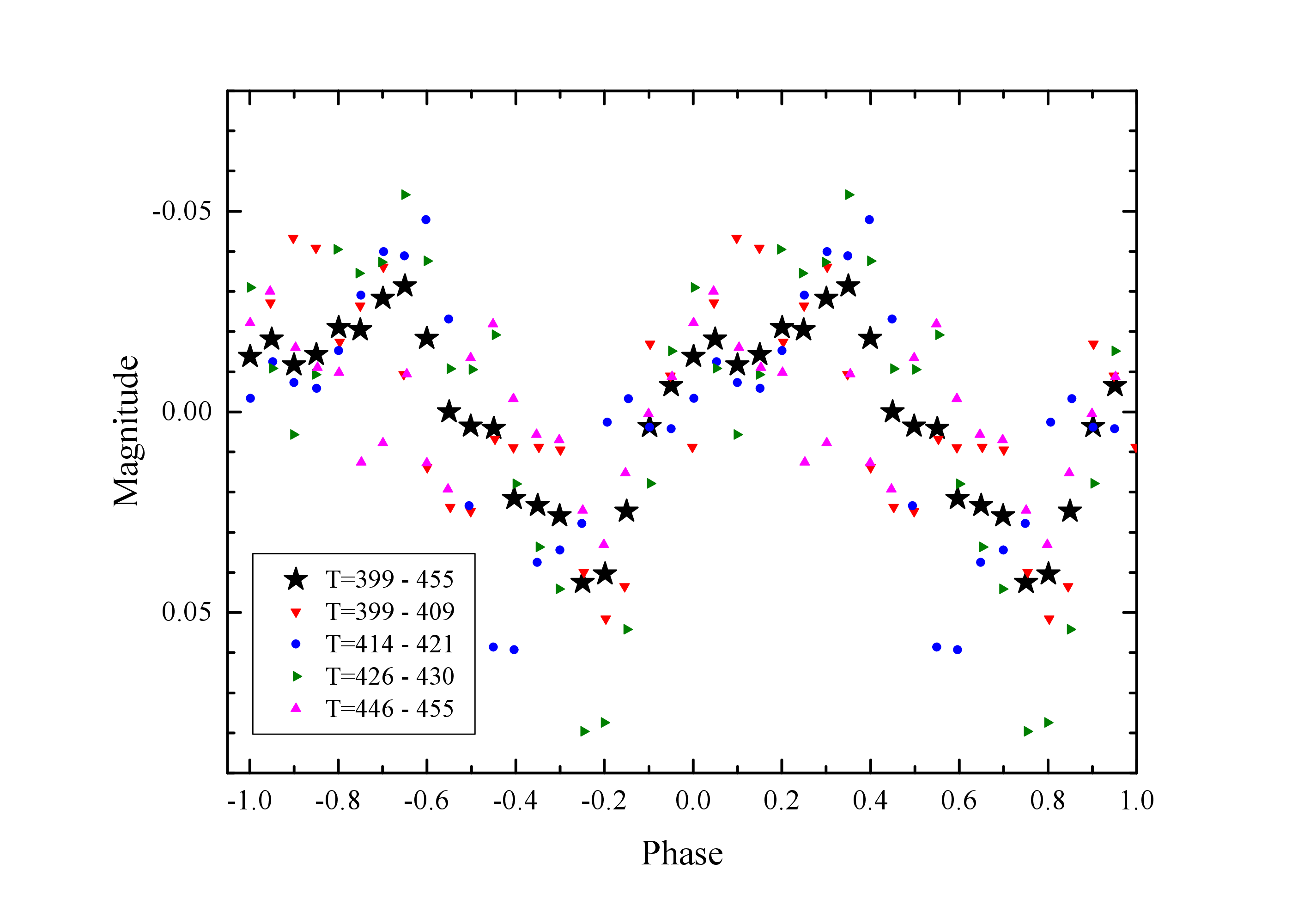}
}
\end{center}
\caption{Left: Lomb--Scargle power spectra for optical data of \SSS\ obtained in 2014 ($T$=399--455, two upper
         spectra) and 2015 ($T$=740--763, a bottom spectrum). The red line represents the spectrum for the original
         2014 photometry, whereas the blue and black lines are for the detrended data.
         The inset shows the enlarged region around the orbital frequency marked by the vertical dashed line.
         Right: the entire set (large stars) and short subsets (small symbols) of the detrended data,
         obtained in 2014 and folded with the period of 112.6~min.}
\label{Fig:PS2014}
\end{figure*}

\subsubsection{Post-superoutburst modulations}
\label{Sec:PostSuperhumps}

A visual inspection of the data revealed the presence of persistent modulations in the light curve
long after the superoutburst. We show below that a period of these modulations is longer than that
of the stage B superhumps, thus also implying their superhump nature.
We traced them until the end of our time series observations in 2015.
In order to explore these modulations, we first studied the observations obtained in 2013 after the
rapid fading phase. We divided this
light curve into short pieces of a 3--6~d duration and analysed them separately. We found that since
at least $T$=53 the modulation had a very stable period. The Lomb--Scargle periodogram of the $V$ data
combined between $T$=53 and 148 shows a very narrow and strong peak at 13.0261 \cd\
(Fig.~\ref{Fig:PostOutburst}, left-hand panel). Folding short subsets at the period 110.547~min
results in modulations of similar amplitude which are all in phase with each other (Fig.~\ref{Fig:PostOutburst},
right-hand panel). The shape of the average profile very much resembles a typical superhump profile (see
e.g. fig.~1 in \citealt{Kato1}). The colour variability of these post-superoutburst superhumps
had changed significantly when comparing with the superhumps observed during the outburst stage
(Fig.~\ref{Fig:SuperhumpsColours}). The period of these post-superoutburst superhumps is about 0.4\%
longer than that of the stage B superhumps.

The modulations were still strong in 2014 ($T$=399--455), although during this season they were
superposed on a longer-term
variability with a time-scale of 8--12~h, which resembles so-called `brightenings' observed in a few
other WZ~Sge-type systems, i.e. in V406~Vir \citep{SDSS1238a,SDSS1238b}. This longer-term variability
produces in the power spectrum an excess of power at low frequencies of 2--3 \cd\ (Fig.~\ref{Fig:PS2014},
left-hand panel, the red spectrum), making it difficult to study the superhump modulations. For this
reason, we first detrended each night's light curve by a second order polynomial. The Lomb--Scargle
periodogram of this detrended data set is dominated by two sharp peaks with frequencies of 12.78 \cd\
(112.6~min) and its first harmonic, and of 18.72 \cd\ (76.9~min, Fig.~\ref{Fig:PS2014}, left-hand panel,
the blue spectrum). The period of the 112.6~min modulation is $\sim$1.8\% longer than
the period of the post-superoutburst superhumps observed in 2013. The modulations seemed very stable.
Folding short subsets of the 2014 light curve at the 112.6~min period results in profiles of similar
amplitude which are all in phase with each other (Fig.~\ref{Fig:PS2014}, right-hand panel).
The origin of the 18.72 \cd\ peak in the power spectrum is unclear.

The most recent time-resolved observations, obtained in 2015 ($T$=740--763), basically confirmed the
photometric behaviour observed in 2014. However, the time series analysis may suggest that superhumps
were finally replaced by orbital variability. There is a peak in the power spectrum (see the inset in
Fig.~\ref{Fig:PS2014}, left-hand panel) coinciding with the orbital frequency of 13.112 \cd, derived
from spectroscopic data in
Section~\ref{Sec:OrbPeriod}. Unfortunately, most of the 2015 data are quite noisy and of a short duration.
The aliasing problem prevents us from claiming a detection of the orbital signal in the optical light curve.

Thus, our observations showed that long after the superoutburst \SSS\ was still exhibiting very stable
superhumps, with a period longer than that of the stage B superhumps. Post-superoutburst superhumps,
also called `late superhumps' \citep{Vogt83}, were observed in several other WZ~Sge-type objects
\citep[see][and references therein]{KatoLateSuperhumps,KatoWZ}. In \SSS, however, they were detectable
for significantly longer time than in any other system. In Section~\ref{Sec:LateSuperhumps} we discuss
these late superhumps in more detail.

\begin{figure}
\begin{center}
\hbox{
\includegraphics[width=8.5cm]{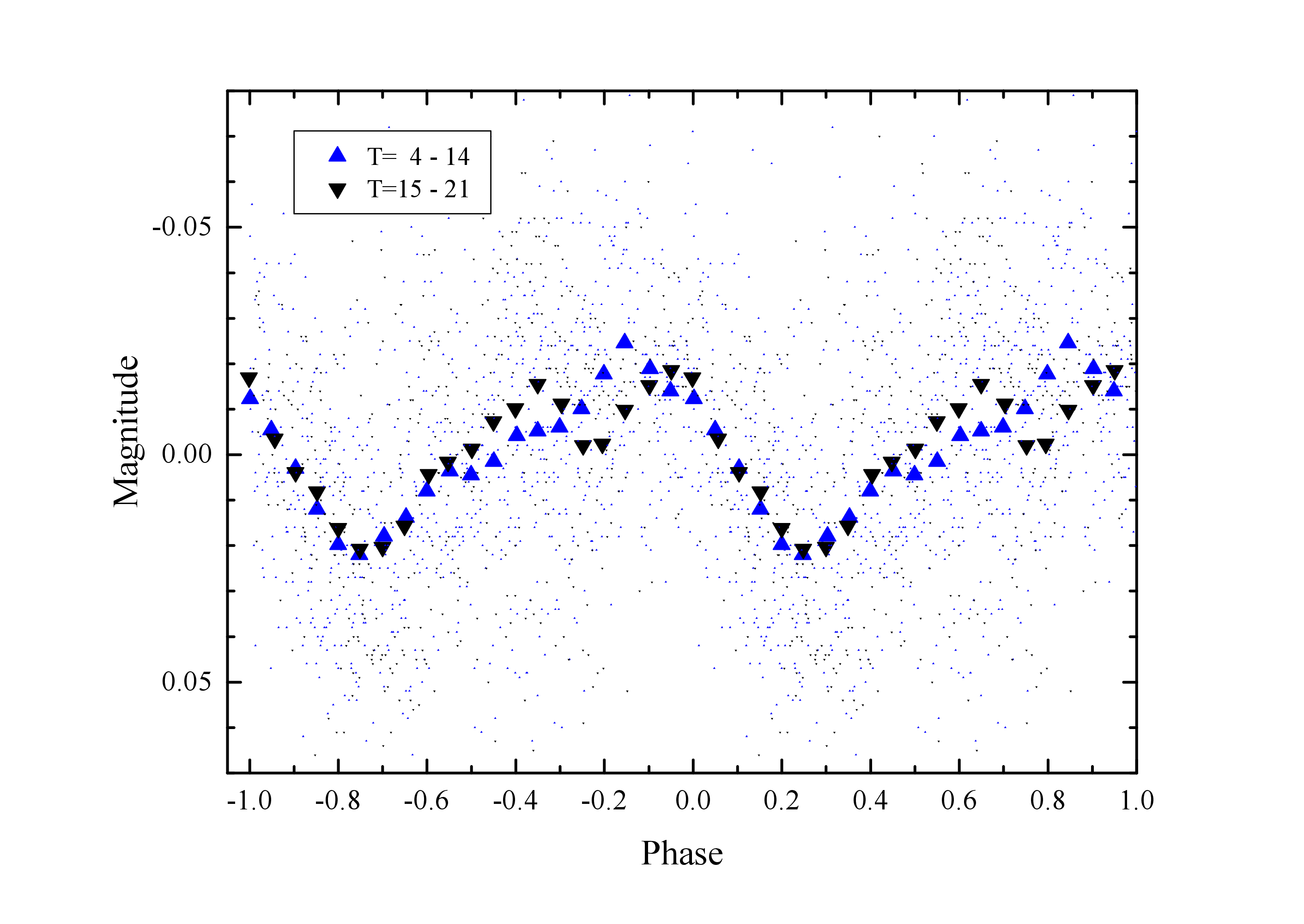}
}
\end{center}
\caption{
The $V$ light curves from the time intervals $T$=4--14 and 15--21 folded with the period of 180.48~min
and averaged in 20 phase bins. The small and large symbols represent the individual observations and
the data averaged in 20 phase bins, respectively. All data are plotted twice for continuity.}
\label{Fig:Folded3h}
\end{figure}

\subsection{An $\sim$180-min variability}
\label{Sec:3h}

The periodograms of most segments of the light curve of \SSS\ show an excess of power around frequencies
$\sim$6--9 \cd\ with the strongest peak at $\sim$8.0 \cd\ (180~min, Figs~\ref{Fig:PSOutburst} and
\ref{Fig:PostOutburst}). This signal was most evident during the outburst stage A. It then weakened,
but was still detectable long after the superoutburst. We note that the global power spectrum
in this frequency range is complex, suggesting a lower degree of coherence, though between $T$=4 and 21
the signal was rather stable. Fig.~\ref{Fig:Folded3h} shows the $V$ light curves from the time intervals
$T$=4--14 and 15--21 folded with the period of 180.48~min and averaged in 20 phase bins. 

A period of 180~min is not directly related to superhump or orbital periods of \SSS\
($P_{orb}$=109.80 min; see Section~\ref{Sec:OrbPeriod}), which are significantly shorter. We note that
there are known a few other short-period CVs which exhibit variabilities with a period of a few hours,
longer than the orbital one (e.g., GW~Lib -- \citealt{GW_Lib}; FS~Aur -- \citealt{FS_Aur0}; V455~And --
\citealt{V455And}). Among those systems only in FS~Aur is such a period very coherent over many years
\citep{FS_Aur}. It was proposed that the precession of a fast-rotating magnetically accreting WD can
successfully explain this phenomenon \citep{FS_Aur2,FS_Aur3}. However, in other objects long period
modulations show a significantly lower degree of coherence \citep{GW_Lib,V455And}, similarly to \SSS.
There is currently no physical explanation for the latter phenomena.

\begin{figure*}
\centering
\includegraphics[width=17.5cm]{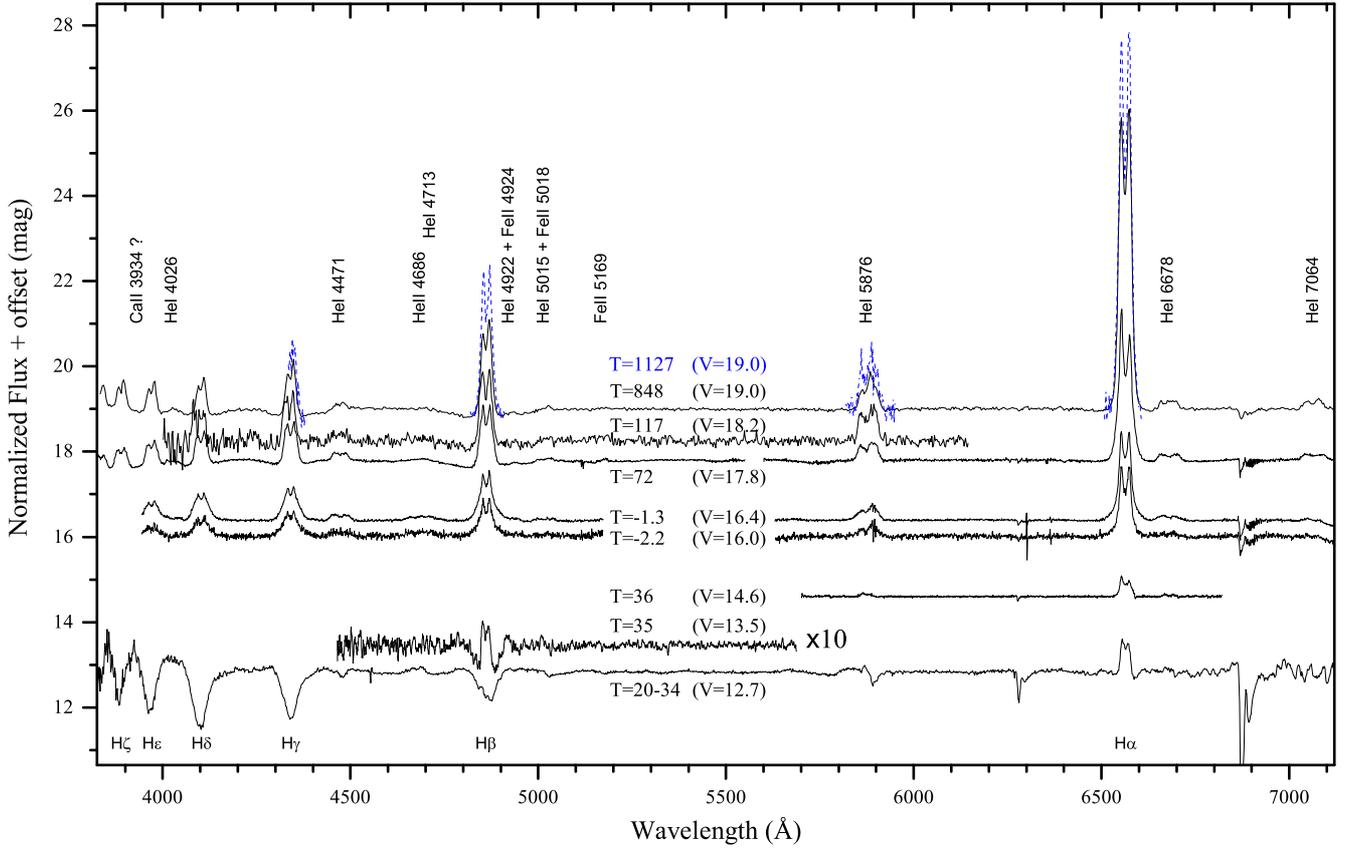}
\caption{Spectral evolution of \SSS\ through all the outburst stages and the following decay.
         All the spectra are normalized to the continuum flux and shifted with respect to each other
         in accordance to the magnitude of the system during the given spectral observation. Thus, the
         outburst spectra ($V$$\approx$12.7) are at the bottom and the quiescent spectra ($V$$\approx$18--19)
         are at the top of the figure. The $T$=1127 spectrum is relatively noisy at the continuum level,
         therefore we only show the emission lines (a dashed blue line).
         Two lower outburst spectra are scaled up by a factor of 10 in relative intensity
         to magnify the spectral features and allow for a better comparison.}
\label{Fig:NormSpec}
\end{figure*}

\begin{figure*}
\begin{center}
\hbox{
\includegraphics[height=7.5cm]{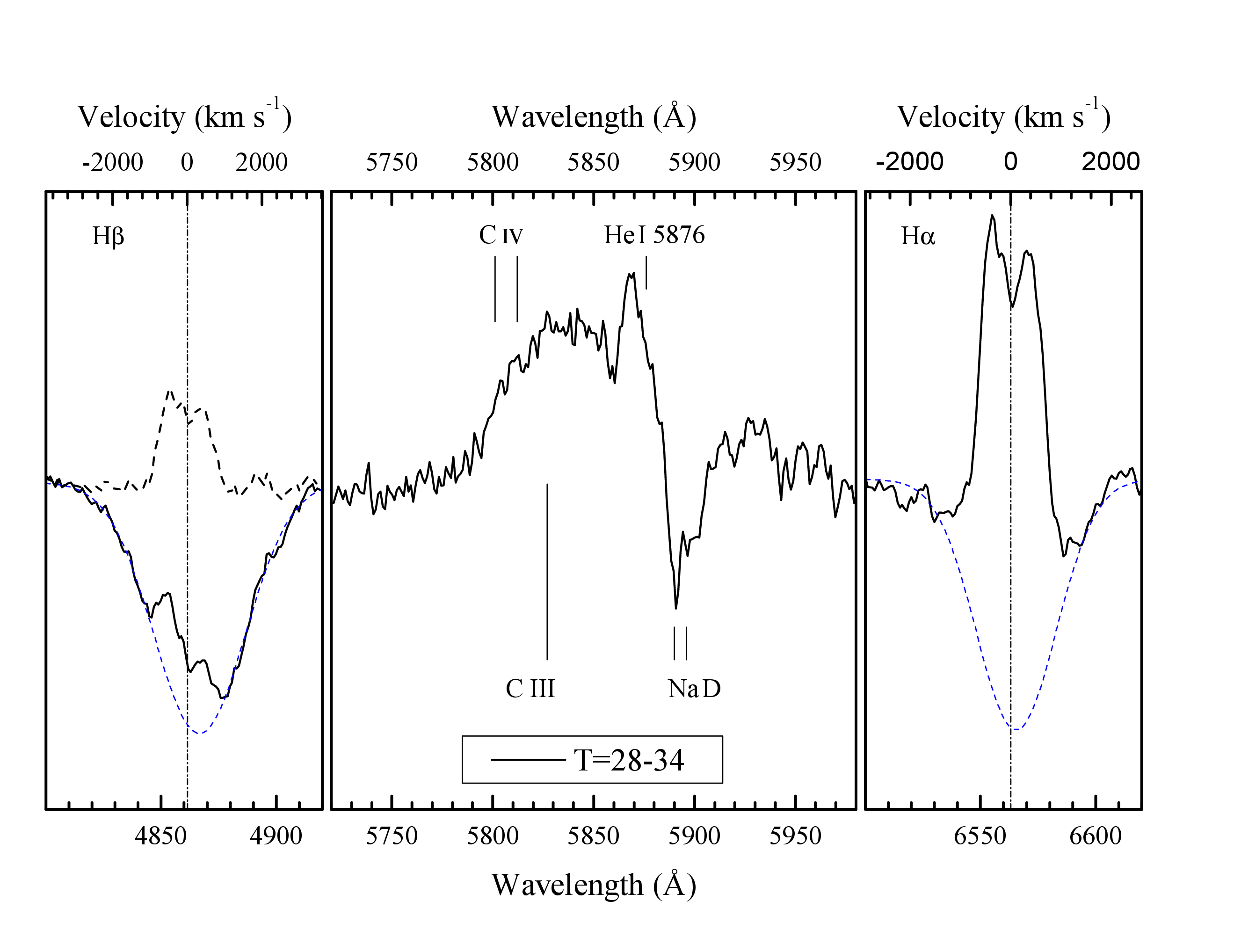}
\hspace{8 mm}
\includegraphics[height=7.5cm]{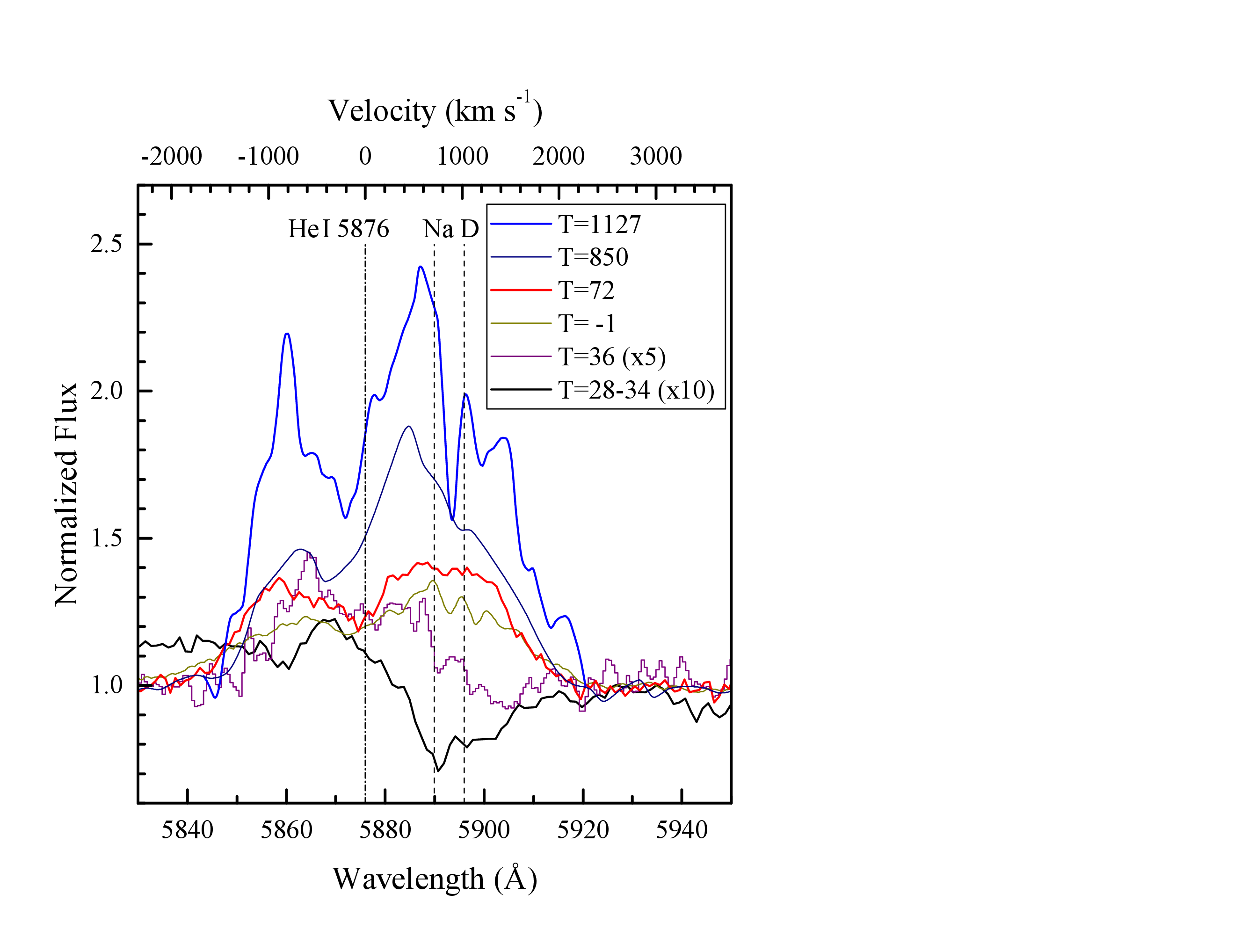}
}
\end{center}
\caption{Left: \Halpha\ and \Hbeta\ line profiles and the region around the \CIII\ / \CIV\ / \HeI\ \l5876 / \NaD\
         complex (black solid line) during the plateau stage of the superoutburst (arbitrary scaling). The dashed
         blue line shows a Gaussian fit to the absorption components of \Halpha\ and \Hbeta. The dashed black line
         shows the profile of \Hbeta\ after being corrected for absorption.
         Right: \HeI\ \l5876 and \NaD\ line profiles from different data sets, shown in actual scale (for a better
         comparison, the $T$=28--34 and 36 profiles have been multiplied by a factor of 10 and 5, respectively).}

\label{Fig:LineProfiles}
\end{figure*}

\section{Optical spectroscopy}

\subsection{Spectral evolution throughout the superoutburst}
\label{Sec:SpecEvolution}

Fig.~\ref{Fig:NormSpec} displays the spectral evolution of \SSS\ through all the outburst stages
and the following decay (the epochs of optical spectra are shown in Fig.~\ref{Fig:LC}).
Note that during the superoutburst plateau the spectra changed very little, hence we show only
the averaged outburst spectrum ($T$=20--34).

At the temporary fading stage ($T$=$-$1 and $-$2), the spectra exhibit strong, very broad and
double-peaked Balmer emission lines and weak emission lines of neutral helium (4471, 5876, 6678 \AA).
There is also a hint of a broad feature around 4700 \AA\ that can be associated with the Bowen blend,
\HeII\ \l4686 and \HeI\ \l4713.

During the plateau stage ($T$=20--34) most of the emission lines were replaced by the broad absorption troughs
with weak emission cores. Among the Balmer series lines, only \Halpha\ was still in emission, even though
the broad absorption wings also showed up around the line. The full width at zero intensity (FWZI) of
the absorption troughs corresponds to a velocity of $\sim$6800 \kms, which is very close to the fairly extreme
width of the emission lines (FWZI$\sim$7000 \kms) observed during the temporary fading. It can be explained
by the broadening effect of rapid rotation of the innermost parts of the optically thick accretion disc.
An interesting feature is that those absorption components are clearly redshifted relative to the emission
core, which are seemingly always centred close to the rest wavelength (Fig.~\ref{Fig:LineProfiles},
left-hand panel). By fitting a Gaussian to the absorption component of \Hbeta\ and \Hgamma\
in nightly averaged outburst spectra, we found a consistent and steady offset of $\sim$180--200 \kms\
from the rest wavelength. At the superoutburst plateau the \CIII/\NIII--\HeII\ complex and \HeI\ \l5876
and \l6678 lines were still in emission. Also, an emission bump centred at $\sim$5830 \AA\ appeared. This
bump is probably a blend of \CIII\ \l5827 and the \CIV\ lines at \l5801 and 5812 \AA, and possibly of
other highly excited species on the shortward side of \HeI\ \l5876. In addition, the spectra show the
appearance of the sodium D doublet (\NaD\ \l5890/5896) in absorption. These lines are very uncommon for
CV spectra. Although the sodium doublet is one of the strongest interstellar lines, it is not expected
to be seen in the spectra of CVs because of their relatively close proximity. Nonetheless, \NaD\ in
absorption was noticed in spectra of several WZ~Sge-type systems at different stages of their superoutbursts
\citep{PattersonEGCnc,NogamiWZSge,NogamiGWLib}, but it was not seen in other dwarf novae showing large
amplitude outbursts \citep[e.g. BZ~UMa;][]{BZ_UMa}.

\begin{table*}
\caption{EWs of the most prominent lines measured from the averaged spectra. Negative values
         are used for \emph{absorption} lines.}
\label{Tab:EW}
\centering
\begin{tabular}{cccccccc}
\hline
Epoch   &   \Halpha    &   \Hbeta     &  \Hgamma    &  \Hdelta    &\HeI\ \l5876 &\HeI\ \l4471 & Magnitude \\
$T$     &              &              &             &             &  + \NaD     &             &   ($V$)   \\
\hline
 --2    & 58.0$\pm$0.9 & 32.5$\pm$0.9 &27.2$\pm$0.6 &18.4$\pm$0.4 &11.4$\pm$0.9 & 5.6$\pm$0.6 & 16.0 \\
 --1    & 81.8$\pm$0.3 & 47.4$\pm$0.4 &34.6$\pm$0.2 &27.7$\pm$0.2 &16.5$\pm$0.3 & 7.3$\pm$0.3 & 16.4 \\
28--34  &  2.0$\pm$0.1 & -3.4$\pm$0.1 &-5.3$\pm$0.1 &-6.0$\pm$0.1 & 0.3$\pm$0.1 &-0.3$\pm$0.1 & 12.7 \\
  35    &   \dots      & -2.1$\pm$0.2 &   \dots     &   \dots     &   \dots     &   \dots     & 13.5 \\
  36    & 12.3$\pm$0.2 &    \dots     &   \dots     &   \dots     & 1.7$\pm$0.4 &   \dots     & 14.6 \\
  72    &126.0$\pm$0.3 & 62.6$\pm$0.3 &35.3$\pm$0.3 &26.2$\pm$1.3 &22.8$\pm$0.9 & 8.2$\pm$0.3 & 17.8 \\
 117    &   \dots      & 83.9$\pm$1.6 &40.7$\pm$2.4 &25.7$\pm$4.2 &38.7$\pm$1.9 & 9.7$\pm$1.4 & 18.2 \\
 850    &277.3$\pm$0.9 & 86.5$\pm$0.2 &38.5$\pm$0.4 &23.9$\pm$0.4 &31.6$\pm$0.7 & 5.2$\pm$1.0 & 19.0 \\
1127    &384.1$\pm$8.5 &112.2$\pm$3.3 &63.0$\pm$5.0 &   \dots     &51.3$\pm$3.0 & 9.9$\pm$2.3 & 19.0 \\
\hline
\multicolumn{8}{l}{
Note: The Magnitude column indicates the approximate magnitude of \SSS\ during the observation.}\\
\multicolumn{8}{l}{
They are taken from simultaneous or interpolated from recent photometric observations.}\\
\end{tabular}
\end{table*}

After the rapid fading the emission bump at $\sim$5830 \AA\ disappeared, but all the absorption lines
transitioned from absorption to emission. \NaD\ also appeared in emission, distorting the red wing
of the double-peaked \HeI\ \l5876 line in all post-outburst spectra, even three years after the
superoutburst (see Fig.~\ref{Fig:LineProfiles}, right-hand panel). We point out that
\NaD\ was also observed in emission in GW~Lib some 40 d after the rapid fading phase
\citep{GWLibVanSpaandonk}. We also note the presence of a narrow, unidentified line at $\sim$5903\,\AA.
This feature seems to be real and not noise or an artefact of the data reduction because it is present
in different spectra and also exhibits a transition from absorption to emission, similarly to \NaD.

As the system flux continued to decline, the emission lines got progressively stronger. In the spectrum
taken during the decay phase on $T$=72, the \Halpha\ has a moderate strength (EW=129\AA) and the Balmer
decrement is relatively flat: \Halpha:\Hbeta:\Hgamma=2.01:1.00:0.56. However, the final spectra taken in
quiescence on $T$=848 and 1127 show the extremely strong \Halpha\ line which peaks at 8 and 11 times the
continuum and has EW=277 and 384 \AA, respectively. To the best of our knowledge, this is the strongest
emission line ever observed in a CV. The Balmer decrement is also very steep in these spectra:
\Halpha:\Hbeta:\Hgamma=3.20$:$1.00$:$0.44 ($T$=848) and 3.42:1.00:0.56 ($T$=1127). This change in the
Balmer decrement from shallow to very steep indicates a transition from optically thick to optically
thin conditions in the accretion disc.
We also note that the spectra obtained on $T$=72 and later show very wide symmetric
absorptions at the wings of the Balmer lines (FWZI of the \Hbeta\ absorption is $\sim$19500
\kms). These absorption features are too broad to be produced in the Keplerian accretion disc and can only be
formed in the high-density photosphere of the WD, indicating a very low accretion luminosity and
mass-accretion rate. Under such conditions one could expect to see the donor star lines in red part of
the spectrum; however there is no evidence at all for atomic lines (e.g. \NaI\ \l8200) or molecular bands
from the donor. We also note that the spectra from all available time-resolved data sets show
no obvious radial velocity variability in the accretion  disc emission lines. In Table~\ref{Tab:EW} we present
the equivalent width (EW) measurements of the most prominent lines measured from the averaged spectra.

\begin{figure}
\centering
\includegraphics[height=6.5cm]{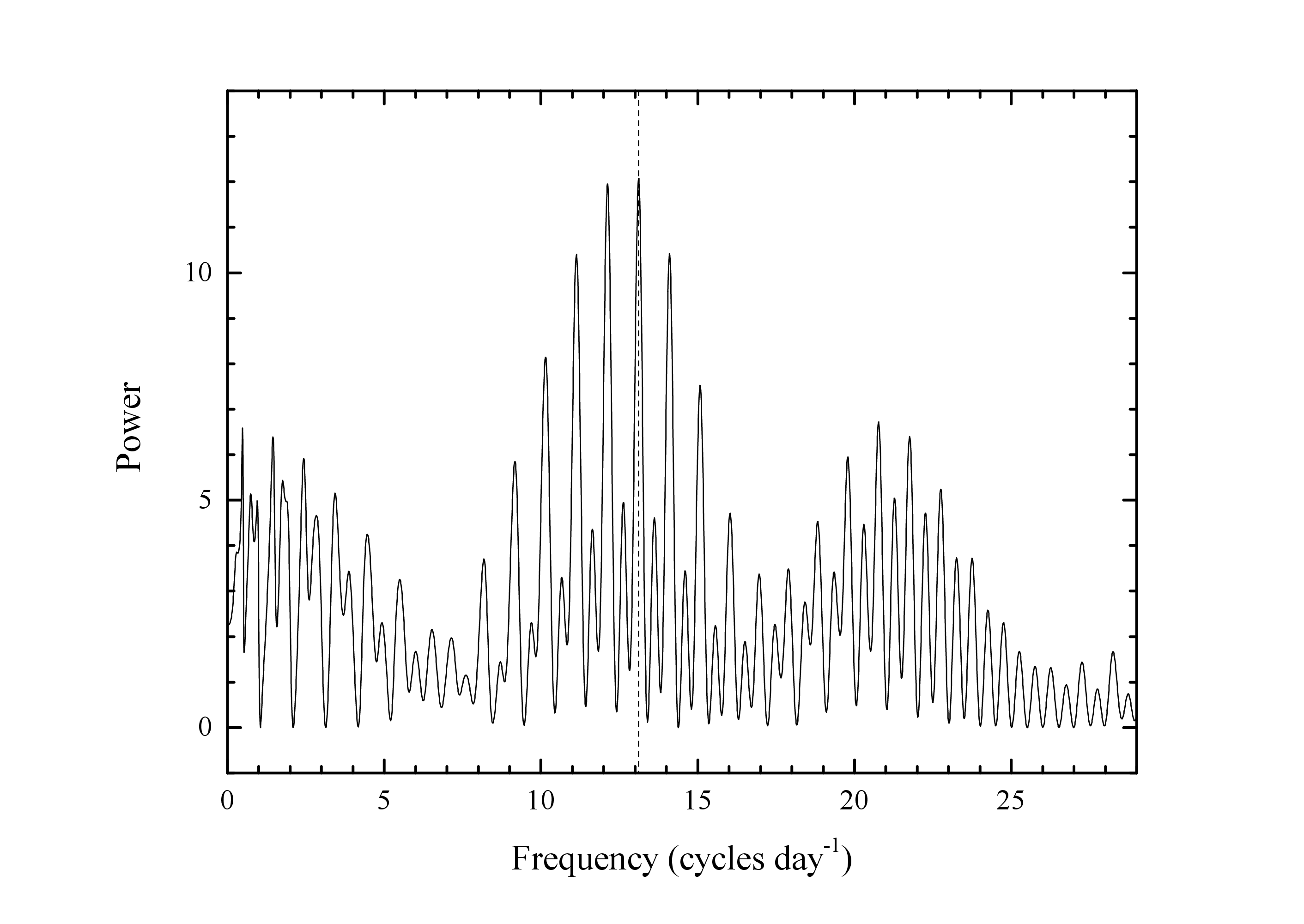}
\caption{Lomb--Scargle periodogram for spectral data of \SSS\ obtained during the $T$=28--30 time interval.
         The vertical dashed line marks the orbital frequency of 13.112 \cd.
         }
\label{Fig:PSradvel}
\end{figure}

\begin{figure}
\begin{center}
\hbox{
\includegraphics[height=8.5cm]{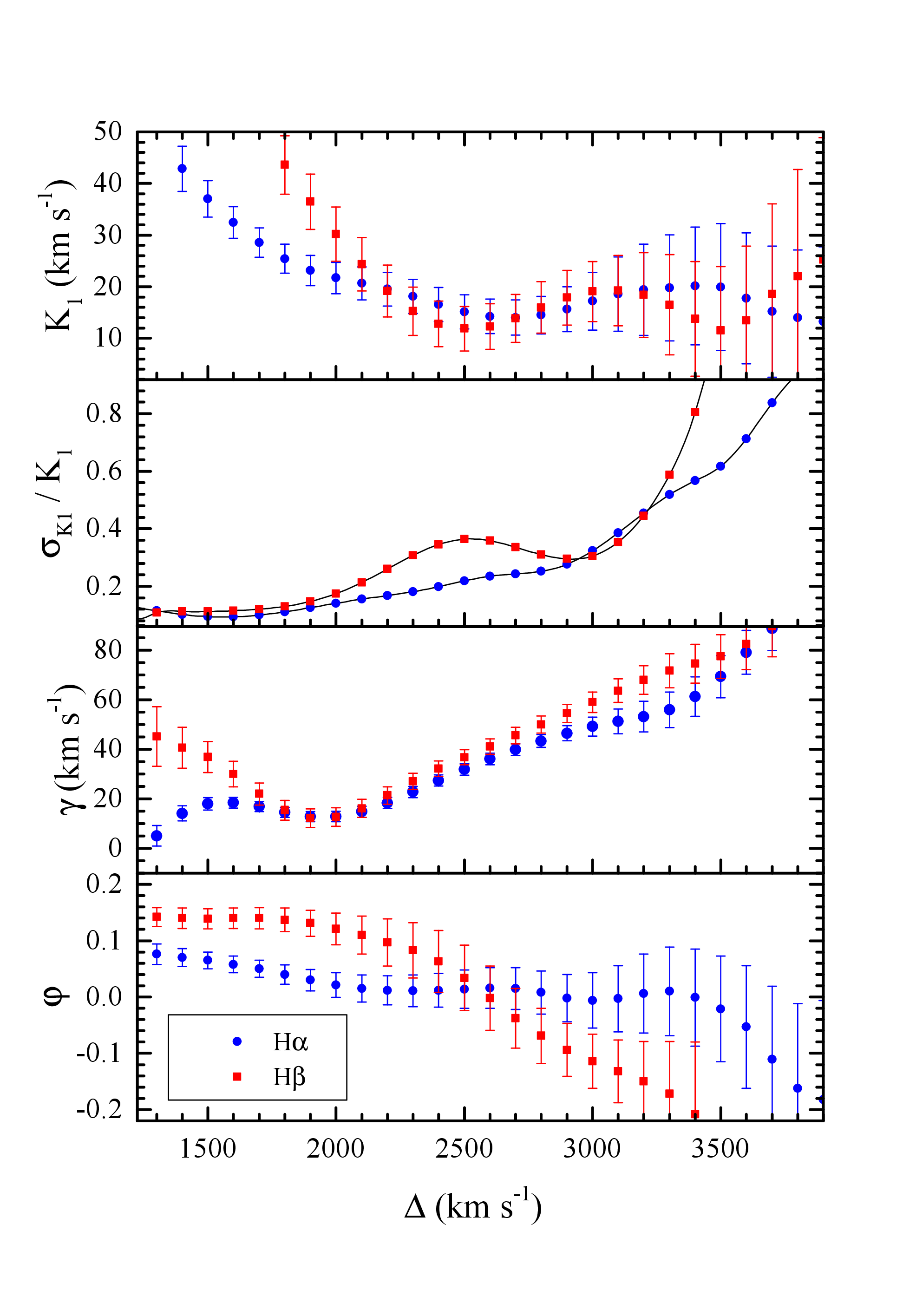}
}
\end{center}
\caption{The diagnostic diagram for the \Halpha\ and \Hbeta\ emission lines from the $T$=72 observations,
         showing the response of the fitted orbital elements to the choice of the double-Gaussian separation.}
\label{Fig:RVdiagram}
\end{figure}

\subsection{Orbital period determination}
\label{Sec:OrbPeriod}

The most direct method of estimating the orbital period of the binary system involves optical spectroscopy.
We used the spectra taken during the superoutburst stage ($T$=28--30) to identify the orbital period of
\SSS. The chief problem was to identify the correct night-to-night alias and only these spectra had long
enough duration while being
taken over consecutive nights for this to be possible. At this stage the lines were in absorption and
asymmetric, but in finding a period, one only requires detection of a periodic variation and any offsets
are immaterial. We measured the radial velocity of H$\beta$, the profile of which showed the strongest
variations, using cross-correlation with a single Gaussian
\citep{Sch:Young}. This has one free parameter, the width of the Gaussian. We tried widths (FWHM) varying
from 1500 to 3000 \kms. For each set of radial velocities, we computed 2000 Lomb--Scargle periodograms
using bootstrap resampling, in each case retaining the period of the highest peak to build up statistics
upon the reliability of the period identification. An example of the periodogram calculated
for the FWHM of 2500 \kms\ is shown in Fig.~\ref{Fig:PSradvel}. The results clustered around two peaks
split by one \cd\ at frequencies of 12.127 and 13.112 \cd.
Although the radial velocities alone do not quite secure the alias, we are sure that in fact the higher
frequency peak of the pair corresponds to the orbital modulation. Indeed, the period of this modulation
is only slightly longer than the period of the superhump signal, as expected for the traditional,
`positive' superhumps which are standard for this class of star. Otherwise, it will lead to the
`negative' superhumps with an unrealistically large difference between orbital and superhump periods.
\footnote{
`Negative' superhumps, which are modulations with a period a few percent shorter than the binary's
orbital period, are sometimes observed in CVs (see e.g. an analysis of the {\it Kepler} data of V344~Lyr
by \citealt{WoodKepler}). In most cases they appear in quiescence and sometimes can survive a normal
outburst \citep{WoodKepler,OsakiKato2013}. However, in rare cases when the negative superhumps were
detected in superoutbursts, they were significantly weaker than normal, positive superhumps \citep{OsakiKato2013}.
Thus, we have no doubt that all the observed superhumps in \SSS\ were of the normal, positive type.}
Naturally, a more complete radial velocity data set is desirable to confirm this reasoning.

We thus determined the orbital period $P_{orb}$ to be 109.80(7) min. This period is in the vicinity
of a possible orbital period of 109.266 min identified by \citet{KatoSSS}. However, we could not confirm
this value from our own analysis of both the CRTS data and our observations, whereas our late photometric
data show a peak in the power spectrum coinciding with the adopted orbital frequency of 13.112 \cd\
(Fig.~\ref{Fig:PS2014}, left-hand panel).

\begin{figure}
\includegraphics[width=8.6cm]{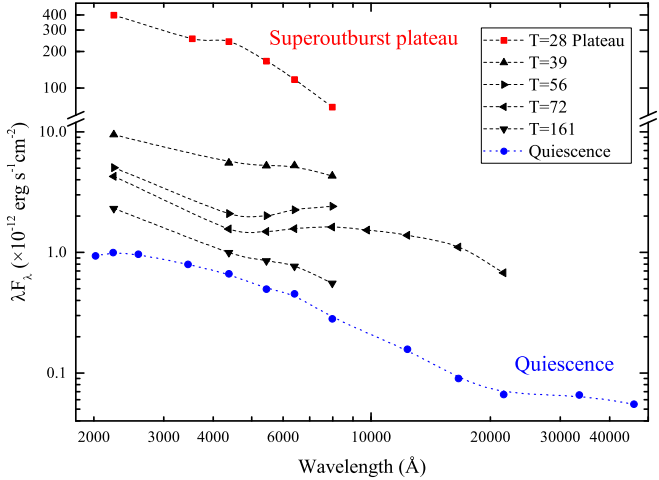}
\caption{The SED of \SSS\ at several key moments of the superoutburst,
         the following decay and quiescence. The SED in quiescence is created from the most recent observations
         taken in 2016 April ($T$=1186--1196).}
\label{Fig:SED1}
\end{figure}

\begin{figure*}
\includegraphics[width=14cm]{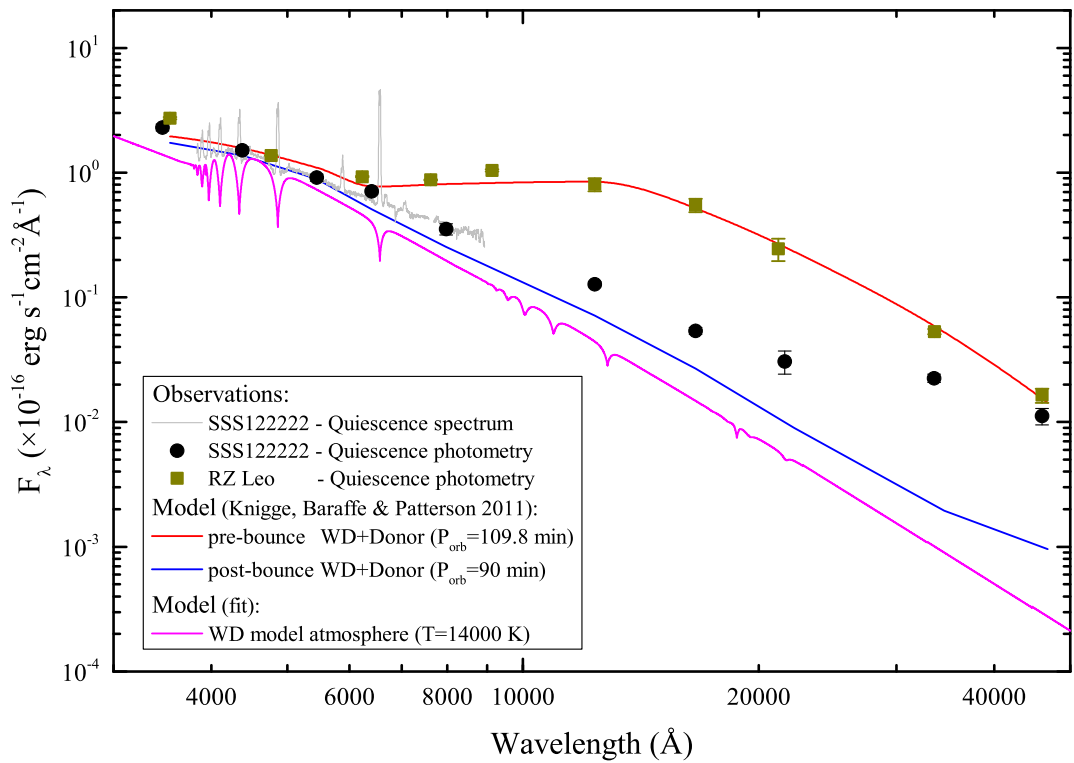}
\caption{Theoretical SEDs of pre- and post-bounce CVs, based on the
         models of \citet{KniggeCVevol}, shown together with the photometric and spectroscopic data of \SSS\
         and RZ~Leo in quiescence (see text for details).}
\label{Fig:SED2}
\end{figure*}

\subsection{Radial velocity study}
\label{Sec:RadVel}

Although our spectra do not show appreciable radial velocity variability, we made
an attempt to measure or put the upper limits on the radial velocity semi-amplitudes of emission lines
using the double-Gaussian method \citep{Sch:Young}. In order to test for consistency in the derived
velocities and the zero phase, we separately used the emission lines \Halpha\ and \Hbeta\ from the
$T$=72 data set. Additionally, to improve the confidence and reliability of the obtained results, we
also created the phase-binned spectra by averaging the individual spectra in 12 phase bins.

All the measurements were made using various Gaussian FWHMs of 50--300 ${\rm km~s^{-1}}$ and different
values of the Gaussian separation $\Delta$ ranging from 1000 \kms\ to 5000 \kms\ in steps of 100 \kms,
following the technique of `diagnostic diagrams' \citep{Shafter1986}. For each value of $\Delta$ we made
a non-linear least-squares fit of the derived velocities to sinusoids of the form
  \begin{equation}  \label{radvelfit}
    V(\varphi,\Delta )=\gamma (\Delta )-K_1(\Delta )\sin \left[ 2\pi \left(
    \varphi-\varphi_0\left( \Delta \right) \right)\right]
  \end{equation}
where $\gamma$ is the systemic velocity, $K_1$ is the semi-amplitude, $\varphi_0$ is the phase of inferior
conjunction of the secondary star and $\varphi$ is the orbital phase calculated relative to epoch
$T_{0}$ = HJD 245\,6372.65160.

The resulting diagnostic diagrams are shown in Fig.~\ref{Fig:RVdiagram}. We found the maximum useful
separation to be $\Delta_{max} \simeq 2800-3000$ \kms\ for both lines until the noise in the line wings
begins to dominate. $K_1$, $\gamma$ and $\varphi_0$
are quite stable over a reasonable range of Gaussian separations around $\Delta_{max}$, thereby helping
us with their choice. Both the tested lines show very consistent results, and we adopted the following
values of the orbital parameters: $K_1=16\pm5$ \kms, $\gamma=44\pm6$ \kms, and ${\varphi}_0=0.00\pm0.06$.
The formal errors are the standard deviations of the least-squares fits. They most likely underestimate
the true errors, as they may not include a priori unknown systematic effects. Moreover, it is well known
that the parameters obtained with the double-Gaussian method are often affected by systematic errors. Thus,
the derived value of $K_1$ should be used with great caution.

\section{The spectral energy distribution}
\label{Sec:SED}

\subsection{The SED evolution}
\label{Sec:SEDevol}

Using our multicolour observations, we have reconstructed the spectral energy distribution (SED) of
\SSS\ in the UV--optical--NIR wavelengths at several key moments of the superoutburst, the following decay
and quiescence (Fig.~\ref{Fig:SED1}). We note, that although the SED in quiescence was
created from the most recent observations obtained in 2016 April ($T$=1186--1196), its UV--optical segment
coincides precisely with the two previous multiband observations ($T$=536 and 737). In addition, we show
the archival pre-outburst observations taken by the Wide-field Infrared Survey Explorer (\textit{WISE})
observatory in the IR band on $T$=$-$920. We also examined NEOWISE-R\footnote{The \textit{WISE} was placed into
hibernation in 2011 February. The survey continued as NEOWISE \citep{NEOWISE}, when the spacecraft was brought
out of hibernation. NEOWISE observations in the 3.4 and 4.6 $\mu$m bands were resumed in December 2013.}
Single Exposure Source Table and found at least six sources which are a very good
positional match to \SSS. All these observations were made between $T$=368 and 906, i.e. when \SSS\ was
already close to or in quiescence. Although the signal-to-noise ratio of this photometry is low ($\sim$3),
all the measurements in the W1 filter (3.4 $\mu$m) are consistent with the pre-outburst WISE observations,
giving the same average magnitude of 16.55. This allowed us to include the preoutburst W1 and W2 (4.6 $\mu$m)
measurements in the current SED in quiescence.

The SED shows a complex transformation through the superoutburst and the following decay. Its
UV--optical part during and after the active stage was evolving in a similar manner to previous observations
of outbursts \citep[see e.g.][]{Verbunt87,Warner}, indicating the evolution of the accretion disc and
boundary layer, and the possible cooling of the accretion-heated WD. On $T$=28, the flux from the $V$ to $I$
bands had a slope $\alpha$=3.29$\pm$0.07 ($f_{\lambda}$$\propto$$\lambda^{-\alpha}$) that is close, but
still less than the Rayleigh--Jeans tail of an accretion disc spectrum. After the rapid fading phase, a
strong red--NIR excess emerged, peaking at 8000--10000 \AA, which disappeared about 100~d later.
In quiescence, the flux in the red-NIR part of the spectrum (from $R$ to $Ks$ bands) is linearly
decreasing (in log--log space), having a slope $\alpha$=2.66$\pm$0.06, and the SED shows no sign of any
excess in the red--NIR wavelength range\footnote{The \textit{WISE} data may indicate the presence of an additional
source of emission at mid-IR wavelengths (3.4--4.6 $\mu$m), resembling the SEDs of several other CVs,
including the prototype system WZ~Sge. This mid-IR excess is usually attributed to dust (see \citealt{WZ_Sge_dust}
and therein), although \citet{HarrisonIR} questioned this interpretation.}.
We note that the appearance and disappearance of the excess
cannot be connected with the donor star, for example with one of its side heated during the outburst,
because strong orbital variability of the light and its colours is expected in this case, which we were
unable to detect in our multicolour data. To the best of our knowledge, such a phenomenon has never been
reported before. We discuss it in Section~\ref{Sec:RedSource}.

\subsection{The SED in quiescence}
\label{Sec:SEDq}

Well into quiescence the spectra of \SSS\ show no sign of the donor star. However, at orbital periods
of $\sim$2~h normal CVs have relatively bright donor stars. Although in most CVs there is still a
substantial contamination of the NIR luminosity by the accretion
process, which makes the detection of their donor stars difficult \citep{Hoard2MASS}, sometimes the donor
can directly be recognized by observing the red--NIR hump in the SED \citep{HarrisonIR}, or after subtracting
the disc contribution in the NIR from the observed data \citep{Ciardi98}. In CVs in which a contamination
of the NIR luminosity by the accretion disc is less significant due to a low mass-transfer rate, the NIR
hump should be more pronounced (see e.g. the SED of the WZ~Sge-type dwarf nova RZ~Leo in
\citealt{Mennickent99}, and also Fig.~\ref{Fig:SED2} and the discussion below) and the donor star can also
be visible through its absorption lines \citep[see e.g.][]{IshiokaIR,Hamilton2011,HarrisonWZ}. On the
other hand, the donor in a period bouncer should be so dim that even the accretion-heated WD alone is
expected to outshine the donor at optical and the NIR \citep{Knigge2006}.

\SSS\ is one such system, in which the contribution of the accretion disc to the continuum is very low,
as evident from extremely strong emission lines and clearly seen absorption lines of the WD. On the
other hand, its orbital period is relatively long, meaning that a near-main-sequence donor star should
be relatively bright. According to the semi-empirical donor star sequence presented by \citet{KniggeCVevol},
a normal donor in \SSS\ should have the absolute magnitudes $J$$\approx$9.3, $H$$\approx$8.7 and
$Ks$$\approx$8.4. In order to compare these with the apparent magnitudes, one needs to
scale the former to the distance of \SSS. The distance to a SU~UMa CV can be estimated
by utilizing the empirical relationships between absolute magnitude and orbital period of dwarf novae at
the superoutburst plateau and at maximum light derived by \citet{PattersonDist}. For \SSS, these relationships
give $\sim$275~pc for an average binary inclination $i$ of 57\degr, which can be adopted as a reasonable value.
Indeed, the broad double-peaked emission line profiles in spectra of \SSS\ suggest a relatively high
inclination. On the other hand, the
extensive photometry and spectroscopy of \SSS\ rule out any significant eclipse, which constrains the
inclination $i$ to be less than about 70\degr\ to avoid an obvious partial eclipse of the disc. Moreover,
the inclination is expected to be even less because the peak-to-peak separation of the emission lines in
\SSS\ is significantly smaller than in highly inclined CVs (compare 940 \kms\ in \SSS\ and, e.g. 1150 \kms\
in the eclipsing CV HT~Cas of similar $P_{orb}$=106~min -- \citealt{NeustroevHT} -- whose inclination is
81\degr). Therefore, we adopted the source distance to be 275~pc with a realistic error on the distance
of about $\pm$75~pc \citep[for a discussion on uncertainties, see][]{PattersonDist}. However, bearing
in mind the unusual properties of \SSS, we caution the reader that this distance estimate may be less accurate.
Thus, the apparent
$JHKs$ magnitudes of a normal donor star in \SSS\ should be $\sim$16.4, 15.8 and 15.5$\pm$0.6 mag,
respectively. This is about two magnitudes brighter than the observed magnitudes of \SSS\
(Table~\ref{Tab:Magnitudes}), even \emph{without} taking into account of any contribution from the
accretion disc and WD, implying thus that the donor star in \SSS\ is of very low luminosity.

We find it instructive to compare the observed SED of \SSS\ in quiescence with the expected SEDs of
a CV at different evolutionary stages. For this, we used the models of \citet{KniggeCVevol} which utilize
a semi-empirical CV evolution track based on the observed mass--radius relationship of their donor stars.
Tables (tables 5--6 and 7--8) from their work provide the predicted absolute magnitudes of the donor star and
the WD along the CV evolution sequence for a CV before and after the binary has reached the period minimum.
For the pre-bounce case, we combined the fluxes of the donor and WD in a CV with $P_{orb}$=109.8~min.
For the post-bounce case we summed the component fluxes of a CV with $P_{orb}$=90~min, the longest calculated
post-bounce period.\footnote{Note, that the donor flux and accordingly its contribution to the total
flux are expected to decrease towards longer post-bounce periods.} In these calculations, no contribution
from an accretion disc was taken into account, thus the calculated  SEDs should be considered as a lower
limit to the actual SEDs. For a comparison, we also calculated a model atmosphere spectrum of a WD with
a mass of 0.9 \Msun\ and
a surface temperature of 14\,000~K, which reproduces the observed broad wings of Balmer line profiles
and the calculated flux of which is in good agreement with the adopted distance.\footnote{Selected
values of temperature and $\log g$ are not unique, slightly different solutions are plausible. Hence,
the adopted parameters should be considered as a first approximation.}

All the calculated SEDs were then scaled to the distance of \SSS, 275 pc. They
are shown in Fig.~\ref{Fig:SED2} together with the photometric and spectroscopic data of \SSS\ in quiescence.
For a comparison, we also show the photometric data of RZ~Leo, another WZ~Sge-type CV, which has a similar
orbital period (109.5~min) and which shares many properties with \SSS\ \citep{Mennickent99,MennickentDiaz}.
In particular, the disc contribution to the spectrum of RZ~Leo is also insignificant and probably even
smaller than in \SSS, as indicated by deep WD absorption lines clearly visible in the SDSS spectrum. We
point out that in contrast to \SSS, the donor star of M5 spectral type is easily seen in the spectrum
of RZ~Leo, confirming the pre-bounce status of the system \citep{MennickentDiaz,Hamilton2011}.
Thus, RZ~Leo represents a good test case for comparing the observed and calculated SEDs. For this analysis,
we used the available photometric measurements of RZ~Leo from the SDSS, 2MASS and \textit{WISE} data bases.
Although the distance to RZ~Leo is estimated to be 260$\pm$60 pc \citep{PattersonDist}, also very close to
\SSS, we rescaled the fluxes to 275 pc.

The observed SEDs of \SSS\ and RZ~Leo show a close resemblance in the optical wavelengths, but a significant
difference in the NIR. RZ~Leo follows the predicted pre-bounce SED surprisingly well, whereas the observed
NIR flux of \SSS\ is much lower (7--10 times in the $JHKs$ bands) than expected from a CV with a normal
donor star; therefore, a main-sequence donor is excluded. However, the observed NIR flux is about two to three
times higher than the sum of the predicted fluxes from the WD and a post-period-minimum donor. This excess
can possibly be explained by a contribution from the accretion disc. Although the optical continuum flux
from the disc in \SSS\ is low, nonetheless it can be more significant at NIR wavelengths \citep{Hoard2MASS}.

\section{Discussion}

The analysis of observations of \SSS\ presented above shows a number of features, many of which
are rare or quite unusual when compared with most of WZ~Sge-type systems and even any other dwarf
novae. Here we merely highlight and discuss a few most interesting points.

\subsection{Mass ratio $q$}
\label{Sec:DiscWZ}

The light curve of \SSS\ is remarkable for its double superoutburst with a deep and prolonged
dip between its two segments, a long duration of the second superoutburst's segment, the longest of
any other known dwarf novae, and its very gentle decline. \citet{Kimura} suggested that such double
superoutbursts which were previously observed in only a few other WZ~Sge-stars may indicate a very
low $q$. \citet{Kato5} presented a survey of the fading rates of about 320 superoutbursts
of $\sim$185 ordinary SU~UMa-type and 45 WZ~Sge-type dwarf novae (for several objects more than one
superoutburst were studied). The fading rate during the superoutburst plateau in \SSS\ was the slowest
among all the studied superoutbursts. It was shown \citep{Osaki89, Cannizzo2010} that the theoretical
fading rate depends mostly on the accretion disc viscosity in the hot state, thus for slower fading rates
one needs to consider a smaller viscosity. \citet{Kato5} suggested that the viscosity should be smaller
in CVs with a low mass ratio, therefore the unusually slow fading rates may also indicate a very low $q$
(see section 5.5 in their paper for a more detailed discussion).

\citet{KatoSSS} attempted
to estimate $q$ using the periods of superhumps during the rising stage and post-superoutburst stage and
found that \SSS\ has a mass ratio less than 0.05. Another approach to finding $q$ is to employ the empirical
relation between the fractional superhump period excess $\epsilon$=($P_{sh}-P_{orb}$)/$P_{orb}$ and the
mass ratio \citep{PattersonEps05}. Because the period of superhumps is variable, it is not clear which
value of $P_{sh}$ should be used. In the literature, the mean superhump period or the period of stage B
superhumps is often used. For \SSS\ the latter period is 110.10~min and $\epsilon$=0.27(7) per cent. \citet{GW_Lib}
noted, however, that \citet{PattersonEps05} calibrated their $\epsilon$--$q$ relation with superhumps
observed during an early bright phase of the outburst. For \SSS\ this corresponds to the second half of
stage A ($P_{sh}$=110.64~min), which gives $\epsilon$=0.76(8) per cent. Thus, the period excess in \SSS\ is one
of the smallest known of all CVs showing superhumps.
For $\epsilon$=0.27 and 0.76 per cent the $\epsilon$--$q$
relation of \citet{PattersonEps05} gives the mass ratio 0.015 and 0.045, respectively. In order to be on
the conservative side and bearing in mind that the $\epsilon$--$q$ relation is poorly calibrated for low
$q$, we adopt the larger value as an estimate of $q$. This is consistent with the value reported by
\citet{KatoSSS}, and also in good agreement with the measured $K_1$. Indeed, by using the WD mass
$M_{wd}$=0.9\,\Msun\ and the inclination $i$=57\degr, adopted in Section~\ref{Sec:SEDq}, the mass function
gives $q$=0.04.

\subsection{Late superhumps}
\label{Sec:LateSuperhumps}

At least 111 d after the rapid fading phase \SSS\ had been still exhibiting very stable modulations
with a period longer than that of the stage B superhumps. This is in general agreement with the previous
observations of other WZ~Sge-type objects
\citep[see][and references therein]{PattersonEGCnc,KatoLateSuperhumps,KatoWZ}, although in \SSS\ these
late superhumps were seen for significantly longer time than in any other WZ~Sge-type system.
Moreover, the stable modulations still persisted even about 420 d after the superoutburst, but their
period has further increased by about 2\%. Unfortunately,
we missed this period transformation and it is unclear if these most recent modulations are evolved
late superhumps or they are another kind of quiescent superhumps. Quiescent superhumps
were observed in several dwarf novae \citep[e.g., in V1159~Ori --][]{Patterson95}. Thanks to the unprecedented
precision and cadence of {\it Kepler} long-term light curves of V344~Lyr and V1504 Cyg, such modulations
were traced throughout the decline phase of superoutbursts to quiescence and through a few following
normal outbursts of these (ordinary) SU~UMa-type stars \citep{StillKepler,WoodKepler,OsakiKato2013}. However,
they seem to not survive the whole superoutburst cycle as they were seen no later than about 40 d after
superoutbursts ended \citep{OsakiKato2013}, a much shorter time interval than in \SSS.

\subsection{A massive cool gas region beyond the 3:1 resonance radius}
\label{Sec:RedSource}

Simultaneously with the start of the rapid fading phase, \SSS\ showed the appearance of the strong red-NIR
excess, which disappeared 100 days later. It is interesting to note that the passage through
the reddest colours had been accompanied by alteration of the decline slope (Fig.~\ref{Fig:DeclineLog}),
by a significant decrease of superhump amplitudes (Fig.~\ref{Fig:O-C}), and perhaps by an emergence of
the late superhumps (Section~\ref{Sec:PostSuperhumps}). This red-NIR excess
and also the appearance of the \NaD\ line in absorption during the plateau stage indicate a fairly cool
region in the outer disc. \citet[][and references therein]{TovmassianZharikov} show that outer parts of
extended accretion discs of period bouncers should indeed contain cool ($\sim$2000 K) material unlike
ordinary dwarf novae in which 5000 K is considered as the disc edge temperature. We suggest that this cool
gas is accumulated in the outermost part of the disc, beyond the 3:1 resonance radius. In the binaries
with low mass ratios such as the WZ~Sge-type stars, the 3:1 resonance radius is located well inside the
Roche lobe of the primary while the tidal truncation radius is almost as large as the Roche lobe. During
energetic superoutbursts after the long period of quiescence, the disc expands in order to transfer the
large amount of angular momentum released by the accreting matter. That can bring substantial mass beyond
the 3:1 resonance radius. It is thought that during a superoutburst a significant fraction (70--80 per cent) of
the disc is accreted \citep{CannizzoKepler}. However, the matter accumulated in the outer parts of the
disc avoids the 3:1 resonance, suffers a lower tidal dissipation of angular momentum and may
be left from being accreted during the main superoutburst \citep{Hellier2001}.

Such a model was originally proposed to explain the repeated rebrightenings observed in EG~Cnc and
some other WZ~Sge-type dwarf novae \citep{KatoWZ}, but later has also been invoked to interpret late
superhumps and other peculiar properties of WZ~Sge-type systems \citep{KatoLateSuperhumps}. It was shown
that `a cool matter reservoir' beyond the 3:1 resonance can supply the matter to the inner disc,
resulting in rebrightenings, whereas the late superhumps should arise near the tidal truncation radius,
from the matter in the reservoir.

In order to develop the model further it is essential to establish the time-scale of the reservoir filling
and depleting. Our observations show that the \NaD\ line has appeared in absorption no later than on $T$=28,
after that the line parameters were very stable until the end of the plateau stage. It suggests that the
cool outer disc was already formed about three weeks after the beginning of the second segment of the
superoutburst. However, the spectra obtained during the temporary fading stage between the superoutburst
segments show the \NaD\ line in emission. It implies that there was already a significant amount of matter
in the reservoir at that time. Perhaps, the matter accumulated there during the first segment of the
superoutburst. The following observations indicate that the reservoir depletion has been happening very
slowly, because even 250 d after the superoutburst the system colours were redder than in quiescence,
and also the late superhumps were observed at least 420 d after the superoutburst. Moreover, the sodium
doublet was observed in emission until the most recent observations, obtained three years after the superoutburst.
It has been steadily increasing the intensity in parallel with \HeI\ \l5876, suggesting a similar excitation
mechanism of both lines.

Considering the very low mass ratio and also the long duration of the superoutburst, one can expect that
a system similar to \SSS\ can accumulate a significant amount of matter in the outer disc. It then seems
to take  years to drain this reservoir.

\subsection{\SSS\ as a period bouncer}
\label{Sec:PeriodBounce}

The orbital period of \SSS, 109.80 min, is one of the longest among the WZ~Sge-type stars \citep{KatoWZ}.
In Section~\ref{Sec:SEDq} we showed that at orbital periods of $\sim$2~h a contribution of the normal donor
star to the NIR segment of the spectrum should dominate in quiescence, but is seemingly entirely absent.
It suggests that the donor star has a low temperature and very low luminosity, that
is consistent with the low $K_1$ and with very small $q\lesssim0.045$ estimated
from the period of superhumps. With such a small mass ratio even if the WD was close to the Chandrasekhar
limit the donor mass must be below the hydrogen-burning minimum mass limit of 0.075 \Msun. This suggests
that \SSS\ might already be evolving away from the period minimum towards longer periods, having passed
through it, with the donor now extremely dim. Using a reasonable
range of system parameters adopted above ($q$=0.015 -- 0.045, $M_{wd}$=0.8 -- 1.0\,\Msun, d=275~pc) and
assuming that the NIR spectrum is dominated by the WD only and that the contribution from the accretion disc
is completely negligible, the residual NIR light sets a conservative upper limit on the temperature of
the donor star of $\sim$2000~K. Moreover, the donor temperature is possibly significantly lower, because
the disc probably contributes at NIR wavelengths \citep{Hoard2MASS}.

The properties of \SSS\ are very similar to those of GD~522, another very strong period-bounce candidate
with relatively long $P_{orb}$ of 103~min \citep{GD552}. However, the longer period of \SSS, together
with its small period excess, suggests that it is by far the most evolved CV known to date.
Assuming that the only angular momentum loss mechanism in period-bounce CVs is that due to gravitational
radiation, the age of \SSS\ should be older than $\sim$6 Gyr \citep{Howell97}.

We should note, however, that there is a separate evolutionary channel leading to brown-dwarf donors
in CVs directly from detached WD/brown-dwarf binaries.
\citet{Politano} found, through population synthesis, that 18 per cent of the total zero-age CVs population are
born with brown-dwarf donors, and 20 per cent of which have orbital periods in the range 78 to 150~min. The
probability of finding that sort of system is not very high, though. According to the standard model
of CV secular evolution, it is at least four times more likely that the donors in period-bounce candidates
became substellar as a result of mass loss during secular evolution rather than that they were born substellar
\citep{Politano}. Moreover, the predicted distribution of orbital periods in CVs with the donors which
were born substellar (fig.~2 in \citealt{Politano}) shows a rapidly dwindling number of such CVs at
longer $P_{orb}$ (as in \SSS\ and GD~522). Still, this alternative evolutionary scenario is possible.

Which evolutionary channel produced \SSS\ and other period-bounce candidates can potentially
be distinguished by the properties of the donor star. Since the donors in period bouncers were born as
normal stars and became substellar during secular evolution, it is not obvious that these objects behave
like ordinary brown dwarfs. In particular, the former objects should have comparatively higher helium
abundance in comparison with normal brown dwarfs. Also, brown dwarfs (at least of a lower mass) do
not burn lithium, whereas normal stars do, thus the lithium test (see e.g. \citealt{Li-test}) can
be applied to distinguish brown dwarfs from substellar objects in period bouncers. Unfortunately, an
accurate derivation of the helium abundance has proved difficult to perform even for normal stars,
whereas even the strongest \LiI\ \l6708 line is very weak and superposed on the broad and relatively
strong \HeI\ \l6678 line. Therefore, we have to agree with \citet{Politano} here that at the present
time we cannot unambiguously discriminate between brown dwarfs and substellar objects which were born
as normal stars and became substellar as a result of mass loss.

\section{Summary}

We have reported extensive UV--optical--NIR photometric and spectroscopic observations of the WZ~Sge-type
dwarf nova \SSS\ during its superoutburst, the following decline and in quiescence.
The observations were taken over three years, starting from 2013 January 5 (four days after the
object's discovery) until 2016 April 17. The principal results of this study can be summarized
as follows.

\begin{enumerate}
\item  The superoutburst showed two segments with an $\sim$5 mag and $\sim$10 d dip between them.
The second segment had a long duration of 33 d and a very gentle decline with a rate of 0.02 mag~d$^{-1}$.

\item The total amplitude of the superoutburst was $\sim$7 mag, its post-outburst decline lasted no less
than 500~d.

\item Superhumps with the average period of $P_{sh}$=110.23(7)~min were clearly visible from our very
first time-resolved observations until at least 420 d after the rapid fading from the superoutburst plateau.
Thus, the late superhumps in \SSS\ were detectable for a much longer time than in any
other WZ~Sge-type system.
As in most SU~UMa stars, $P_{sh}$ in \SSS\ changed slightly during the superoutburst and the following
decline.

\item Multicolour photometry showed redder colours of superhumps at their light maximum, although the
character of colour evolution of superhumps has also changed at different superoutburst stages.

\item Simultaneously with the start of the decline, \SSS\ showed the appearance of a strong NIR excess
resulting in very red colours, which reached the extreme values ($B$$-$$I$$\simeq$1.4) about 20 d later.
The colours then became bluer again, but it took at least 250 d to acquire a stable level.
We interpreted it as the appearance of a massive cool gas region in the outermost part of the disc,
beyond the 3:1 resonance radius.

\item After the disappearance of the NIR excess, the observed IR flux of \SSS\ in quiescence was much
lower (7--10 times in the $JHKs$ bands) than is expected from a CV with a near-main-sequence donor.

\item A time series analysis of the spectral data revealed an orbital period of 109.80(7)~min. Thus,
the fractional superhump period excess is $\lesssim$0.8\%, indicating a very low mass ratio $q\lesssim$0.045.
This is consistent with the low radial velocity semi-amplitude of the accretion disc emission lines,
$K_1=16\pm5$ \kms.

\item A small $q$, very low luminosity and the lack of spectral signatures of the donor star strongly
suggest its brown-dwarf-like nature and that \SSS\ has already evolved away from the period minimum
towards longer periods, although the scenario with the donor star that has not become substellar as
a result of mass loss, but was born substellar cannot be ruled out.

\item A conservative upper limit on the temperature of the donor star is estimated to be 2000~K.

\item The long orbital period suggests that \SSS\ is by far the most evolved CV known to date.

\end{enumerate}

\section*{Acknowledgements}

TRM and DS are supported by the STFC under grant ST/L000733.
KLP and JPO acknowledge the support of the UK Space Agency. SGP acknowledges financial support from FONDECYT
in the form of grant number 3140585. GT and SVZ acknowledge PAPIIT grants IN-108316/IN-100614 and CONACyT
grants 166376, 151858 and CAR 208512 for resources provided towards this research. VS thanks Deutsche
Forschungsgemeinschaft (DFG)  for  financial  support  (grant  WE  1312/48-1).
This publication makes use of data products from the Wide-field Infrared Survey Explorer, which is a
joint project of the University of California, Los Angeles, and the Jet Propulsion Laboratory/California
Institute of Technology, funded by the National Aeronautics and Space Administration.
We acknowledge the use of observations made with the NASA Galaxy Evolution Explorer.
GALEX is operated for NASA by the California Institute of Technology under NASA contract NAS5-98034.
We acknowledge with thanks the variable star observations from the AAVSO International Database contributed
by observers worldwide and used in this research.
The results presented in this paper are based on observations collected at the European Southern Observatory
under programme IDs 290.D-5192 and 095.D-0837. We are thankful to the anonymous referee whose comments helped
greatly to improve the paper.


\appendix

\section{Observations and data reduction}
\label{Sec:Appendix}

\subsection{Optical photometric observations}

\begin{table*}
\caption{Magnitudes of comparison stars in the field of \SSS. The $UBV(RI)_C$ magnitudes and their errors were
         determined in this work, while the $JHKs$ magnitudes were taken from the 2MASS All-Sky Catalogue of Point
         Sources \citep{2MASS}.}
\begin{center}
\begin{tabular}{ccccccccc}
\hline
 Star &    $U$       &       $B$      &       $V$      &       $R_c$    &       $I_c$    &       $J$      &       $H$      &      $Ks$      \\
\hline
 1  & 15.07$\pm$0.03 & 14.87$\pm$0.02 & 14.14$\pm$0.02 & 13.65$\pm$0.02 & 13.19$\pm$0.03 & 12.58$\pm$0.02 & 12.15$\pm$0.02 & 12.06$\pm$0.02 \\
 2  & 16.63$\pm$0.05 & 16.76$\pm$0.03 & 16.28$\pm$0.03 & 15.89$\pm$0.03 & 15.43$\pm$0.04 & 15.13$\pm$0.06 & 14.86$\pm$0.06 & 14.58$\pm$0.09 \\
 3  & 14.69$\pm$0.03 & 16.20$\pm$0.02 & 15.65$\pm$0.02 & 15.20$\pm$0.02 & 14.80$\pm$0.03 & 14.28$\pm$0.03 & 13.94$\pm$0.04 & 13.86$\pm$0.06 \\
 4  & 16.55$\pm$0.05 & 16.37$\pm$0.02 & 15.77$\pm$0.02 & 15.37$\pm$0.02 & 14.98$\pm$0.04 & 14.52$\pm$0.03 & 14.30$\pm$0.05 & 14.08$\pm$0.07 \\
 5  & 17.73$\pm$0.11 & 17.10$\pm$0.04 & 16.41$\pm$0.03 & 15.85$\pm$0.03 & 15.28$\pm$0.04 & 14.74$\pm$0.04 & 14.22$\pm$0.05 & 14.32$\pm$0.07 \\
\hline
\end{tabular}
\end{center}
\label{Tab:Comparisons}
\end{table*}

\begin{figure}
\includegraphics[width=8.5cm]{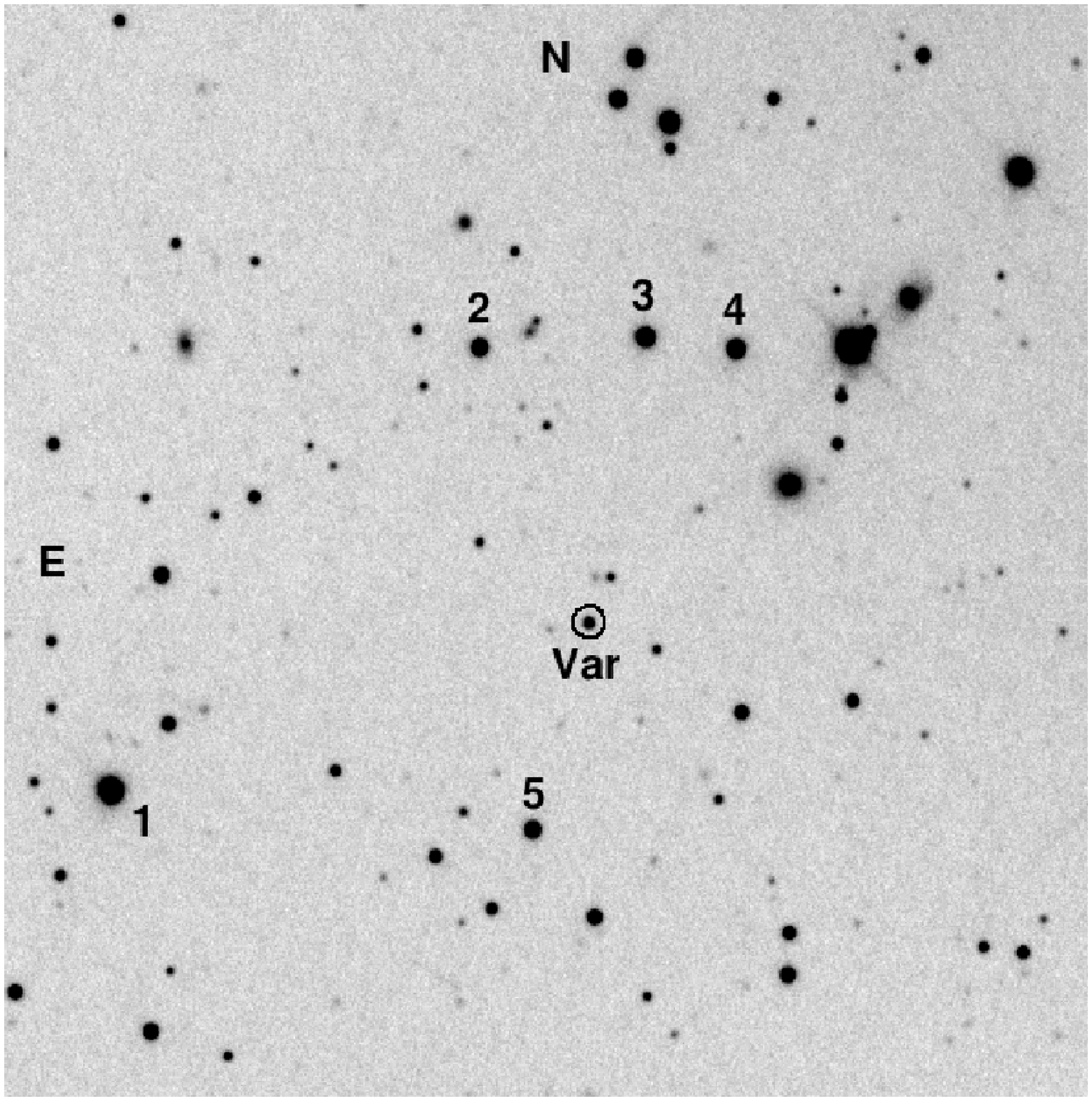}
\vspace{5 mm}
\caption{The 6\arcmin$\times$6\arcmin\ field of view of \SSS, oriented such that up
is North and left is East. The object's location is indicated by a ring. Calibrated
comparison stars are labelled as 1, 2, 3, 4 and 5, and their magnitudes are listed in
Table~\ref{Tab:Comparisons}.}
\label{Fig:Chart}
\end{figure}

Most of the optical time-resolved photometric observations were taken by three observatories located in
Chile and New Mexico. The ROAD in Chile houses a 0.40-m f/6.8 ODK
telescope from Orion Optics UK, equipped with a CCD camera from Finger Lakes Instrumentation,
an FLI ML16803 CCD, which holds a Kodak KAF-16803 CCD chip with 4k$\times$4k pixels. All the measurements
made with this telescope between 2013 January 8 and 2013 March 15 were unfiltered (the CCD response is
close to that of the $V$ filter).
The exposure times were 60~s during the outburst stage, but were increased to 120~s
when the source faded. Some of these data were already presented by \citet{KatoSSS}. The observations at
New Mexico Skies in Mayhill, New Mexico, were performed from 2013 January 11 to 2013 February 26 with the
0.35-m Celestron C14 robotic telescope and an SBIG ST-10XME CCD camera with Johnson--Cousins $BV(RI)_C$
Astrodon Photometric filters. The images were usually taken in a sequence $B$--$V$--$R$--$I$.
During the bright state the exposure times were 15~s for the $V\,R\,I$ filters and 30~s for $B$.
They were increased to 240 and 400~s, respectively, when the source declined in brightness.
Depending on the weather conditions, we monitored the star for 4--8 h per night. Since 2013 March 5,
we started using two 0.4-m robotic PROMPT telescopes located in Chile \citep{PROMPT}. After 2013 March 17,
only the PROMPT telescopes were utilized. Most of the PROMPT observations in 2013 were performed in the
$V$ filter, whereas the data obtained in 2014 and 2015 are mostly unfiltered (the CCD response
is close to that of the $R$ filter) though occasionally we also took $BV(RI)_C$ data. The exposure
times were 180~s.
In addition, during 2013 February 5--7 we obtained three nights of time-resolved observations with the
0.84~m telescope at the Observatorio Astron\'{o}mico Nacional at San Pedro M\'{a}rtir (OAN SPM) in Mexico.
These observations in the Johnson $V$ filter were performed with exposure times of 30~s, simultaneously
with the spectroscopic observations at the same site (see Section~\ref{SpecObsSec}).

In order to establish the zero-points and calibrate the comparison stars, on 2013 February 5 we obtained
$UBV(RI)_C$ images of eight Landolt standards \citep{Landolt}. These standards were observed several times
at different airmasses throughout the night. The comparison stars which were checked for variability are
indicated in Fig.~\ref{Fig:Chart}, and their corresponding magnitudes are listed in Table~\ref{Tab:Comparisons}.
The errors are formal 1-$\sigma$ uncertainties returned by the fit and are likely to underestimate
the real uncertainties.

The data reduction  was performed using the \iraf\ environment and the software AIP4Win v.~2.4.0
\citep{AIP4Win}. All images were corrected for bias and flat-fielded in the standard way. We performed
aperture photometry of our object and several comparison stars. Reducing the unfiltered observations, we
used the $V$ and $R$ comparison star magnitudes for the ROAD and PROMPT data, respectively. Our tests
showed that such an approach could induce measurement errors no more than a few hundredth of magnitude,
in comparison with filtered photometry. The typical accuracy of source measurements during the bright
state was about 0.02--0.03 mag, which then worsened to 0.05--0.15 mag when the source declined in
brightness.

In order to simplify the analysis of the long-term light curve, we also formed 1-d averages of all the
observations, reducing thus the scatter from both random errors and stochastic and short-term variability.
On average, each night of observations during the bright state contains about 68 $V$-band and/or
unfiltered data points and 55 data points in the $B$, $R$ and $I$ bands. Post-outburst nightly observations
consist on average of 14, 52, 36 and 14 exposures in the $B$, $V$, $R$ (including unfiltered) and $I$
filters. The median values of the standard error of these 1-d averages are better than 0.01 and 0.05 mag
in all the filters for the bright-state and post-outburst observations, respectively.
We also note that the $BVRI$ images obtained in 2014, 2015 and 2016 are rather noisy. In order to improve
the confidence of the colour measurements we first aligned and summed all the nightly images for each
filter and then measured the average magnitudes.

Table~\ref{ObsPhotTab} provides the journal of the optical time-resolved photometric observations. In
order to help the reader to trace long-term brightness evolution of \SSS, we also show its 1-d averaged
$V$ magnitudes. For the nights when no data in this filter were taken, the $V$ magnitudes were obtained
from the $R$ magnitudes and $V-R$ colour indices interpolated to the night in question.

\subsection{NIR observations}

A near-infrared observation of \SSS\ was conducted on 2016 April 7 using the IRSF at
Sutherland, South Africa. The IRSF consists of a 1.4~m telescope and SIRIUS \citep{SIRIUS}.
The SIRIUS camera can obtain $J$, $H$ and $Ks$
band images simultaneously, with three 1K $\times$ 1K HgCdTe detectors. It has a square field of
view of 7\farcm7$\times$7\farcm7 and a 0\farcs45 pixel$^{-1}$ pixel scale.
During our observations, the exposure time of a frame was 60~s and a total of 130 frames with 10-point
dithering were taken during 150~min. Data reduction was carried out with a pipeline for
SIRIUS\footnote{http://irsf-software.appspot.com/yas/nakajima/sirius.html}, which includes a correction
for non-linearity, dark subtraction, and flat-fielding. The IR magnitudes were calibrated using the same
comparison stars as were chosen for the optical photometry (Fig.~\ref{Fig:Chart}, Table~\ref{Tab:Comparisons}).
Their magnitudes and errors were taken from the 2MASS All-Sky Catalogue of Point Sources \citep{2MASS}.

\subsection{{\it Swift} UVOT observations}

The {\it Swift} X-ray satellite \citep{Swift} started observing \SSS\ on 2013 January 6. Data were collected
using both the XRT telescope and the UVOT with the $uvm2$ filter in position.
Following the initial 4~ks exposure, 1--2~ks observations were obtained
approximately every 1--3 d until 2013 July 1, with occasional gaps in the schedule caused by high priority
{\it Swift} observations of other targets. On 2014 June 26 and on 2015 January 16 we performed two additional
observations. A final data set was collected on 2016 April 17. This UVOT observation was carried out in all six
filters: $uvw2$ (central wavelength 1928~\AA), $uvm2$ (2246~\AA), $uvw1$ (2600~\AA), $u$ (3465~\AA),
$b$ (4392~\AA) and $v$ (5468~\AA).

The data were processed and analysed using {\sc heasoft} 6.16, together with the most recent
version of the calibration files. The UVOT magnitudes were calculated using a standard
5 arcsec extraction region, with the background estimated from four nearby circular source-free
regions. Table~\ref{Tab:SwiftLog} gives details of the {\it Swift} observations.

\subsection{Spectroscopic observations}
\label{SpecObsSec}

Table~\ref{Tab:SpecObs} provides the journal of the optical spectroscopic observations.
The first two sets of spectroscopic data were obtained at the end of the nights starting 2013 January 4 and 5 with
the 4.2-m William Herschel Telescope on La Palma. The observations were taken with the dual arm spectrograph
ISIS, using the 600 lines/mm gratings in the blue and red arms to cover the wavelength intervals 3945--5170\AA\
at 0.88 \AA/pixel and 5614--7140\AA\ at 0.98 \AA/pixel, respectively; the FWHM resolution in each arm was
2.0~\AA. The total exposure times were 29 and 92~min, respectively, with the majority of exposure 5~min in
duration. We used a slit width of 1.2 arcsec. The weather was clear on both nights, and the seeing was
around $\sim 1.2$ arcsec on the second night, however the southerly position of \SSS, meaning that all
data were acquired at high airmass, which, along with poor seeing on the first night, cost significant
throughput; given the narrow slit, no flux calibration was attempted. The observations were reduced using the
{\sc starlink} software packages {\sc figaro}, {\sc kappa} and {\sc pamela} to extract the spectra optimally
\citep{MarshPamela}. The wavelength scales were linearly interpolated in time from bracketing comparison arc
spectra. We fitted 30 and 28 lines with cubic polynomials (four coefficients) obtaining rms scatters of 0.015
and 0.021\AA\ in the blue and red arms, respectively.

Twelve observations were obtained at the OAN SPM in Mexico on the 2.1-m telescope. The data on 2013
January 28 were obtained with the Echelle spectrograph \citep{Echelle}.
This instrument gives a resolution of 0.234 \AA\ pixel$^{-1}$ at H$\alpha$ using the UCL camera and a
CCD-Tek chip of 1024$\times$1024 pixels with a 24~$\mu$m$^2$ pixels size. The spectra cover 25 orders
and span the spectral range 3915--7105~\AA. A total of 12 spectra were obtained with 600~s individual
exposures. The reduction procedure was performed using \iraf. Comparison spectra of Th--Ar lamps were
used for the wavelength calibration. The absolute flux calibration of the spectra was achieved by taking
nightly echellograms of the standard stars HD93521 and HR153. Further observations with this telescope
were conducted on 2013 February 5--8, on February 10--13, on May 5 and on 2016 February 9 and 10
with the Boller and Chivens (B\&Ch) spectrograph, equipped with a 13.5~$\mu$m ($2174\times2048$) Marconi
E2V-4240 CCD chip. Spectra were obtained in both single-shot and time-resolved manner with different
instrumental setups and exposure times. For details see Table~\ref{Tab:SpecObs}.
Several Cu--He--Ne--Ar lamp exposures were taken during the observations of the target for wavelength
calibrations, and the standard spectrophotometric stars Feige~110, HR3454 and G191-B2B \citep{Oke}
were observed for flux calibrations. The data reduction was also performed using the \iraf\ environment.
All the nights of the observations at OAN SPM were photometric with exception for the first half of the
night of 2013 February 6 when a part of the output spectra was ruined by clouds. An unaffected subset of
this night's data includes five spectra taken in the beginning and 30 spectra obtained in the end of the
observations, with a total exposure time of 103 min.

Another set of spectroscopic data was taken with the MagE spectrograph \citep{MagE} on the Magellan Clay
Telescope on 2013 February 5. These observations were gathered on the same night as one of the OAN SPM
observations, but they do not overlap. The object was observed with the 1.0 arcsec slit, for a total of
91 min, broken up into 22 blocks of 300-600 s, separated by Th--Ar lamp measurements. Seeing varied
from 0.9 to 1.5 arcsec over the course of the observations. The standard spectrophotometric star
Feige 67 was observed with the same setup. So were various calibration frames such as  Xe-flash frames,
quartz lamp
frames and Th--Ar frames from flat-fielding and wavelength calibration purposes. Data were reduced with
the Carnegie MagE pipeline \citep[first described in][]{MagReduct}. The orders of the combined spectrum
were normalized and merged to produce final one-dimensional spectra for further analysis.

On 2013 March 21 \SSS\ was observed with the medium-resolution spectrograph X-shooter \citep{Xshooter}
mounted at the Cassegrain focus of VLT-UT2 at Paranal. X-shooter comprised three detectors: the
UVB arm, which gives spectra from 0.3 to 0.56 $\mu$m, the VIS arm which covers 0.56--1 $\mu$m and the
NIR arm which covers 1--2.4 $\mu$m. We used slit widths of 1.0 arcsec, 0.9 arcsec and 0.9 arcsec in
X-shooter's three arms and 2\t2 binning in the UVB and VIS arms resulting in a spectral resolution of
3000--4000 across
the entire spectral range. The reduction of the raw frames was conducted using the standard pipeline
release of the X-shooter Common Pipeline Library (CPL) recipes (version 6.5.1) within ESORex, the ESO
Recipe Execution Tool, version 3.11.1. The standard recipes were used to optimally extract and wavelength
calibrate each spectrum. The instrumental response was removed by observing the spectrophotometric
standard star EG274 and dividing it by a flux table of the same star to produce the response function.
The wavelength scale was also heliocentrically corrected. During our observations we nodded the object
along the slit in order to improve the NIR arm reduction. However, we decided to reduce the data in
STARE mode to increase the phase resolution of the data. While this had no effect on the UVB and VIS
arm reductions, the sky subtraction was poorer in the NIR arm using this method; nevertheless, the
improved phase resolution gained by this method outweighed the poorer sky subtraction.

On 2015 May 5 we performed the observations with the FORS2 spectrograph mounted at
the Cassegrain focus of VLT-UT1 at Paranal. In order to maximize the wavelength coverage, we took two
spectra using the GRIS\_300V+10 grism and two using the GRIS\_300I+11 grism, providing an overall wavelength
coverage of 3830--10110 \AA. The exposure times were 300~s. We used a 1 arcsec slit and 2\t2 binning of the
MIT CCD, resulting in a resolution of 12 \AA\ at the central wavelength. The spectra were reduced using
{\sc kappa}, {\sc figaro} and {\sc pamela}. The wavelength scale was derived
from Hg--Cd and He--Ne--Ar arc lamp exposures taken in the morning following the observations, with the
telescope pointing at zenith. To account for any flexure, we adjusted the wavelength scale using the
known wavelengths of strong night-sky lines. Although the sky was clear at the time of the target
observations, the morning twilight was cloudy, so no spectroscopic flux standard was taken. Instead,
we used archival spectra of LTT3864, taken with the same instrument setup, to derive an approximate
flux calibration.

\begin{table*}
\caption{Journal of photometric observations of \SSS. The table also provides the 1-d averaged
$V$ magnitudes of the source (see text for details).}
\begin{tabular}{||cccc||cccc||cccc||}
\hline
HJD start& Duration & Band &   $V$    &HJD start& Duration & Band &   $V$    &HJD start& Duration & Band &   $V$    \\
2450000+ &   (h)    &      &  (mag)   &2450000+ &   (h)    &      &  (mag)   &2450000+ &   (h)    &      &  (mag)   \\
\hline
6300.726 &   2.66   &   V  &  16.79   &6356.589 &   5.19   & BVRI &  17.51   &6461.574 &   0.63   &   V  &  18.38   \\
6301.723 &   2.72   &   V  &  16.67   &6357.622 &   5.87   & BVRI &  17.47   &6509.502 &   1.21   & BVRI &  18.40   \\
6303.735 &   2.42   &   V  &  14.43   &6358.632 &   3.33   &   V  &  17.51   &6670.658 &   0.70   & BVRI &  18.82   \\
6304.006 &   0.17   &   V  &  13.74   &6360.627 &   6.37   & BVRI &  17.70   &6675.695 &   1.39   &   R  &  18.87   \\
6304.715 &   2.93   &   V  &  12.60   &6361.615 &   6.32   &  VRI &  17.61   &6676.666 &   4.22   &   R  &  18.86   \\
6305.713 &   3.03   &   V  &  12.37   &6362.633 &   3.79   &   V  &  17.65   &6677.740 &   1.34   &   R  &  19.05   \\
6306.772 &   1.63   &   V  &  12.20   &6363.576 &   5.09   &   V  &  17.74   &6686.735 &   2.65   &   R  &  18.85   \\
6307.707 &   3.69   &   V  &  12.28   &6364.550 &   5.64   &   V  &  17.62   &6687.623 &   5.75   &   R  &  18.89   \\
6308.772 &   6.45   &   V  &  12.55   &6365.525 &   6.92   &   V  &  17.69   &6688.620 &   3.92   &   R  &  18.91   \\
6309.728 &   7.63   & BVRI &  12.58   &6366.625 &   3.73   &   V  &  17.68   &6699.591 &   4.30   &   R  &  18.94   \\
6310.801 &   5.93   & BVRI &  12.46   &6367.520 &   6.95   &   V  &  17.70   &6708.695 &   4.45   &   R  &  18.94   \\
6312.922 &   2.94   & BVRI &  12.31   &6368.559 &   5.51   &   V  &  17.77   &6709.568 &   4.54   &   R  &  18.91   \\
6314.689 &   8.58   & BVRI &  12.39   &6371.604 &   7.06   & BVRI &  17.83   &6710.572 &   5.62   &   R  &  18.94   \\
6315.917 &   3.06   & BVRI &  12.41   &6372.708 &   3.19   & BVRI &  17.81   &6711.608 &   5.68   &   R  &  18.87   \\
6316.789 &   2.24   &   V  &  12.40   &6377.630 &   6.11   & BVRI &  17.86   &6712.616 &   0.15   &   R  &  18.87   \\
6317.680 &   4.87   &   V  &  12.41   &6378.559 &   4.34   &   V  &  17.87   &6713.594 &   6.95   &   R  &  18.91   \\
6318.678 &   4.96   &   V  &  12.41   &6380.751 &   3.06   &   V  &  17.92   &6714.711 &   4.20   &   R  &  18.82   \\
6319.675 &   5.01   &   V  &  12.42   &6383.512 &   4.13   &   V  &  18.01   &6715.552 &   5.59   &   R  &  18.78   \\
6320.673 &   4.61   &   V  &  12.44   &6384.510 &   1.99   &   V  &  17.98   &6716.659 &   2.80   & BVRI &  18.88   \\
6321.719 &   4.01   &   V  &  12.48   &6385.549 &   2.95   &   V  &  18.09   &6717.745 &   3.42   &   R  &  18.75   \\
6323.665 &   9.06   & BVRI &  12.52   &6386.518 &   4.68   &   V  &  18.05   &6719.539 &   3.15   &   R  &  18.77   \\
6324.662 &   9.40   & BVRI &  12.56   &6388.541 &   4.22   &   V  &  17.96   &6720.535 &   2.73   &   R  &  18.80   \\
6325.660 &   5.52   &   V  &  12.57   &6389.507 &   5.23   & BVRI &  17.95   &6721.565 &   3.66   &   R  &  18.69   \\
6326.660 &   5.48   &   V  &  12.61   &6393.596 &   5.56   &   V  &  18.03   &6722.656 &   3.10   & BVRI &  18.86   \\
6327.658 &   9.36   & BVRI &  12.65   &6394.546 &   4.87   &   V  &  18.02   &6723.534 &   2.17   & BVRI &  18.90   \\
6328.890 &   3.71   &UBVRI &  12.69   &6395.520 &   5.27   & BVRI &  18.05   &6726.525 &   3.67   &   R  &  18.83   \\
6329.888 &   4.07   & BVRI &  12.72   &6397.515 &   4.29   &   V  &  18.07   &6727.525 &   3.67   &   R  &  18.83   \\
6330.873 &   4.04   & BVRI &  12.77   &6398.512 &   4.21   &   V  &  18.07   &6728.588 &   3.66   &   R  &  18.79   \\
6331.883 &   3.85   & BVRI &  12.81   &6401.729 &   3.52   &   V  &  18.00   &6729.523 &   3.77   &   R  &  18.71   \\
6334.861 &   4.03   & BVRI &  12.96   &6402.501 &   0.52   &   V  &  18.13   &6730.527 &   3.83   &   R  &  18.74   \\
6336.625 &   9.74   & BVRI &  14.55   &6403.712 &   3.78   &   V  &  18.10   &6746.508 &   4.23   &   R  &  18.81   \\
6337.729 &   7.35   & BVRI &  15.90   &6416.726 &   2.59   &   V  &  18.12   &6749.530 &   5.38   &   R  &  18.72   \\
6338.624 &   6.38   &   V  &  16.16   &6417.486 &   8.26   &   V  &  18.25   &6750.609 &   3.72   & BVRI &  18.96   \\
6339.645 &   9.34   & BVRI &  16.45   &6418.487 &   7.77   & BVRI &  18.24   &6752.503 &   4.40   &   R  &  18.85   \\
6340.642 &   9.23   & BVRI &  16.65   &6423.780 &   0.83   &   V  &  18.29   &6755.519 &   4.31   &   R  &  18.86   \\
6341.786 &   5.92   & BVRI &  16.75   &6424.483 &   5.61   &   V  &  18.14   &6772.558 &   3.72   & BVRI &  18.94   \\
6342.637 &   9.31   & BVRI &  16.86   &6427.481 &   7.33   &   V  &  18.22   &6827.546 &   3.70   & BVRI &  19.01   \\
6343.634 &   6.34   &   V  &  16.96   &6428.480 &   7.67   &   V  &  18.11   &6836.548 &   3.70   & BVRI &  18.96   \\
6344.631 &   6.41   &   V  &  17.13   &6431.538 &   5.47   &   V  &  18.25   &7036.750 &   0.40   &   R  &  19.02   \\
6345.778 &   5.87   & BVRI &  17.17   &6432.478 &   4.85   &   V  &  18.29   &7037.763 &   3.20   & BVRI &  19.02   \\
6346.843 &   4.22   &   V  &  17.25   &6433.479 &   4.66   &   V  &  18.33   &7040.826 &   0.59   &   R  &  19.00   \\
6347.679 &   5.28   &   V  &  17.28   &6434.477 &   6.11   &   V  &  18.36   &7041.809 &   1.01   &   R  &  18.95   \\
6348.666 &   5.71   &   V  &  17.33   &6436.476 &   5.32   & BVRI &  18.31   &7043.645 &   2.67   &   R  &  19.10   \\
6349.656 &   7.36   &   V  &  17.43   &6442.540 &   0.51   & BVRI &  18.24   &7060.599 &   2.56   &   R  &  18.88   \\
6351.650 &   6.09   &   V  &  17.40   &6446.474 &   4.72   & BVRI &  18.34   &7061.627 &   1.29   &   R  &  18.71   \\
6352.716 &   4.22   &   V  &  17.55   &6448.476 &   6.43   & BVRI &  18.31   &7062.594 &   3.16   &   R  &  19.01   \\
6353.601 &   6.90   & BVRI &  17.40   &6458.557 &   0.99   & BVRI &  18.41   &7063.591 &   1.39   &   R  &  18.85   \\
6354.596 &   7.06   & BVRI &  17.43   &6459.633 &   0.21   &   V  &  18.43   &7403.120 &   0.77   & BVRI &  19.03   \\
6355.546 &   4.83   &   V  &  17.49   &6460.588 &   2.73   & BVRI &  18.46   &7497.323 &   3.12   & BVRI &  19.01   \\
\hline
\end{tabular}
\label{ObsPhotTab}
\end{table*}

\begin{table*}
\caption{Details of the {\it Swift} observations of \SSS.}
\begin{center}
\begin{tabular}{llccllcc}
\hline
HJD mid  &  Obs. ID       &Exp.Time&       $uvm2$     &HJD mid  &    Obs. ID     &Exp.Time&       $uvm2$     \\
2450000+ &                & (ksec) &        (mag)     &2450000+ &                & (ksec) &        (mag)     \\
\hline

6299.385 & 00032666001/1  &  1.92 &  14.83~$\pm$~0.03 &6348.749 & 00032666027    &  0.26 &  15.76~$\pm$~0.05 \\
6299.447 & 00032666001/2  &  2.04 &  14.80~$\pm$~0.02 &6349.397 & 00032666027    &  0.65 &  15.65~$\pm$~0.03 \\
6301.381 & 00032666002    &  1.98 &  --               &6352.003 & 00032666028    &  1.04 &  15.99~$\pm$~0.03 \\
6303.115 & 00032666003/1  &  1.45 &  14.90~$\pm$~0.03 &6354.805 & 00032666029    &  0.95 &  15.81~$\pm$~0.03 \\
6303.199 & 00032666003/2  &  0.53 &  15.26~$\pm$~0.03 &6357.950 & 00032666030    &  0.94 &  15.74~$\pm$~0.03 \\
6305.329 & 00032666004/1  &  1.80 &  10.87~$\pm$~0.02 &6364.289 & 00032666032    &  0.99 &  15.83~$\pm$~0.03 \\
6305.395 & 00032666004/2  &  0.17 &  10.84~$\pm$~0.03 &6369.639 & 00032666034    &  1.00 &  16.06~$\pm$~0.03 \\
6307.199 & 00032666005    &  1.80 &  10.64~$\pm$~0.02 &6372.710 & 00032666035    &  1.06 &  15.97~$\pm$~0.03 \\
6309.406 & 00032666006/1  &  1.04 &  --               &6376.443 & 00032666036    &  1.11 &  16.08~$\pm$~0.03 \\
6309.472 & 00032666006/2  &  0.91 &  --               &6378.519 & 00032666037    &  1.13 &  16.01~$\pm$~0.03 \\
6310.997 & 00032666007/1  &  1.62 &  10.87~$\pm$~0.02 &6381.599 & 00032666038    &  0.90 &  16.05~$\pm$~0.04 \\
6311.391 & 00032666007/2  &  0.49 &  10.86~$\pm$~0.02 &6384.658 & 00032666039    &  1.02 &  16.21~$\pm$~0.04 \\
6313.198 & 00032666008/1  &  0.87 &  --               &6387.530 & 00032666040    &  0.91 &  16.15~$\pm$~0.04 \\
6313.411 & 00032666008/2  &  1.22 &  --               &6391.320 & 00032666041    &  1.02 &  16.20~$\pm$~0.04 \\
6315.071 & 00032666009/1  &  1.68 &  10.75~$\pm$~0.02 &6394.210 & 00032666042    &  0.94 &  16.27~$\pm$~0.04 \\
6315.133 & 00032666009/2  &  0.55 &  10.75~$\pm$~0.02 &6397.075 & 00032666043    &  0.91 &  16.23~$\pm$~0.04 \\
6317.004 & 00032666010/1  &  0.93 &  10.80~$\pm$~0.02 &6399.962 & 00032666044    &  0.99 &  16.26~$\pm$~0.04 \\
6317.071 & 00032666010/2  &  1.01 &  10.85~$\pm$~0.02 &6407.992 & 00032666046    &  1.03 &  16.28~$\pm$~0.04 \\
6318.944 & 00032666011/1  &  1.32 &  10.74~$\pm$~0.02 &6416.000 & 00032666047    &  1.05 &  16.37~$\pm$~0.04 \\
6319.007 & 00032666011/2  &  0.72 &  10.76~$\pm$~0.02 &6418.949 & 00032666048    &  0.96 &  16.40~$\pm$~0.04 \\
6322.354 & 00032666012    &  1.12 &  10.90~$\pm$~0.02 &6421.980 & 00032666049    &  1.02 &  16.41~$\pm$~0.04 \\
6324.555 & 00032666013    &  1.01 &  10.87~$\pm$~0.02 &6424.929 & 00032666050    &  0.76 &  16.45~$\pm$~0.04 \\
6327.706 & 00032666014    &  1.05 &  11.02~$\pm$~0.02 &6428.338 & 00032666051    &  1.02 &  16.41~$\pm$~0.04 \\
6330.836 & 00032666015    &  1.00 &  11.18~$\pm$~0.02 &6430.733 & 00032666052    &  0.99 &  16.51~$\pm$~0.04 \\
6333.839 & 00032666016    &  1.09 &  11.34~$\pm$~0.02 &6434.017 & 00032666053/1  &  0.82 &  16.37~$\pm$~0.04 \\
6337.197 & 00032666017    &  0.28 &  13.31~$\pm$~0.03 &6434.081 & 00032666053/2  &  0.19 &  16.53~$\pm$~0.08 \\
6338.122 & 00032666018    &  0.48 &  14.72~$\pm$~0.03 &6436.953 & 00032666054    &  1.09 &  16.52~$\pm$~0.04 \\
6338.385 & 00032666019    &  0.49 &  14.95~$\pm$~0.03 &6439.958 & 00032666055    &  1.06 &  16.60~$\pm$~0.04 \\
6338.647 & 00032666020    &  1.03 &  14.96~$\pm$~0.03 &6442.913 & 00032666056    &  0.86 &  16.44~$\pm$~0.04 \\
6339.385 & 00032666021    &  0.98 &  15.03~$\pm$~0.03 &6445.833 & 00032666057    &  0.86 &  16.58~$\pm$~0.04 \\
6339.918 & 00032666022    &  1.05 &  15.10~$\pm$~0.03 &6449.463 & 00032666058    &  0.46 &  16.56~$\pm$~0.05 \\
6340.450 & 00032666023    &  1.04 &  15.03~$\pm$~0.03 &6451.455 & 00032666059    &  0.73 &  16.54~$\pm$~0.04 \\
6341.850 & 00032666024/1  &  1.17 &  15.41~$\pm$~0.03 &6457.962 & 00032666060    &  1.03 &  16.60~$\pm$~0.04 \\
6341.919 & 00032666024/2  &  1.59 &  15.27~$\pm$~0.03 &6464.171 & 00032666062    &  1.04 &  16.63~$\pm$~0.14 \\
6341.986 & 00032666024/3  &  2.00 &  15.37~$\pm$~0.03 &6467.190 & 00032666063    &  1.04 &  16.26~$\pm$~0.04 \\
6342.045 & 00032666024/4  &  0.20 &  15.46~$\pm$~0.05 &6475.068 & 00032666065    &  1.04 &  16.54~$\pm$~0.04 \\
6342.927 & 00032666025/1  &  0.60 &  15.51~$\pm$~0.03 &6835.086 & 00032666066    &  2.95 &  17.52~$\pm$~0.04 \\
6342.983 & 00032666025/2  &  0.60 &  15.29~$\pm$~0.03 &7038.833 & 00032666068    &  1.91 &  17.63~$\pm$~0.09 \\
6345.652 & 00032666026/1  &  0.21 &  15.49~$\pm$~0.05 &7496.135 & 00032666069    &  2.06 &  17.53~$\pm$~0.09 \\
6345.930 & 00032666026/2  &  0.99 &  15.58~$\pm$~0.03 &         &                &       &                   \\

\hline
\multicolumn{8}{l}{
Note: The Obs. ID is a unique identifier given to every observation taken with {\it Swift}. The Obs. ID
       column also shows the}\\
\multicolumn{8}{l}{
snapshot number after the slash; a snapshot is a time interval spent continuously observing the same sky
       position. The $uvm2$}\\
\multicolumn{8}{l}{
column indicates the measured $uvm2$ magnitude of \SSS\ during the corresponding snapshot.}\\

\end{tabular}
\label{Tab:SwiftLog}
\end{center}
\end{table*}


\begin{table*}
\begin{center}
\caption{Log of spectroscopic observations of \SSS.}
\begin{tabular}{clccccc}
\hline\hline
 HJD Start & Telescope/      &  \l~range        & Exp.Time & Number   & Duration & Magnitude ($V$) \\
 2450000+  & Instrument      &     (\AA)        &  (s)     & of exps. &  (h)     & \\
\hline
 6297.765  & WHT / ISIS      &  3915--7105      &  600     &  5       & 2.00     & 16.0 \\
 6298.741  & WHT / ISIS      &  3915--7105      &  600     & 13       & 2.00     & 16.4 \\
 6320.942  & 2.1~m / Echelle &  3775--7350      &  600     & 12       & 2.11     & 12.4 \\
 6328.832  & Magellan / MagE &  3040--8280      & 180--300 & 22       & 1.52     & 12.7 \\
 6328.894  & 2.1~m / B\&Ch   &  3775--7275      &  180     & 60       & 3.48     & 12.7 \\
 6329.893  & 2.1~m / B\&Ch   &  3775--7275      &  180     & 44       & 3.37     & 12.7 \\
 6330.946  & 2.1~m / B\&Ch   &  4555--6930      &  300     & 38       & 2.06     & 12.8 \\
 6331.880  & 2.1~m / B\&Ch   &  3720--7220      &  300     & 1        & 0.08     & 12.8 \\
 6333.871  & 2.1~m / B\&Ch   &  3720--7220      &  300     & 1        & 0.08     & 12.9 \\
 6334.872  & 2.1~m / B\&Ch   &  3720--7220      &  300     & 1        & 0.08     & 13.0 \\
 6335.984  & 2.1~m / B\&Ch   &  4465--5685      &  600     & 12       & 1.90     & 13.5 \\
 6336.902  & 2.1~m / B\&Ch   &  5700--6820      &  600     & 18       & 3.66     & 14.6 \\
 6372.715  & VLT / X-Shooter &  3000--24760     & 130--160 & 36       & 1.89     & 17.8 \\
 6417.697  & 2.1~m / B\&Ch   &  3780--6145      &  600     & 12       & 1.89     & 18.2 \\
 7148.483  & VLT / FORS2     &  3830--10100     &  300     &  4       & 0.46     & 19.0 \\
 7427.959  & 2.1~m / B\&Ch   &  4330--6700      & 1200     &  3       & 1.01     & 19.0 \\
 7428.933  & 2.1~m / B\&Ch   &  4330--6700      &  600     & 13       & 2.24     & 19.0 \\

\hline

\multicolumn{7}{l}{
Note: The Magnitude column indicates the approximate magnitude of \SSS\ during the observation.}\\
\multicolumn{7}{l}{
They are taken from simultaneous or interpolated from recent photometric observations.}\\

\end{tabular}
\label{Tab:SpecObs}
\end{center}
\end{table*}

\label{lastpage}
\end{document}